\documentclass[usenatbib,fleqn]{mnras}

\usepackage{newtxtext,newtxmath}

\usepackage[T1]{fontenc}
\usepackage{ae,aecompl}

\usepackage{graphicx}	
\usepackage{amsmath}	

\usepackage{amssymb}	
\usepackage{bm}
\usepackage{enumitem}
\usepackage{fontawesome}
\usepackage{siunitx}
\usepackage[mathlines]{lineno}
\usepackage{longtable}

\newcommand{\Ho}{$H_0$}
\newcommand{\Ds}{D_{\rm s}}
\newcommand{\Dds}{D_{\rm ds}}
\newcommand{\Dd}{D_{\rm d}}
\newcommand{\Ddt}{D_{\Delta t}}
\newcommand{\vd}{\sigma_{\rm los}}
\newcommand{\rmd}{\mathrm{d}}
\newcommand{\lcdm}{$\Lambda\text{CDM}$}
\newcommand{\Hunit}{km s\textsuperscript{$-$1} Mpc\textsuperscript{$-$1}}
\newcommand{\Hval}{$74.2_{-3.0}^{+2.7}$}
\newcommand{\lensname}{DES J0408$-$5354}

\newcommand{\Ddtval}{$3382_{-115}^{+146}$}
\newcommand{\Ddval}{$1711_{-280}^{+376}$}

\newcommand{\edit}[0]{}
\newcommand{\editn}[1]{#1}
\newcommand{\editnn}[1]{#1}
\newcommand{\editt}[1]{#1}
\newcommand{\editth}[1]{#1}
\newcommand{\editfr}[1]{#1}
\newcommand{\editfv}[1]{#1}
\newcommand{\editsx}[1]{#1}
\newcommand{\editref}[0]{}

\usepackage{etoolbox}

\usepackage{eso-pic}

\AddToShipoutPictureBG*{%
  \AtPageUpperLeft{%
    \hspace{0.75\paperwidth}%
    \raisebox{-3.5\baselineskip}{%
      \makebox[0pt][l]{\textnormal{DES 2019-0475}}
}}}%

\AddToShipoutPictureBG*{%
  \AtPageUpperLeft{%
    \hspace{0.75\paperwidth}%
    \raisebox{-4.5\baselineskip}{%
      \makebox[0pt][l]{\textnormal{FERMILAB-PUB-19-523}}
}}}%

\makeatletter

\patchcmd{\NAT@citex}
  {\@citea\NAT@hyper@{%
     \NAT@nmfmt{\NAT@nm}%
     \hyper@natlinkbreak{\NAT@aysep\NAT@spacechar}{\@citeb\@extra@b@citeb}%
     \NAT@date}}
  {\@citea\NAT@nmfmt{\NAT@nm}%
   \NAT@aysep\NAT@spacechar\NAT@hyper@{\NAT@date}}{}{}

\patchcmd{\NAT@citex}
  {\@citea\NAT@hyper@{%
     \NAT@nmfmt{\NAT@nm}%
     \hyper@natlinkbreak{\NAT@spacechar\NAT@@open\if*#1*\else#1\NAT@spacechar\fi}%
       {\@citeb\@extra@b@citeb}%
     \NAT@date}}
  {\@citea\NAT@nmfmt{\NAT@nm}%
   \NAT@spacechar\NAT@@open\if*#1*\else#1\NAT@spacechar\fi\NAT@hyper@{\NAT@date}}
  {}{} 

\makeatother

\title[${\rm H_0}$ from strong lens system \lensname]{STRIDES: a 3.9 per cent measurement of the Hubble constant from the strong lens system \lensname}

\author[A. J. Shajib et al.]
{\parbox{\textwidth}
{
A. J. Shajib,$^{1}$\thanks{E-mail: ajshajib@astro.ucla.edu} 
S. Birrer,$^{1, 2}$
T. Treu,$^{1, 3}$
A. Agnello,$^{4}$
E. J. Buckley-Geer,$^{5}$
J. H. H. Chan,$^{6}$
L. Christensen,$^{4}$
C. Lemon,$^{6}$
H. Lin,$^{5}$
M. Millon,$^{6}$ 
J. Poh,$^{7, 8}$
C. E. Rusu,$^{9}$
D. Sluse,$^{10}$
C. Spiniello,$^{11,12}$
G. C.-F. Chen,$^{13}$
T. Collett,$^{14}$
F. Courbin,$^{6}$
C. D. Fassnacht,$^{13}$
J. Frieman,$^{5,8}$
A. Galan,$^{6}$
D. Gilman,${}^{1}$
A. More,$^{59}$
T. Anguita,$^{15}$
M. W. Auger,$^{60,61}$
V. Bonvin,$^{6}$
R. McMahon,$^{60,61}$
G. Meylan,$^{6}$
K. C. Wong,$^{16}$
T.~M.~C.~Abbott,$^{17}$
J.~Annis,$^{5}$
S.~Avila,$^{18}$
K.~Bechtol,$^{19,20}$
D.~Brooks,$^{21}$
D.~Brout,$^{22}$
D.~L.~Burke,$^{23,24}$
A.~Carnero~Rosell,$^{25,26}$
M.~Carrasco~Kind,$^{27,28}$
J.~Carretero,$^{29}$
F.~J.~Castander,$^{30,31}$
M.~Costanzi,$^{32,33}$
L.~N.~da Costa,$^{26,34}$
J.~De~Vicente,$^{25}$
S.~Desai,$^{35}$
J.~P.~Dietrich,$^{36,37}$
P.~Doel,$^{21}$
A.~Drlica-Wagner,$^{5,8}$
A.~E.~Evrard,$^{38,39}$
D.~A.~Finley,$^{5}$
B.~Flaugher,$^{5}$
P.~Fosalba,$^{30,31}$
J.~Garc\'ia-Bellido,$^{40}$
D.~W.~Gerdes,$^{38,39}$
D.~Gruen,$^{23,24,41}$
R.~A.~Gruendl,$^{27,28}$
J.~Gschwend,$^{26,34}$
G.~Gutierrez,$^{5}$
D.~L.~Hollowood,$^{42}$
K.~Honscheid,$^{43,44}$
D.~Huterer,$^{39}$
D.~J.~James,$^{45}$
T.~Jeltema,$^{42}$
E.~Krause,$^{46}$
N.~Kuropatkin,$^{5}$
T.~S.~Li,$^{5,8,47,48}$
M.~Lima,$^{26,49}$
N.~MacCrann,$^{43,44}$
M.~A.~G.~Maia,$^{26,34}$
J.~L.~Marshall,$^{50}$
P.~Melchior,$^{47}$
R.~Miquel,$^{29,51}$
R.~L.~C.~Ogando,$^{26,34}$
A.~Palmese,$^{5}$
F.~Paz-Chinch\'{o}n,$^{27,28}$
A.~A.~Plazas,$^{47}$
A.~K.~Romer,$^{52}$
A.~Roodman,$^{23,24}$
M.~Sako,$^{22}$
E.~Sanchez,$^{25}$
B.~Santiago,$^{26,53}$
V.~Scarpine,$^{5}$
M.~Schubnell,$^{39}$
D.~Scolnic,$^{54}$
S.~Serrano,$^{31,55}$
I.~Sevilla-Noarbe,$^{25}$
M.~Smith,$^{56}$
M.~Soares-Santos,$^{57}$
E.~Suchyta,$^{58}$
G.~Tarle,$^{39}$
D.~Thomas,$^{14}$
A.~R.~Walker,$^{17}$
Y.~Zhang$^{5}$
} \\
\vspace{0.0cm} \\
\parbox{\textwidth}
{Affiliations are listed at the end of the paper
}}

\date{Accepted 2020 March 18. Received 2020 February 24; in original form 2019 October 14}

\pubyear{2019}

\begin{document}

\label{firstpage}
\pagerange{\pageref{firstpage}--\pageref{lastpage}}
\maketitle

\begin{abstract}
We present a blind time-delay cosmographic analysis for the lens system \lensname. \editt{This system is extraordinary for the presence of two sets of multiple images at different redshifts, which provide the opportunity to obtain more information at the cost of increased modelling complexity with respect to \editfv{previously analysed} systems.} We perform detailed modelling of the mass distribution for this lens system using three band \textit{Hubble Space Telescope} imaging. We combine the measured time delays, line-of-sight central velocity dispersion of the deflector, and statistically constrained external convergence with our lens models to estimate two cosmological distances. We measure the ``effective'' time-delay distance corresponding to the redshifts of the deflector and the lensed quasar $\Ddt^{\rm eff}=$ \Ddtval\ Mpc and the angular diameter distance to the deflector $\Dd=$ \Ddval\ Mpc, \editfv{with covariance between the two distances}. From these constraints on the cosmological distances, we infer the Hubble constant \Ho = \Hval\ \Hunit\ assuming a flat \lcdm\ cosmology \editfv{and a uniform prior for $\Omega_{\rm m}$ as $\Omega_{\rm m} \sim \mathcal{U}(0.05, 0.5)$}. \editfv{This measurement gives the most precise constraint on \Ho\ to date from a single lens.} Our measurement is consistent with \editsx{that obtained from} the previous sample of six lenses \editsx{analysed} by the \editth{\Ho\ Lenses in COSMOGRAIL's Wellspring (H0LiCOW)} collaboration. \editt{It is also consistent with measurements of \Ho\ based on the local distance ladder, reinforcing the tension with the inference from early Universe probes, for example, \editsx{with 2.2$\sigma$ discrepancy} from the cosmic microwave background measurement.} 
\end{abstract}


\section{Introduction}

The concordance $\Lambda$ cold dark matter (\lcdm) cosmology explains the accelerated expansion of the Universe by incorporating the cosmological constant $\Lambda$ \citep{Riess98, Perlmutter99}. The \lcdm\ model is very successful in predicting observations covering a \edit{large} range of physical scales -- from the scale of sound horizon at the recombination epoch, down to the structure formation at the megaparsec scale \citep[e.g.][]{Alam17, PlanckCollaboration18, Abbott18b}. The Hubble constant, \Ho, plays a central role in cosmology\edit{, including \editfv{in} the \lcdm\ model}. The Hubble constant is not only crucial to determine the age of the Universe, it also normalizes the distances to distant galaxies. As a result, \editsx{a precise} understanding of the galaxy formation and evolution, and the Universe as a whole, closely depends on the precise knowledge of the Hubble constant.

\editt{Recently, a \editfv{significant} tension has been reported between the measurements of \editfv{the} Hubble constant using early-Universe and late-Universe probes \citep[e.g.][]{PlanckCollaboration18, Riess19, Wong19}}. Among others, the most precise \edit{constraints on the Hubble constant come from extrapolating the cosmic microwave background (CMB) observation at the early-Universe, and from} the measurement based on the cosmic distance ladder calibrated with parallax distances, Cepheids, and type Ia supernovae (SNIae). Assuming a \lcdm\ cosmology, the \textit{Planck} measurement gives $H_0 = 67.4 \pm 0.5$ \Hunit \citep{PlanckCollaboration18}. The \editn{Supernovae, $H_0$, for the Equation of State of dark energy (SH0ES) team} measures $H_0 = 74.03 \pm 1.42$ \Hunit\ by calibrating the SNIa distance ladder using Cepheids and parallax distances \citep{Riess19}. These two measurements are at 4.4$\sigma$ tension. \editfv{A} cosmic distance ladder measurement \editn{from the Carnegie--Chicago Hubble project} calibrated by the tip of the red giant branch (TRGB) stars reports $H_0 = 69.8 \pm 1.9$ \Hunit, consistent with both of the above values on opposite sides \citep{Freedman19}. \editn{However, the SH0ES team finds $H_0=72.4\pm1.9$ \Hunit using the TRGB stars to calibrate the SNIae distance ladder \citep{Yuan19}.} Additional probes, all consistent with the tension at varying degrees of significance are summarized by \citet{Verde19}. This tension between the early-Universe and the late-Universe probes can be due to unknown systematics in any or all of the probes. However, if systematics can be ruled out as the source of this tension, then this tension would require extension of the \lcdm\ model. \editt{In order to reach a conclusion on the tension and whether new physics is needed, it is paramount to have multiple independent measurements of the Hubble constant, \editth{each with sufficient precision on its own to resolve the discrepancy ($<2$ per cent). In parallel it is also crucial to investigate in detail all possible sources of systematic uncertainties in each method.}}

Time-delay cosmography measures \Ho\ and other cosmological parameters independently of both the CMB or other high-redshift observations and the local probes such as the ones  using the cosmic distance ladder \citep{Refsdal64}. The time-delay between the arrival time of photons at multiple images of a strong-lensing system (hereafter, lens) depends on the three angular diameter distances -- between the observer and the deflector, between the deflector and the source, and between the observer and the source. A combination of these three angular diameter distances gives the \editn{so-called ``time-delay distance'' \citep{Suyu10}. This} time-delay distance is inversely proportional to \Ho\ and thus measuring this distance \editsx{directly} constrains \Ho.

To measure the time delay between the arrivals of photon at different lensed images \editn{that were emitted at the same time}, we require a time-variable source. Although \citet{Refsdal64} originally proposed using strongly lensed supernovae as a time-variable source to measure the time-delay, only a few such supernovae have been discovered so far \citep[e.g.][]{Kelly15, Goobar17, Grillo18}. \editth{Even though the number of lensed supernova is still too small to be a competitive cosmological probe, the \editsx{re-appearance} of supernova Refsdal as predicted provides an important validation of the method \citep{Treu16b}.}

Strongly lensed quasars have provided time-variable sources in larger numbers. As a result, these objects have been predominantly used to measure \Ho\ from their time-delays \citep[e.g.][]{Schechter97, Treu02b, Suyu10}. Although some of the early measurements had shortcomings in the data quality or the analysis technique, both of these aspects have tremendously improved over the past decade \citep[for a review with historical perspective, see][]{Treu16}. \editth{The key breakthroughs in the past two decades have been: \textbf{(i)} high cadence monitoring to determine the time delays \citep[e.g.][]{Fassnacht02, Tewes13}, \textbf{(ii)} high resolution images of the \editsx{lensed arcs from the} quasar host galaxy \editref{and pixel-based lens modelling} to constrain the \editsx{lens mass distribution} \citep{Suyu10}, \textbf{(iii)} adding stellar kinematics  of the deflector \citep{Treu02b}, and \textbf{(iv)} statistical analysis of the line of sight to constrain the external convergence \citep{Suyu10, Greene13, Rusu17}.} \editfv{Implementing these improvements}, the \editth{\Ho\ Lenses in COSMOGRAIL's Wellspring (H0LiCOW)} collaboration measure $H_{0} = 73.3_{-1.8}^{+1.7}$ \Hunit\ using six lens systems \citep{Suyu10, Suyu13, Suyu14, Wong17, Bonvin17, Birrer19, Rusu19, Chen19, Wong19}.

To reach 1 per cent precision in the Hubble constant with time-delay cosmography, a sample of $\sim$40 lenses is necessary \citep{Shajib18}. \edit{To have such a large sample of \editsx{strongly lensed quasars} available in the first place, the STRong-lensing Insights into Dark Energy Survey collaboration \citep[\href{http://strides.astro.ucla.edu/}{STRIDES};][]{Treu18} has discovered numerous new lenses from the Dark Energy Survey (DES) footprint, in cases combining data from other large-area sky surveys \citep[e.g.][]{Agnello15b, Nord16, Ostrovski17, Agnello18d, Anguita18, Lemon19}. The STRIDES is an external collaboration of the DES. The DES data \editsx{are} particularly useful in discovering new lenses due to its combination of uniform depth and coverage of area in the Southern hemisphere that is not covered by the Sloan Digital Sky Survey (SDSS).} \editref{Additionally, thanks to new data mining and machine learning based techniques, new lenses have been discovered also from other photometric surveys -- such as the VLT Survey Telescope-ATLAS (VST-ATLAS), Kilo-Degree Survey Strongly lensed Quasar Detection project (KiDS-SQuaD) \citep[e.g.,][]{Agnello15, Agnello18b, Spiniello18}.}

In this paper, we \editsx{present a blind analysis of} \editfv{the} lens system \lensname\ and infer \Ho\ from its time delays \edit{\citep{Courbin18}}. \editfv{This lens was discovered in the DES footprint \citep{Lin17, Diehl17}.} \editth{This paper \editsx{sets down two goals underlying our analysis}. First, we aim to increase the statistical precision of the \Ho\ determination by presenting \editsx{results from} the analysis of a new lens system. Second, this system is being analysed independently and in parallel by two teams using two different codes in order to estimate potential systematics arising from modelling choices and software.} \editth{This paper presents the first of these two independent and blind analyses for \lensname. To facilitate meaningful comparison between independent modelling teams, the participating teams agreed beforehand on a set of baseline models with minimal but sufficient specifications. The teams are free to extend on the baseline models for exploring different sources of systematics as they see fit. This additional exploration by a team proceeds independently while keeping the cosmographic inferences blind. In this way, we aim to check on systematics that can potentially arise from different codes through comparison of the baseline models from different teams, and also from different model choices within one team's analysis. In this paper, we infer \Ho\ using the lens modelling software \textsc{lenstronomy}, which is publicly available online at Github\footnote{\faGithub\ \url{https://github.com/sibirrer/lenstronomy}} \editfv{\citep{Birrer15, Birrer18}}. \editfv{A second independent team uses the software \textsc{glee} to analyse the same lens system \citep{Suyu10b}}. In a future paper, cosmographic inference based on \edit{this second analysis} and a comparison between the two analyses will be presented (Y\i ld\i r\i m et al., in preparation). \editn{Both of the independent modelling works \editsx{use} the results from a companion paper, which analyses the lens environment to detect galaxy groups and estimate the external convergence using the DES data, and measure the stellar kinematics of the central deflector galaxy from spectroscopic observations \citep{BuckleyGeer20}.}}

\editth{Our concerted effort to analyse a system independently but based on the same data, and with some overlap in modelling choices, is an important step forward in estimating the modelling errors with respect to previous works.} \editth{Previous efforts by the H0LiCOW and Strong-lensing High Angular Resolution Programme (SHARP) collaborations took some step in this direction by assigning different lead investigators and softwares to the analysis of the six lenses \editref{\citep{Lagattuta12, Suyu17}.}}
\editth{
The lens systems B1608+656, RXJ1131--1231, HE 0435--1223, WFI2033--4723, and PG 1115+080 were analysed using the lens modelling software \textsc{glee}, whereas the systems \editfv{RXJ1131--1231 and} SDSS 1206+4332 \editfv{were} analysed using the software \textsc{lenstronomy} \editfv{\citep{Suyu10, Suyu13, Birrer16, Birrer19, Wong17, Rusu19, Chen19}}. In total, four \editfv{different lead} investigators modelled these six lenses, \editfv{even though there was overlap between the team members}. The two softwares used in the modelling differ in various aspects. For example, \textsc{lenstronomy} performs source-reconstruction using a basis set of shapelets, whereas \textsc{glee} performs a pixelized source-reconstruction with regularization. \textsc{lenstronomy} is a publicly available open-source software, whereas \textsc{glee} is not.}

 
In order to preserve the blindness of the analysis, this paper and the companion describing the analysis of the environment and line of sight used to compute the external convergence were internally reviewed by the \editsx{STRIDES collaboration and the DES strong-lensing working group} before unblinding.  Once both the analyses and manuscripts met the approval of the internal reviewers and co-authors, \editsx{unblinding} happened on 2019 September 25. \editsx{After unblinding,} the only changes to the manuscript were the addition of the unblinded measurements, \editsx{discussion on the unblinded results in Section~\ref{sec:discussion}}, minor editing for clarity, grammar, and typos \editsx{after the DES collaboration-wide review}, and the addition of the plot showing the galaxy group's convergence described in Appendix \ref{app:impact_group}.

This paper is organized as follows. In Section \ref{sec:analysis_framework}, we \edit{lay out} the necessary formalism and describe the analysis framework. We present the data sets used in our analysis in Section \ref{sec:datasets}. Next in Section \ref{sec:model_ingredients}, we describe the different mass and light profiles that are used in the lens modelling. We present the various lens model choices in Section \ref{sec:model_choices}. We report the results from the lens modelling and the cosmographic inference in Section \ref{sec:results}. Finally, we discuss the results and summarize the paper in Section \ref{sec:discussion}. \editref{We provide summaries of the uncertainty budget in our inferred \Ho, systematic checks, adopted models and parameter priors in Appendices \ref{app:systematic_summary} and \ref{app:param_prior}.} \editref{The reported uncertainties in this paper are computed from 16\textsuperscript{th} and 84\textsuperscript{th} percentiles of the posterior probability distribution.}

\section{Framework of the cosmographic analysis} \label{sec:analysis_framework}
In this section, we outline our cosmographic analysis using strong-lensing time delays. We briefly lay out the strong-lensing time-delay formalism in Section \ref{sec:tdsl}, discuss the lensing degeneracies in Section \ref{sec:msd}, present an overview of the kinematic analysis in Section \ref{sec:kinematic_analysis}, describe the cosmological analysis in Section \ref{sec:cosmo_analysis}, and formalize the underlying Bayesian inference framework of our \editfv{analysis} in Section \ref{sec:bayesian_framework}.

\subsection{Strong-lensing time delay} \label{sec:tdsl}
\editref{The framework described in this subsection was developed in previous studies -- e.g., see \citet{Schneider92, Blandford92} -- and was applied in previous studies to the measure \Ho\ from time delays \citep[e.g.,][]{Suyu10, Wong17, Birrer19}.}

The time delay $\Delta t_{\rm XY}$ \editsx{between} arrival of photons \editsx{at} two images, indexed with X and Y, of a multiply-imaged quasar by a single deflector is given by
\begin{linenomath}\begin{equation} \label{eq:time_delay}
	\begin{split}
	\Delta t_{\rm XY} &= \frac{1+z_{\rm d}}{c} \frac{\Dd \Ds }{\Dds} \left[ \frac{(\btheta_{\rm X} - \bbeta)^2}{2} - \frac{(\btheta_{\rm Y} - \bbeta)^2}{2} - \psi(\btheta_{\rm X}) + \psi(\btheta_{\rm Y})  \right].
	\end{split}
\end{equation}\end{linenomath}
Here, the three angular diameter distances are $\Dd$: between the observer and the deflector, $\Ds$: between the observer and the source, and $\Dds$: between the deflector and the source. Additionally, $z_{\rm d}$ is the deflector redshift, \editfv{$c$ is the speed of light}, $\btheta$ is the image position, $\bbeta$ is the source position, and $\psi$ is the deflection potential. The deflection potential is defined such \editsx{that} its gradient gives the deflection field $\balpha \equiv \nabla \psi$. Then, the deflection potential relates to the convergence $\kappa$ as $\kappa = \nabla^2 \psi / 2$. We define the Fermat potential $\phi$ as
\begin{linenomath}\begin{equation} \label{eq:single_plane_fermat_pot}
	\phi(\btheta) \equiv \frac{(\btheta - \bbeta)^2}{2} - \psi(\btheta),
\end{equation}\end{linenomath}
and the time-delay distance as
\begin{linenomath}\begin{equation} \label{eq:single_plane_Ddt}
	\Ddt \equiv (1+z_{\rm d}) \frac{\Dd \Ds }{\Dds}.
\end{equation}\end{linenomath}
Then, we can express equation (\ref{eq:time_delay}) in a more compact form as
\begin{linenomath}
\begin{equation}
	\Delta t_{\rm XY} = \frac{\Ddt}{c} \left( \phi(\btheta_{\rm X})  - \phi(\btheta_{\rm Y}) \right) \equiv \frac{\Ddt}{c} \Delta \phi_{\rm XY} .
\end{equation}
\end{linenomath}

If multiple deflectors \editfv{are} \edit{present at close angular proximity} at different redshifts, then we need to use the multilens-plane formalism \edit{to describe the lensing effect with sufficient accuracy}. The time delay between two images for the case of lensing with $P$ lens planes can be obtained by tracing the lensed light-ray backward from the image plane to the source plane as
\begin{linenomath}\begin{equation}
	\begin{split}
		\Delta t_{\rm XY} &= \sum_{i=1}^P \frac{D_{\Delta t, i,i+1}}{c} 
		\left[ \frac{(\btheta_{{\rm X},i} - \btheta_{{\rm X},i+1})^2}{2} - \frac{(\btheta_{{\rm Y},i} - \btheta_{{\rm Y},i+1})^2}{2} \right. \\
		& \qquad \qquad \qquad \qquad \left. - \zeta_{i,i+1} \left\{ \psi_{i}(\btheta_{{\rm X},i}) - \psi_{i}(\btheta_{{\rm Y}, i}) \right\}  \right]
	\end{split}
\end{equation}\end{linenomath}
(cf. equation 9.17 of \citealt{Schneider92}). Here, the first lens plane is the nearest to the observer and the $(P+1)$-th plane refers to the source plane. The time-delay distance $D_{\Delta t,i,j}$ between a pair of planes is defined as
\begin{linenomath}\begin{equation}
	D_{\Delta t,i,j} \equiv \frac{1 + z_{i}}{c}\frac{D_i D_j}{D_{ij}}, \quad i < j,
\end{equation}\end{linenomath}
where $D_{i}$ is the angular diameter distance from the observer to the $i$th plane and $D_{ij}$ is the angular diameter distance between the $i$th and $j$th planes. The rescaling factor $\zeta_{i,j}$ is defined as
\begin{linenomath}\begin{equation}
	\zeta_{i, j} \equiv \frac{D_{ij}D_{\rm s}}{D_j D_{i{\rm s}}}, \quad i < j.
\end{equation}\end{linenomath}
In this \editfv{multilens-plane} case, we can define the time-delay distance between the central deflector plane and the source plane as the effective time-delay distance $D_{\Delta t}^{\rm eff} \equiv D_{\Delta t,{\rm d},{\rm s}}$ that normalizes the multilens-plane time delay as
\begin{linenomath}\begin{equation} \label{eq:multi_time_delay}
\begin{split}
		\Delta t_{\rm XY} &= \frac{D_{\Delta t}^{\rm eff}}{c}	 \sum_{i=1}^P \frac{1+z_i}{1+z_{\rm d}} \frac{D_i D_{i+1} D_{\rm ds}}{D_{\rm d} D_{\rm s}D_{i\ i+1} } 
			\left[ \frac{(\btheta_{{\rm X},i} - \btheta_{{\rm X},i+1})^2}{2} \right. \\ 
			& \qquad \qquad \qquad \qquad - \frac{(\btheta_{{\rm Y},i} - \btheta_{{\rm Y},i+1})^2}{2}  \\
			& \qquad \qquad \qquad \qquad \left. - \zeta_{i,i+1} \left\{ \psi_{i}(\btheta_{{\rm X},i}) - \psi_{i}(\btheta_{{\rm Y}, i}) \right\}  \right]\\
			&= \frac{D_{\Delta t}^{\rm eff}}{c} \Delta \phi_{\rm XY}^{\rm eff}.	
	\end{split}
\end{equation}\end{linenomath}
Here, we defined the effective Fermat potential for the multilens-plane case as
\begin{linenomath}\begin{equation} \label{eq:multi_fermat_potential}
	\begin{split}
		\phi^{\rm eff} (\btheta) &\equiv \sum_{i=1}^{P} \frac{1+z_i}{1+z_{\rm d}} \frac{D_i D_{i+1} D_{\rm ds}}{D_{\rm d} D_{\rm s}D_{i\ i+1} } \left[ \frac{(\btheta_{i} - \btheta_{ i+1})^2}{2} - \zeta_{i,i+1} \psi_{i}(\btheta_{i})  \right].
	\end{split}
\end{equation}\end{linenomath}
In equation (\ref{eq:multi_time_delay}), the effective Fermat potential difference $\Delta \phi^{\rm eff}_{\rm XY}$ only contains the distance ratios. Thus, this term does not depend on \Ho. \editfv{However, the distance ratios weakly depend on the relative expansion history, thus on the density parameter $\Omega$ in the context of \lcdm}. Only the effective time-delay distance $D_{\Delta t}^{\rm eff}$ depends on \Ho\ in equation (\ref{eq:multi_time_delay}). For the single lens plane case with $P=1$, the effective Fermat potential $\phi^{\rm eff}$ and the effective time-delay distance $D_{\Delta t}^{\rm eff}$ naturally take the form of their single-lens-plane equivalents $\phi$ and $D_{\Delta t}$ \editnn{from equations (\ref{eq:single_plane_fermat_pot}) and (\ref{eq:single_plane_Ddt})}, respectively.

\subsection{Mass-sheet degeneracy} \label{sec:msd}
\editn{For lensing, the imaging observables such as the flux ratios and the relative astrometry} are invariant with respect to the mass-sheet transformation \citep[MST;][]{Falco85}. If we transform the convergence and the source plane as
\begin{linenomath}\begin{equation} \label{eq:mst}
	\begin{split}
	&\kappa(\btheta) \to \kappa_{\vartheta} (\btheta) = \vartheta \kappa (\btheta) + 1 - \vartheta, \\
	&\bbeta \to \bbeta^\prime = \vartheta \bbeta,
	\end{split}
\end{equation}\end{linenomath}
then the lensing observables except the time delay remain invariant. This invariance under the MST is called the mass-sheet degeneracy (MSD). \editref{Notably, the MST also rescales the magnification, thus the MSD can be broken with standard candles \citep{Bertin06}.}

We can express the physically existent ``true'' mass distribution as
\begin{linenomath}\begin{equation} \label{eq:kappa_true}
	\kappa_{\rm true} = \kappa_{\rm lens} + \kappa_{\rm ext}	,
\end{equation}\end{linenomath}
where, $\kappa_{\rm lens}$ is the convergence of the central deflector including nearby satellites and perturbers, and $\kappa_{\rm ext}$ is the convergence from projecting all the line-of-sight inhomogeneities onto the lens plane. If we impose the condition $\lim_{\theta \to \infty} \kappa_{\rm lens} = 0$, then we have $\lim_{\theta \to \infty} \kappa_{\rm true} = \kappa_{\rm ext}$. As a result, we can interpret the external convergence $\kappa_{\rm ext}$ as the convergence far from or ``external'' to the central deflector. However, as we cannot constrain $\kappa_{\rm ext}$ only from the lensing observables due to the MSD, we aim to constrain a model $\kappa_{\rm model}'$ that captures all the lensing effects of $\kappa_{\rm true}$. By taking $\vartheta = 1/(1-\kappa_{\rm ext})$ in equation (\ref{eq:mst}), we can obtain an MST of $\kappa_{\rm true}$ as
\begin{linenomath}\begin{equation} \label{eq:kappa_model_prime}
	\kappa_{\vartheta} = 	\frac{1}{1-\kappa_{\rm ext}}(\kappa_{\rm lens} + \kappa_{\rm ext}) - \frac{\kappa_{\rm ext}}{1 - \kappa_{\rm ext}} = \frac{\kappa_{\rm lens}}{1 - \kappa_{\rm ext}} = \kappa'_{\rm model}.
\end{equation}\end{linenomath}
Here, we name this $\kappa_{\vartheta}$ as $\kappa'_{\rm model}$ because it captures all the lensing effect of $\kappa_{\rm true}$ by the virtue of MST. If we can constrain $\kappa_{\rm model}'$, then we can obtain $\kappa_{\rm true}$ simply through a MST with $\vartheta = 1 - \kappa_{\rm ext}$ where $\kappa_{\rm ext}$ is separately constrained by studying the lens environment. However, the lens model $\kappa_{\rm model}$ that we actually constrain can potentially be an internal MST of $\kappa_{\rm model}'$ given by
\begin{linenomath}\begin{equation} \label{eq:kappa_model}
	\kappa_{\rm model}' = \vartheta_{\rm int} \kappa_{\rm model} + 1 - \vartheta_{\rm int}.
\end{equation}\end{linenomath}
The internal MST factor $\vartheta_{\rm int}$ only changes the \editn{shape} of the mass profile, but it does not add any physical mass to the model within the Einstein radius. Note that both $\kappa_{\rm model}$ and $\kappa^{\prime}_{\rm model}$ can satisfy $\lim_{\theta\to\infty}\kappa=0$ by construction. In that case, $\vartheta_{\rm int}$ is not a constant over the whole plane and we have the condition $\lim_{\theta\to\infty} \vartheta_{\rm int} = 1$ \citep{Schneider14}. This condition implies that $\vartheta_{\rm int}$ does not physically add an infinite background-mass-sheet. 
%
%
 With such a $\vartheta_{\rm int}$, both models $\kappa_{\rm model}$ and $\kappa_{\rm model}'$ can reproduce the lensing observables that are indistinguishable within the noise level in the data. Finally, combining equations (\ref{eq:kappa_true}), (\ref{eq:kappa_model_prime}), and (\ref{eq:kappa_model}), we write the relation between the ``true'' convergence $\kappa_{\rm true}$ and the modelled convergence $\kappa_{\rm model}$ as
 \begin{linenomath}\begin{equation}
  	\kappa_{\rm true} = (1 - \kappa_{\rm ext}) \left[\vartheta_{\rm int}\kappa_{\rm model} + 1 -\vartheta_{\rm int} \right] + \kappa_{\rm ext}.
 \end{equation}\end{linenomath}
 Using different but equally plausible model parametrizations -- e.g. power-law profile, composite profile -- we explore different model families related by equation (\ref{eq:kappa_model}). To alleviate the MSD within a model family by constraining $\vartheta_{\rm int}$, we \editth{utilize} non-lensing observables, e.g. kinematics of the deflector galaxy. Kinematics probes the 3D deprojection of $\kappa_{\rm lens}$ for a given combination of $\kappa_{\rm model}$ and $\kappa_{\rm ext}$. Moreover, \editsx{the} addition of the kinematic information also constrains the angular diameter distance to the deflector $D_{\rm d}$ \citep{Paraficz09, Jee15}. As a result, the uncertainty on the estimated \Ho\ \editth{is improved by kinematics} in two ways: 
 \begin{enumerate}[leftmargin=15pt]
 	\item by alleviating the MSD, and
 	\item by adding extra constraint on cosmology through $D_{\rm d}$ 
 \end{enumerate}
 \citep{Birrer16, Jee16, Shajib18}. In the next subsection, we outline the kinematic analysis framework.
 
\subsection{Kinematic analysis} \label{sec:kinematic_analysis}
The kinematic observable is the luminosity-weighted line-of-sight stellar velocity dispersion $\vd$. To model the 3D mass distribution consistent with the observed velocity dispersion, we adopt the spherical solution of the Jeans equations. \editref{Although the true mass distribution is non-spherical, the assumption of spherical symmetry is sufficient given the 10--25 per cent uncertainty in our kinematic data \citep[Section \ref{sec:velocity_dispersion_data};][]{Sonnenfeld12}.} We can express the spherical Jeans equation as
\begin{linenomath}\begin{equation} \label{eq:jeans}
	\frac{\rmd \left( l(r)\ \sigma_{\rm r}^2 \right)}{\rmd r} + \frac{2 \beta_{\rm ani}\ l(r) \  \sigma_{\rm r}^2}{r} = - l(r)\ \frac{\rmd \Phi}{\rmd r}.
\end{equation}\end{linenomath}
Here, $l(r)$ is the 3D luminosity density of the stars, $\sigma_{\rm r}$ is the intrinsic radial velocity dispersion, and $\beta_{\rm ani}(r)$ is the anisotropy parameter relating $\sigma_{\rm r}$ with the tangential velocity dispersion $\sigma_{\rm t}$ as
\begin{linenomath}\begin{equation}
	\beta(r) \equiv 1 - \frac{\sigma_{\rm t}^2}{\sigma_{\rm r}^2}.
\end{equation}\end{linenomath}
By solving equation (\ref{eq:jeans}), we can obtain the luminosity-weighted, line-of-sight velocity dispersion as
\begin{linenomath}\begin{equation} \label{eq:los_vel_dis}
	\vd^2(R) = \frac{2G}{I(R)} \int_R^{\infty} \mathcal{K}_{\beta} \left(\frac{r}{R} \right) \frac{l(r)\ M(r)}{r} \ \rmd r,
\end{equation}\end{linenomath}
where $M(r)$ is the enclosed mass within radius $r$ \editref{\citep[equation (A15)--(A16) of][]{Mamon05}}. Here, the function $\mathcal{K}_{\beta}(\varrho)$ depends on the parametrization of $\beta(r)$. We adopt the Osipkov--Merritt parametrization \edit{of} the anisotropy parameter given by
\begin{linenomath}\begin{equation} \label{eq:anisotropy_param_def}
	\beta_{\rm ani}(r) = \frac{r^2}{r^2 + r_{\rm ani}^2},
\end{equation}\end{linenomath}
where $r_{\rm ani}$ is the anisotropy scale radius \citep{Osipkov79,Merritt85b,Merritt85}. For this parametrization, the function $\mathcal{K}_{\beta}$ takes the form
\begin{linenomath}\begin{equation}
\begin{split}
	\mathcal{K}_{\beta} (u) &= \frac{u_{\rm ani}^2 + 1/2}{(u_{\rm ani}+1)^{3/2}}	\left( \frac{u^2+u_{\rm ani}^2}{u} \right) \tan^{-1} \left( \sqrt{\frac{u^2-1}{u^2_{\rm ani}+1}} \right)  \\
	&\qquad\qquad\qquad - \frac{1/2}{u_{\rm ani}^2 + 1} \sqrt{1 - \frac{1}{u^2}},
\end{split}	
\end{equation}\end{linenomath}
where $u_{\rm ani} = r_{\rm ani}/R$ \citep{Mamon05}.

The enclosed mass $M(r)$ is computed from the 3D mass profile. \edit{For the convergence and surface brightness profiles that cannot be straightforwardly deprojected into three dimension, we decompose them into concentric Gaussian components \citep{Bendinelli91, Emsellem94, Cappellari02, Shajib19b}. We then deproject the Gaussian components into 3D Gaussians to compute the enclosed mass $M(r)$ and 3D light density profile $l(r)$.}
 
\subsection{Cosmological distances} \label{sec:cosmo_analysis}

In this section, we effectively follow \citet{Birrer16, Birrer19} to jointly infer $\Ddt$ and $\Dd$. From the modelled convergence profile $\kappa_{\rm model}^{\prime}$ of the deflector, we derive the time-delay distance $\Ddt^{\prime}$ particular to the deflector's line of sight. We need to correct $\Ddt^{\prime}$ for the external convergence $\kappa_{\rm ext}$ to obtain the true time-delay distance $\Ddt$. From equations (\ref{eq:time_delay}) and (\ref{eq:kappa_model_prime}), we can express the true time-delay distance $\Ddt$ as
\begin{linenomath}\begin{equation} \label{eq:corrected_time_delay_distance}
	\Ddt = \frac{\Ddt^{\prime}}{1 - \kappa_{\rm ext}}.
\end{equation}\end{linenomath}
%


We can express $\sigma_{\rm los}$ in terms of parameters characterizing the 2D mass and light distributions and relevant angular diameter distances as
\begin{linenomath}\begin{equation} \label{eq:vel_dis_to_distance}
	\sigma_{\rm los}^2 = \frac{\Ds}{\Dds}\ c^2\ J(\xi_{\rm lens},\ \xi_{\rm light},\  \beta_{\rm ani}),
\end{equation}\end{linenomath}
where $\xi_{\rm lens}$ is the set of mass parameters, $\xi_{\rm light}$ is the set of light distribution parameters, $c$ is the speed of light, and the function $J$ captures all the dependencies from the mass profile, the light profile, and the orbital anisotropy \citep{Birrer16}. \editnn{The parameters in the argument of the function $J$ are expressed in angular units, thus they do not depend on the cosmology.} Then from equation (\ref{eq:time_delay}), we have
\begin{linenomath}\begin{equation}
	\frac{\Dd \Ds}{\Dds} = \frac{c\ \Delta t_{\rm XY}}{(1+z_{\rm d})\ \Delta \phi^{\rm eff}_{\rm XY}(\xi_{\rm lens})}	.
\end{equation}\end{linenomath}
Combining this equation with equation (\ref{eq:vel_dis_to_distance}), we can write
\begin{linenomath}\begin{equation}
	\Dd = \frac{c^3\ \Delta t_{\rm XY}\ J(\xi_{\rm lens}, \xi_{\rm light}, \beta_{\rm ani})}{(1+z_{\rm d})\ \sigma_{\rm los}^2\ \Delta \phi^{\rm eff}_{\rm XY}(\xi_{\rm lens}) }
\end{equation}\end{linenomath}
\citep{Birrer16}. As a result, we can estimate the angular diameter distance $\Dd$ to the deflector by combining the kinematics with the lensing observables. Therefore, we can infer two cosmological distances, $\Ddt$ and $\Dd$, at specific redshifts relevant to the lens system. Thus, we can constrain the Hubble constant and other cosmological parameters from the distance--redshift relation for a given cosmology. In the next section, we describe the combined Bayesian framework to infer the Hubble constant from the observables.

\subsection{Bayesian inference framework} \label{sec:bayesian_framework}

\editref{We tabulate the notations used in this section in Table \ref{tab:notation_list} as a quick reference for the readers.} At the top level, the two cosmological distances \editnn{$\Ddt$ and $\Dd$} contain all the cosmographic information. We express the set of cosmological distances using the notation $D$, which is a function $D(\omega; C)$ of the set of cosmological parameters $\omega$ for a given cosmology $C$. We denote the set containing all the observables as $O \equiv \{O_{\rm img}, O_{\Delta t}, O_{\rm kin}, O_{\rm env} \}$, where $O_{\rm img}$ contains the imaging data of the lens system, $O_{\Delta t }$ contains the observed time delays, $O_{\rm kin}$ contains the spectra the of the deflector to estimate the kinematics, and $O_{\rm env}$ contains photometric and spectroscopic survey data of the lens environment to estimate the external convergence. Then from Bayes' theorem, we can write
\begin{linenomath}\begin{equation} \label{eq:cosmo_posterior}
\begin{split}
	p(\omega \mid O, C) &\propto p(O \mid \omega, C)\ p(\omega \mid C) \\
	&= p \left(O \mid D(\omega; C)\right) \ p(\omega \mid C),
\end{split}
\end{equation}\end{linenomath}
where the probability density $p(\omega \mid O, C)$ is called the posterior of $\omega$, \edit{the probability density $p(O \mid \omega, C)$ is called the likelihood of $O$} given $\{\omega, C\}$, and the probability density $p(\omega \mid C)$ is called the prior for $\omega$. In the last line of the above equation, we have changed $\{\omega, C\}$ into $D(\omega; C)$ in the likelihood term, as it allows us to break down the computation of the likelihood into two steps. First, we compute the likelihood $p(O \mid D)$ of the observed data for given cosmological distances marginalizing over various model choices and their respective parameters. Then, we can fold in the prior of the cosmological parameters $p(\omega \mid C)$ to obtain the posterior $p (\omega \mid O, C)$. As the different pieces of the data \editref{in $O$} are independent, we can break up the likelihood into likelihoods of each observable type as
\begin{linenomath}\begin{equation}
	p(O \mid D) = p(O_{\rm img} \mid D) \ p(O_{\Delta t} \mid D) \ p(O_{\rm kin} \mid D) \ P(O_{\rm env} \mid D).
\end{equation}\end{linenomath}
\editnn{When computing these likelihood functions,} we adopt a combination of model choices. We denote the model choice containing the mass model parameters $\xi_{\rm lens}$ and deflector light model parameters $\xi_{\rm light}$ as $M$. In addition, we have to make specific choices for the parametrization $\xi_{\rm source}$ of the source light distribution and the parametrization $\xi_{\rm pert}$ of the mass profiles of the line-of-sight perturbers. We denote the model choice encompassing $\xi_{\rm source}$ and $\xi_{\rm pert}$ as $S$. We also marginalize over the external convergence $\kappa_{\rm ext}$ and the parameters $\xi_{\beta}$ characterizing $\beta_{\rm ani}$. Adding it all together, we can marginalize all the specific model parameters to express the total likelihood given the distances as
\begin{linenomath}\begin{equation}
\begin{split}
	&p(O \mid D, M) = \int p(O_{\rm img} \mid \xi_{\rm lens}, \xi_{\rm light}, \xi_{\rm source}, \xi_{\rm pert}, M, S) \\
		&\qquad \times p(O_{\Delta t} \mid \Delta t( \xi_{\rm lens}, \xi_{\rm light}, \xi_{\rm pert}, \kappa_{\rm ext}; M, S)) \\
	 	&\qquad \times p(\xi_{\rm source}, \xi_{\rm pert} \mid S)\ p(S)\ \\
	 	&\qquad \times p(O_{\rm kin} \mid \sigma_{\rm los} ( \xi_{\rm lens}, \xi_{\rm light}, \kappa_{\rm ext}, \xi_{\beta}; M)) \ p(\xi_{\beta} \mid \xi_{\rm lens}, \xi_{\rm light}, M) \\
	 	&\qquad \times p(O_{\rm env} \mid \kappa_{\rm ext})\ p(\kappa_{\rm ext}) \\
	 	&\qquad \times p(\xi_{\rm lens}, \xi_{\rm light} \mid M)\\
	 	&\qquad \times \rmd \xi_{\rm source}\ \rmd \xi_{\rm pert}\ \rmd S \ \rmd \xi_{\beta}\ \rmd \kappa_{\rm ext} \ \rmd \xi_{\rm lens} \ \rmd \xi_{\rm light}.
\end{split}
\end{equation}\end{linenomath}
Here, we omitted some model parameters and model specifications in the conditional statements of the likelihoods where the corresponding likelihood does not depend on them. Breaking up the likelihood \editnn{as} above allows us to partially separate the computation of the likelihoods for different observable types before marginalizing over the model parameters. We first describe the imaging likelihood and marginalization over relevant models and model parameters in Section \ref{sec:lens_model_evidence}, then we explain the derivation of the joint posterior combining time delay and kinematics likelihoods with the lens model posterior in Section \ref{sec:joint_td_kin_lens_posterior}.

\renewcommand{\arraystretch}{1.2}
\begin{table}
\caption{\label{tab:notation_list}
\editref{List of notations used in Section \ref{sec:bayesian_framework}.}
}
\begin{tabular}{ll}
	\hline
	Notation & Definition \\
	\hline
	$D$ & $\equiv\{D_{\rm D_{\Delta t}, D_{\rm d}} \}$, set of cosmological distances \\
	$C$ & a given cosmological model \\
	$\omega$ & set of cosmological parameters in $C$ \\
	$O_{\rm img}$ & imaging data of the lens systems \\
	$O_{\Delta t}$ & observed time delay \\
	$O_{\rm kin}$ & spectra of the deflector to measure the stellar kinematics \\
	$O_{\rm env}$ & data to estimate the external convergence \\
	$O$ & $\equiv\{ O_{\rm img}, O_{\Delta t}, O_{\rm kin}, O_{\rm env}\}$ \\
	$M$ & mass and light model for the central deflector \\
	$\xi_{\rm lens}$ & mass model parameters in $M$ \\
	$\xi_{\rm light}$ & light model parameters in $M$ \\
	$S$ & model specifying the source light and the perturber mass \\
	$\xi_{\rm source}$ & source light parameters in $S$ \\ 
	$\xi_{\rm pert}$ & perturber mass parameters in $S$ \\
	$\kappa_{\rm ext}$ & external convergence \\
	$\beta_{\rm ani}$ & anisotropy parameter from equation (\ref{eq:anisotropy_param_def}) \\
	$\xi_{\beta}$ & model parameters characterizing $\beta_{\rm ani}$ \\
	$\sigma_{\rm los}$ & line-of-sight velocity dispersion \\
	$\Xi$ & $\equiv \{ \xi_{\rm lens}, \xi_{\rm light}, \xi_{\rm source}, \xi_{\rm pert} \}$ \\
	$\Delta \phi^{\rm eff}_{\rm XY}$ & effective Fermat potential difference from equation (\ref{eq:multi_fermat_potential}) \\
	$\mathcal{Z}$ & Bayesian model evidence \\
	$\nu$ & non-linear parameters in $\Xi$ \\
	$\blambda$ & vector containing linear parameters in $\Xi$ \\
	$\hat{\blambda}$ & maximum likelihood estimator of $P(O_{\rm img} \mid \nu, \blambda, M, S)$ \\
	$\bm{\mathsf{K}}_{{\blambda\blambda}}$ & covariance matrix of $\blambda$ \\
	$\bm{w}$ & uniform prior width of $\blambda \sim U(-\bm{w}/2, \bm{w}/2)$  \\ 
	$\xi_{\rm micro}$ & parameters relevant to the microlensing time-delay effect \\ 
	\hline
\end{tabular}
\end{table}

\subsubsection{Lens model posterior and evidence from imaging likelihood} \label{sec:lens_model_evidence}

We can first obtain the posterior of the lens model parameters $\Xi \equiv \{ \xi_{\rm lens}, \xi_{\rm light}, \xi_{\rm source}, \xi_{\rm pert} \}$ as
\begin{linenomath}\begin{equation} \label{eq:modelling_posterior}
p(\Xi \mid O_{\rm img}, M, S) =  \frac{p(O_{\rm img} \mid \Xi, M, S)\ p(\Xi \mid M, S)}{p(O_{\rm img} \mid M, S)}.
\end{equation}\end{linenomath}
Here, the term in the denominator $\mathcal{Z} 
\equiv p(O_{\rm img} \mid M, S)$ is the evidence for \edit{the imaging data ${O}_{\rm img}$ given the model $\{M, S\}$}. We first change the variables $\{\xi_{\rm lens},\xi_{\rm pert} \} \to \{\xi_{\rm lens}, \Delta \phi^{\rm eff}_{\rm XY} \}$ in equation (\ref{eq:modelling_posterior}) to be able to marginalize over parameters related to the line-of-sight galaxies while retaining their effect on the Fermat potential difference $\Delta \phi_{\rm XY}$. As the Jacobian \editref{determinant} $\lvert \rmd \{\xi_{\rm lens}, \Delta \phi_{\rm XY}^{\rm eff} \} / \rmd \{\xi_{\rm lens},\xi_{\rm pert} \} \rvert$ cancels out from both sides, we have
\begin{linenomath}\begin{equation} \label{eq:modelling_posterior_pot}
	\begin{split}
	p(\xi_{\rm lens}, &\xi_{\rm light}, \xi_{\rm source}, \Delta \phi_{\rm XY}^{\rm eff} \mid O_{\rm img}, M, S) \\
	=&\ p(O_{\rm img} \mid \xi_{\rm lens}, \xi_{\rm light}, \xi_{\rm source}, \Delta \phi_{\rm XY}^{\rm eff}, M, S) \\
	&\ \times \frac{p(\xi_{\rm lens}, \xi_{\rm light}, \xi_{\rm source}, \Delta \phi_{\rm XY}^{\rm eff} \mid M, S)}{p(O_{\rm img} \mid M, S)}.
	\end{split}
\end{equation}\end{linenomath}
We can marginalize this posterior over $\xi_{\rm source}$ and $S$ as
\begin{linenomath}\begin{equation} \label{eq:modelling_posterior_2}
\begin{split}
	p(\xi_{\rm lens}, &\xi_{\rm light}, \Delta \phi_{\rm XY}^{\rm eff} \mid O_{\rm img}, M) \\ 
	&= \int p(\xi_{\rm lens}, \xi_{\rm light}, \xi_{\rm source}, \Delta \phi_{\rm XY}^{\rm eff} \mid O_{\rm img}, S, M) \\
	&\qquad\qquad \times p(S)\ \rmd \xi_{\rm source} \ \rmd S.
\end{split}
\end{equation}\end{linenomath}
\editsx{Since} the term inside the integral contains the evidence term $\mathcal{Z}$, the integral over the model space $S$ automatically weights the models $\{S\}$ according to their evidence ratios. \edit{As we can only discretely sample models $\{S_n\}$ from the model space $S$, the integral in equation (\ref{eq:modelling_posterior}) becomes a discrete sum as
\begin{linenomath}
	\begin{equation} \label{eq:discrete_model_posterior}
	\begin{split}
		p(\xi_{\rm lens}, &\xi_{\rm light}, \Delta \phi_{\rm XY}^{\rm eff} \mid O_{\rm img}, M) \\ 
	&= \sum_{n} \Delta S_n \int p(\xi_{\rm lens}, \xi_{\rm light}, \xi_{\rm source}, \Delta \phi_{\rm XY}^{\rm eff} \mid O_{\rm img}, S_n, M) \\
	&\qquad\qquad \times p(S_n)\ \rmd \xi_{\rm source}.
	\end{split}
	\end{equation}		
\end{linenomath}
Here, the term $\Delta S_n$ can be interpreted as the model space volume represented by the model $S_n$, thus it can account for sparse sampling from the model space.
}

In our model, we have both linear and non-linear parameters. \editnn{The linear parameters are the amplitudes of the surface brightness profiles that we treat with a basis set in our model \citep{Birrer15}.} We denote the linear parameters using the vector $\blambda$ and non-linear parameters using the set $\nu$. Hence, the lens model parameters can alternatively be expressed as $\Xi \equiv \{ \nu, \blambda \}$. We can write the evidence integral as
\begin{linenomath}\begin{equation}
	\mathcal{Z} = \int p(O_{\rm img} \mid \nu, \blambda, M, S)\ p(\blambda \mid M, S)\ p( \nu \mid M, S)\ \rmd \blambda\ \rmd \nu.
\end{equation}\end{linenomath}
We can first marginalize over the linear parameters to get the likelihood $P(O_{\rm img} \mid \nu, M, S)$ in terms of only the non-linear parameters as 
\begin{linenomath}\begin{equation} \label{eq:evidence_linear_non_linear}
	p(O_{\rm img} \mid \nu, M, S) = \int p(O_{\rm img} \mid \nu, \blambda, M, S) \ p(\blambda \mid M, S) \ \rmd \blambda.
\end{equation}\end{linenomath}
If $\hat{\blambda}$ is the maximum-likelihood estimator of $P(O_{\rm img} \mid \nu, \blambda, M, S)$ for given $\{\nu, M, S\}$, then we can approximate the likelihood \editref{using up to the second-order terms in the Taylor expansion in} the vicinity of $\hat{\blambda}$ as
\begin{linenomath}\begin{equation}
\begin{split}
	p(O_{\rm img} \mid \nu, \blambda, M, S) \approx&\ p(O_{\rm img} \mid \nu, \hat{\blambda}, M, S)  \\
	&\qquad \times \exp \left[ - \frac{1}{2} \Delta \blambda^{\mathsf{T}} \bm{\mathsf{K}}_{{\blambda\blambda}}^{-1} \Delta\blambda  \right],
\end{split}
\end{equation}\end{linenomath}
where \editref{$\Delta \blambda = \blambda - \hat{\blambda}$}, and $\bm{\mathsf{K}}_{{\blambda\blambda}}$ is the covariance matrix of $\blambda$ \citep[\editref{equation [12] of}][]{Birrer15}. \editnn{We can directly obtain $\hat{\blambda}$ given the set of non-linear parameters $\nu$ by solving a set of linear equations.} If we take a uniform prior $U(-\bm{w}/2, \bm{w}/2)$ for $\blambda$, then from equation (\ref{eq:evidence_linear_non_linear}) we have
\begin{linenomath}\begin{equation}
	p(O_{\rm img} \mid \nu, M, S) \approx p(O_{\rm img} \mid \nu, \hat{\blambda}, M, S) \frac{\left[ (2\pi)^n \det \left( \bm{\mathsf{K}}_{{\blambda\blambda}}\right) \right]^{1/2}}{\prod_i^n w_i},
\end{equation}\end{linenomath}
where $n = \dim(\blambda)$ is the number of linear parameters. Then, we can express the evidence as
\begin{linenomath}\begin{equation} \label{eq:full_evidence_eqn}
	\mathcal{Z} \approx \frac{\left[ (2\pi)^n \det \left( \bm{\mathsf{K}}_{{\blambda\blambda}} \right) \right]^{1/2}}{\prod_i^n w_i} \int p(O_{\rm img} \mid \nu, \hat{\blambda}, M, S)\ p( \nu \mid M, S)\ \rmd \nu.
\end{equation}\end{linenomath}
%

\subsubsection{Joint posterior combining time delay and kinematics likelihoods} \label{sec:joint_td_kin_lens_posterior}

Next, we can fold in the time-delay likelihood to update the posterior and marginalize over the Fermat potential $\Delta \phi_{\rm XY}^{\rm eff}$ as 
\begin{linenomath}\begin{equation}
	\begin{split}
		p(D, &\xi_{\rm lens}, \xi_{\rm light}, \kappa_{\rm ext} \mid O_{\rm img}, O_{\Delta t}, O_{\rm env}, M) \\
		&\propto \int p(O_{\Delta t} \mid \Delta t(D, \xi_{\rm lens}, \xi_{\rm light}, \Delta \phi_{\rm XY}^{\rm eff}, \kappa_{\rm ext}; O_{\rm img}, O_{\rm env}, M))\\
		& \qquad \times p(\xi_{\rm lens}, \xi_{\rm light}, \Delta \phi_{\rm XY}^{\rm eff} \mid O_{\rm img}, M) \\
		& \qquad \times p(\kappa_{\rm ext} \mid O_{\rm env}) \ p(D) \ \rmd \Delta \phi_{\rm XY}^{\rm eff}.
	\end{split}
\end{equation}\end{linenomath}
\editnn{\citet{Tie18} introduce a possible microlensing time-delay effect due to the asymmetric magnification of a quasar accretion disc\editref{ -- assuming the lamppost model \citep{Shakura73} -- }due to microlensing by the foreground stars in the deflector galaxy.} \editref{Note, the time-delay measurement from the quasar light curves accounts for the long-term variation in the microlensing magnification pattern. \citet{Tie18}'s microlensing time-delay effect is due to the non-uniform weighting of the quasar accretion disc brightness by the microlensing magnification pattern, thus this effect depends on the gradient of the magnification across the accretion disc. The long-term change in the magnification pattern is not necessarily correlated with the gradient of the magnification across the accretion disc.} For the case that marginalizes over this microlensing time-delay effect, the above equation becomes
\begin{linenomath}\begin{equation}
	\begin{split}
		p(&D, \xi_{\rm lens}, \xi_{\rm light}, \kappa_{\rm ext} \mid O_{\rm img}, O_{\Delta t}, O_{\rm env}, M) \\
		&\propto \int p(O_{\Delta t} \mid \Delta t(D, \xi_{\rm lens}, \xi_{\rm light}, \xi_{\rm micro}, \kappa_{\rm ext}, \Delta \phi_{\rm XY}^{\rm eff}; O_{\rm img}, O_{\rm env}, M)) \\
		& \qquad \times p(\xi_{\rm lens}, \xi_{\rm light}, \Delta \phi_{\rm XY}^{\rm eff} \mid O_{\rm img}, M) \\
		& \qquad \times p(\kappa_{\rm ext} \mid O_{\rm env})\ p(\xi_{\rm micro}) \ p(D) \ \rmd \Delta \phi_{\rm XY}^{\rm eff}\ \rmd \xi_{\rm micro},
	\end{split}
\end{equation}\end{linenomath}
where $\xi_{\rm micro}$ is the set of parameters relevant to the microlensing time-delay effect, e.g. parameters related to the properties of the black hole and the accretion disc  \citep[][\editref{Section \ref{sec:microlensing}}]{Chen18}.

Then, we can update the posterior once again by folding in the kinematic likelihood as
\begin{linenomath}\begin{equation}
	\begin{split}
		p(D, \xi_{\rm lens},& \xi_{\rm light}, \kappa_{\rm ext}, \xi_{\beta} \mid O_{\rm img}, O_{\Delta t}, O_{\rm kin}, O_{\rm env}, M) \\
		 =&\ p(O_{\rm kin} \mid \sigma_{\rm los}(D, \xi_{\rm lens}, \xi_{\rm light}, \kappa_{\rm ext}, \xi_{\beta}; M)) \\
		 &\ \times  p(\xi_{\beta} \mid \xi_{\rm lens}, \xi_{\rm light}, M) 
		 \\
		  &\ \times p(D, \xi_{\rm lens}, \xi_{\rm light}, \kappa_{\rm ext} \mid O_{\rm img}, O_{\Delta t}, M).
	\end{split}
\end{equation}\end{linenomath}

Now, we can marginalize over the model parameters to obtain the posterior of \editnn{the cosmological distances} $D$ as
\begin{linenomath}\begin{equation}
	\begin{split}
		p(D &\mid O, M) = p(D \mid O_{\rm img}, O_{\Delta t}, O_{\rm kin}, O_{\rm env}, M) \\
		 &= \int p(D, \xi_{\rm lens}, \xi_{\rm light}, \kappa_{\rm ext}, \xi_{\beta} \mid O_{\rm img}, O_{\Delta t}, O_{\rm kin}, O_{\rm env}, M)  \\
		 & \qquad \qquad \times \rmd \xi_{\beta} \ \rmd \kappa_{\rm ext}\ \rmd \xi_{\rm lens} \ \rmd \xi_{\rm light}.
	\end{split}
\end{equation}\end{linenomath}

Finally, we can marginalize over the deflector mass model choices as
\begin{linenomath}\begin{equation}
	p(D \mid O) = \sum_{M} p(D \mid O, M)\ p(M \mid O).
\end{equation}\end{linenomath}
A particular choice of mass model $M$ breaks the MSD \citep{Schneider14}. However, we cannot ascertain that our adopted mass model choice represents the true mass distribution. As a result, we cannot weigh different mass models according to their evidence ratios as a higher evidence value may just be a fluke from breaking the MSD near a better fit of the data. Therefore, we take $p(M \mid O) = 1$ to equally weight different deflector mass model choices.

As the likelihood $p(O \mid D)$ follows the proportionality relation
\begin{linenomath}\begin{equation}
	p(O \mid D) \propto \frac{p(D \mid O)}{p(D)},
\end{equation}\end{linenomath}
we can then use the distance posterior $p(D \mid O)$ to obtain the posterior of \editsx{the cosmological parameters} $p(\omega \mid O, C)$ from equation (\ref{eq:cosmo_posterior}).

\section{The Lens System and Data Sets} \label{sec:datasets}
In this paper, we perform cosmographic analyses of the lens systems \lensname. This lens was discovered and confirmed by \citet{Lin17} from a large sample of potential galaxy--galaxy lenses \editnn{in the DES footprint} \citep{Diehl17}. \citet{Agnello17} acquired follow-up data and modelled the system presenting evidence for a faint perturber G2 near one of the quasar images, which was later confirmed by the deeper and higher resolution imaging from the \textit{Hubble Space Telescope} \editth{\citep[\textit{HST};][]{Shajib19}}. 

The necessary data sets and ancillary measurements for cosmographic analysis are
\begin{enumerate}[leftmargin=15pt]
	\item high-resolution imaging of the lens system,
	\item \editth{spectroscopy of the lens components to measure} \editnn{redshifts},
	\item measured time-delays between the images,
	\item LOS velocity dispersion of the central deflector galaxy, and
	\item estimate of the external convergence.
\end{enumerate}
Each type of data set or ancillary measurement is described in the following subsections.

\subsection{\textit{HST} imaging of the lens system}
\textit{HST} Wide-Field Camera 3 (WFC3) imaging was obtained under the program GO-15320 \citep[PI: Treu;][]{Shajib19}. The images were taken in three filters: F160W in infrared (IR), F814W and F475X in ultraviolet--visual (UVIS). For each filter, four exposures -- two short and two long -- were taken to cover the large dynamic range in brightness encompassing the bright quasar images and the fainter extended host galaxy. For the IR band, we chose a 4-point dither pattern and STEP100 readout sequence for the MULTIACCUM mode. For the UVIS bands, we adopted a 2-point dither pattern. The total exposure times for the three filters are 2196.9 s in F160W, 1428 s in F814W, and 1348 s in F475X.

The data in each band \editsx{were} reduced with the standard \textsc{astrodrizzle} package \editnn{\citep{Avila15}}. The final pixel scale after drizzling is 0.08 arcsec in the IR band, and 0.04 arcsec in the UVIS band. We estimate the background level in the reduced image from each band using \textsc{sextractor} and subtract it from the reduced image \citep{Bertin96}.

\begin{figure*}
	\includegraphics[width=0.75\textwidth]{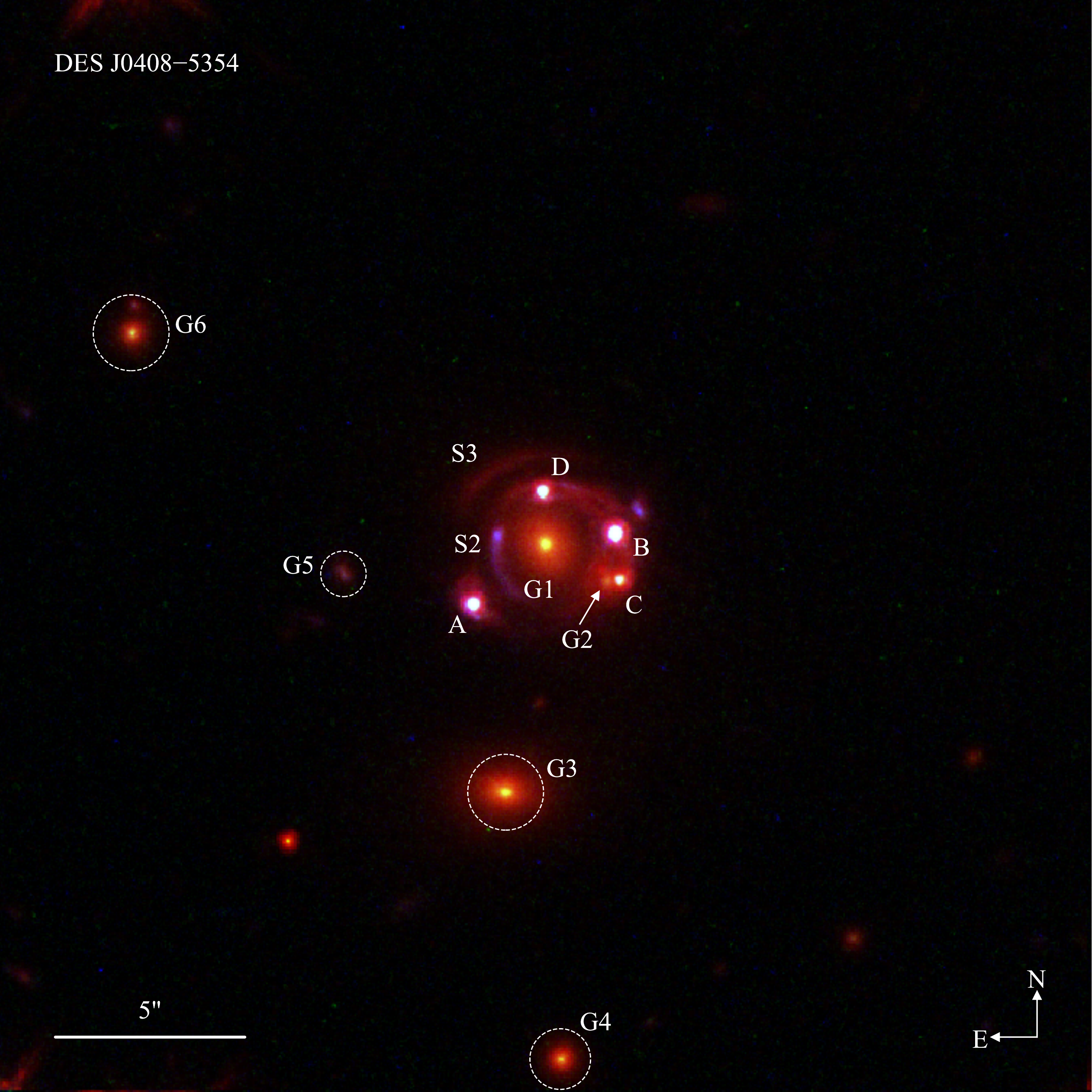} 
	\caption{ 
	RGB color composite of the lens systems \lensname. The three \textit{HST} filters used to create the RGB image are F160W (red), F814W (green), and F475X (blue). The relative amplitudes between the three filters are adjusted in this figure for better visualization by achieving a higher contrast. We label different components of the lens system. G1 is the main deflector galaxy and G2 is its satellite galaxy. In addition to the lensed arcs from the extended quasar host galaxy, this lens system has extra source components S2 and S3. The source component S2 is doubly imaged and forms an extended arc inside the Einstein radius. S3 forms another fainter extended arc on the North-East outside the Einstein radius without a noticeable counterimage. Four nearby perturbers G3--G6 along the line of sight are marked with the dashed, white circles.
	\label{fig:cutouts}
	}
\end{figure*}

Fig. \ref{fig:cutouts} shows the color-composite image for the lens system and its surrounding. 
The central deflector galaxy G1 has a visible satellite galaxy G2. The four prominent nearby galaxies along the line of sight are marked with G3, G4, G5, and G6. \editsx{Note that the naming convention of these galaxies is different in \citet{Lin17} and \citet{Agnello17}.}

The lens has multiple lensed arcs from additional source components, S2 and S3. The lensed arc S2 lies inside the Einstein radius and it has a noticeable counterimage on the North--West of image B. Another faint lensed arc S3 lies on the East of image D. We could not identify the counterimage of S3 from visual inspection.

\subsection{\editth{Spectroscopic observations of} the lens components}

The central deflector G1 sits at the redshift $z_{\rm d} = 0.597$ and the quasar sits at redshift $z_{\rm QSO} = 2.375$ \citep{Lin17}. \editnn{\citet{BuckleyGeer20} \editth{measure} redshifts for the nearby line-of-sight galaxies G3--G6 \editth{from spectroscopic observations using the Magellan and the Gemini telescopes} \editsx{obtaining} $z_{\rm G3}=0.769$, $z_{\rm G4}=0.771$, $z_{\rm G5}=1.032$, and $z_{\rm G6}=0.594$. \editref{The redshifts are precise up to the specified decimal point.}}

We \editnn{measure} the redshift of S2 $z_{\rm S2} = 2.228$ from the integral-field spectroscopy observations of \lensname\ with the Multi-Unit Spectroscopic Explorer (MUSE, on the ESO VLT UT4). The MUSE observations of the lens and its immediate neighbourhood, within approximately $45$ arcsec, were carried out in Period 102 during two nights on 2019 January 11 and 13 [run 0102.A-0600(E), PI Agnello]. The observations were executed in wide-field mode with adaptive-optics (AO) corrections, so that the multiple images and galaxies in this lens could be properly deblended. The AO wide field mode of MUSE results in a wavelength coverage from 4700 to 5803 \si{\angstrom}, and 5966--9350 \si{\angstrom} at a spectral resolution of $R\sim$ 1700--3400. Each observation block contains four exposures, with the main target placed in four different quadrants of the instrument's field of view.  An approximately 15 arcsec $\times$ 15 arcsec region centred on the lens was exposed for 4h, with a dither-and-rotation pattern that minimized artefacts due to the multiple instrument slicers and channels. We reduced the data cubes using the standard \textsc{esorex} pipeline recipes and flux calibrated \editsx{them} using observations of standard stars obtained on the two nights. Offsets between 20 individual exposures were determined from cross-correlations of white light images created from individual \editsx{data cubes}. We cleaned strong sky-line residuals from the final combined data cube using \textsc{zap} \citep{Soto16}. The setup results in a final data cube with a full field of view of 92 arcsec $\times$ 95 arcsec. For this work, we analysed a 8 arcsec $\times$ 8 arcsec `mini-cube' centred around the lens. We use three stars in the field as reference point source function (PSF) cubes. We decompose the `mini-cube' as a superposition of four Moffat profiles for the quasar images, and a convolved de Vaucouleurs profile for the deflector light distribution. By means of this procedure, all component spectra could be reliably separated and the quasar shot noise on the deflector spectra was minimized. \editref{We use Mg \textsc{ii} emission lines to measure S2's redshift and velocity dispersion. As S2 and the quasar are at different redshift, the quasar's Mg \textsc{ii} contamination does not overlap with S2's Mg \textsc{ii} lines. Also, given the large systematic uncertainty on the velocity dispersion described in the next paragraph, residual AGN contamination is not a dominant source of bias.}

\editfv{We also measure the line-of-sight velocity dispersions of \editn{G3--G6} and S2 from the MUSE spectra (Table \ref{tab:perturber_properties}). We adopt an uncertainty of 20 km s\textsuperscript{$-$1} on the measured velocity dispersion to account for the typical systematic uncertainty for kinematics extracted from MUSE spectra \citep{Guerou17}.} \editref{The estimated PSF from the stars in the MUSE observation can be different than the PSF of the quasar due to different SED within a filter. However, the impact in the estimated velocity dispersion from this potential PSF mismatch is subdominant to this conservative estimate of the systematic uncertainty.}

\renewcommand{\arraystretch}{1.5}
\begin{table}
\caption{\label{tab:perturber_properties}
	Redshift and stellar velocity dispersion for the line-of-sight galaxies G3--G6 and S2. \editsx{The relative offsets of the observed centroids for G3--G6 are computed from the coordinate RA 04:08:21.71 and Dec $-$53:53:59.34}. The tabulated uncertainties for the velocity dispersions are statistical uncertainties. However, we adopt a 20 km s\textsuperscript{$-$1} uncertainty for each measurement to account for the typical systematic uncertainty for kinematics obtained from MUSE spectra \citep{Guerou17}. \editref{The redshifts are precise up to the specified decimal point.}
	}
\begin{tabular}{lcccc}
\hline
Galaxy & $\Delta$RA & $\Delta$Dec & Redshift & \editth{Stellar} velocity dispersion \\
& (arcsec) & (arcsec) & & (km s\textsuperscript{$-$1})  \\
\hline
G3 & \phantom{-0}1.08 & \phantom{0}$-$6.52 & 0.769 & 226 $\pm$ 7\phantom{0} \\
G4 & \phantom{0}-0.40 & $-$13.58 & 0.771 & 153 $\pm$ 10  \\ 
G5 & \phantom{-0}5.34 & \phantom{0}$-$0.78 & 1.032 & \phantom{0}56 $\pm$ 2\phantom{0} \\ 
G6 & \phantom{-}10.90 & \phantom{0$-$}5.53 & 0.594 & \phantom{0}63 $\pm$ 7\phantom{0} \\
S2 & -- & -- & 2.228 & \phantom{0}46 $\pm$ 9\phantom{0} \\
\hline
\end{tabular}
\end{table}

\subsection{Time delays}

\citet{Courbin18} present the measured time delays between the images of \lensname. This system was monitored to obtain light-curves of the lensed images using the MPIA 2.2 m telescope at La Silla observatory between 2016 October 1 and 2017 April 8. The system was observed almost daily except for 14 consecutive nights between 2016 December 10 and 2016 December 24, and for one week in 2017 January due to bad weather and technical problems.  Additional monitoring was carried out using the 1.2 m the \textit{Leonhard Euler} 1.2 m Swiss Telescope (\textit{Euler}) between 2016 July and 2017 April. The mean observation cadence with \textit{Euler} is 5 d. From these light-curves of the lensed images, the measured time delays are $\Delta t_{\rm AB} = -112.1 \pm 2.1$ d, $\Delta t_{\rm AD} = -155.5 \pm 12.8$ d, and $\Delta t_{\rm BD} = -42.4 \pm 17.6$ d \editnn{(see Fig. \ref{fig:cutouts} for the naming of the images)}. 
The time delays relative to image C could not be measured due the close proximity of a satellite galaxy \edit{as} it is difficult to deblend the quasar flux from the satellite's in the ground-based monitoring data.

\subsection{Velocity dispersion \editth{of the central deflector}} \label{sec:velocity_dispersion_data}
\citet{BuckleyGeer20} \editfv{measure} the velocity dispersion of G1. The velocity dispersion is measured with four different observing setups: two mask setups with the Magellan telescope, one with the Gemini telescope, and one with the MUSE spectra. The specifics and the measured values from these four setups are tabulated in Table \ref{tab:vel_dis}. \editn{We estimate the systematic uncertainty $\sigma_{\sigma_{\rm los}}^{\rm sys}$ in the measured velocity dispersion to add the reported statistical uncertainty $\sigma_{\sigma_{\rm los}}^{\rm stat}$. We \editsx{infer a systematic uncertainty of} \editfr{17 km s\textsuperscript{$-$1}} from the \editfr{variance} in the estimated velocity dispersions when different settings -- e.g. the stellar population library, the stellar templates, \editnn{the wavelength region} -- are varied in the kinematic fitting. We form a covariance matrix for the velocity dispersion measurements with ${(\sigma_{\sigma_{\rm los}}^{\rm sys}})^2+({\sigma_{\sigma_{\rm los}}^{\rm stat}})^2$ for the diagonal terms and $({\sigma_{\sigma_{\rm los}}^{\rm sys}})^2$ for off-diagonal terms, as the source of the systematic \editnn{in the kinematic fitting} is common between all the measurements.}

\begin{table*}
\caption{\label{tab:vel_dis}
	Measurements of velocity dispersion from three different setups from \citet{BuckleyGeer20}. \editref{The quoted uncertainties are only statistical, see Section \ref{sec:velocity_dispersion_data} for the estimated systematic uncertainty.}
	}
\begin{tabular}{lccccc}
\hline
Instrument and setup & Aperture dimension & Aperture rotation & Seeing & Moffat PSF exponent & Velocity dispersion \\
& (arcsec $\times$ arcsec) & (deg E of N) & (arcsec) & &  (km s$^{-1}$) \\
\hline
Magellan mask A & \phantom{00}1$\times$1 & 99 & 0.68 & $-$2.97 & 230 $\pm$ 37 \\ 
Magellan mask B & \phantom{00}1$\times$1 & 99 & 0.76 & $-$3.20 & 236 $\pm$ 42 \\ 
Gemini mask A2 &  0.75$\times$1 & 0  & 0.52 & $-$3.06 & 220 $\pm$ 21  \\
MUSE & \phantom{00}1$\times$1 & 0 & 0.61 & $-$1.55 & 227 $\pm$ 9\phantom{0} \\
\hline
\end{tabular}
\end{table*}

\subsection{Estimate of the external convergence}

\editth{\citet{BuckleyGeer20} present the distribution of the external convergence $\kappa_{\rm ext}$ for \lensname. This analysis is based on the weighted galaxy number counts approach of \citet{Greene13}, which was further developed by \citet{Rusu17}, \citet{Birrer19}, and \citet{Rusu19}. In brief, weighted number counts are computed in $45$ arcsec- and $120$ arcsec-radii apertures centred on the lensing system, up to a depth of $I=22.5$ mag, using simple physical weights robust to measurement errors, such as the inverse of the distance to the lens and photometric/spectroscopic redshifts. Analogous number counts are computed in a large number of same-size apertures and depth in a cosmological survey, in this case DES, so as to measure the over/underdensity of the \lensname\ line of sight relative to the median line of sight through the Universe, in terms of weighted number count ratios. In this case, the line of sight was found to be underdense, and a combination of weighted number count ratios was used as constraint to select statistically similar lines of sight from the Millennium Simulation \citep{Springel05}. Using the external convergence maps from \citet{Hilbert09} corresponding to each Millennium Simulation line of sight, \editsx{we construct a probability distribution function of $\kappa_{\rm ext}$.}} \editref{This probability distribution function of $\kappa_{\rm ext}$ is provided in Section \ref{sec:ext_convergence_sampling} (specifically in Fig. \ref{fig:k_ext}).}

\section{Lens model ingredients} \label{sec:model_ingredients}
In this section, we describe the mass and light profiles used to construct the lens model in our analysis.

\subsection{Central deflector's mass profiles}
To model the main deflector's mass distribution, we adopt two sets of profiles: (i) power-law, and (ii) composite mass profile.

\subsubsection{Power-law mass profile}
We adopt the power-law elliptical mass distribution \citep{Barkana98}. This profile is described by
\begin{linenomath}\begin{equation}
	\kappa_{\rm PL}(\theta_1, \theta_2) = \frac{3 - \gamma}{2} \left[ \frac{\theta_{\rm E}}{\sqrt{ q_{\rm m} \theta_1^2 + \theta_2^2/q_{\rm m} }} \right]^{\gamma - 1},
\end{equation}\end{linenomath}
where $\gamma$ is the power-law slope, $\theta_{\rm E}$ is the Einstein radius, and $q_{\rm m}$ is the axis ratio. The coordinates $(\theta_1,\ \theta_2)$ are in the frame that is aligned with the major and minor axes. This frame is rotated by a position angle $\varphi_{\rm m}$ from the frame of on-sky coordinates.

\subsubsection{Composite mass profile}
In the composite mass profile, we adopt separate mass profiles for the baryonic and the dark components of the mass distribution. 

For the dark component, we choose a Navarro--Frenk--White (NFW) profile with ellipticity defined in the potential. The spherical NFW profile in 3D is given by
\begin{linenomath}\begin{equation}
	\rho_{\rm NFW}(r) = \frac{\rho_s}{\left(r/r_{\rm s}\right) \left(1 + r/r_{\rm s}\right)^2},
\end{equation}\end{linenomath}
where $r_{\rm s}$ is the scale radius, and $\rho_{\rm s}$ is the normalization \citep{Navarro97}. 

For the baryonic mass profile, we adopt the Chameleon convergence profile. The Chameleon profile approximates the S\'ersic profile within a few per cent in the range 0.5--3$R_{\rm eff}$, where $R_{\rm eff}$ is the effective or half-light radius of the S\'ersic profile. The Chameleon profile is the difference between two non-singular isothermal ellipsoids given by
\begin{linenomath}\begin{equation} \label{eq:chameleon}
	\begin{split}
	\kappa_{\rm Chm} (\theta_1, \theta_2) &= \frac{\kappa_0}{1 + q_{\rm m}} \left[ \frac{1}{\sqrt{\theta_1^2+\theta_2^2/q_{\rm m}^2 + 4w_{\rm c}^2/(1+q_{\rm m}^2)}} \right. \\
	 & \quad \quad \quad \quad \left. - \frac{1}{\sqrt{\theta_1^2+\theta_2^2/q_{\rm m}^2 + 4w_{\rm t}^2/(1+q_{\rm m}^2)}} \right]
\end{split}
\end{equation}\end{linenomath}
\citep{Dutton11, Suyu14}. This profile is convenient to compute lensing properties using closed-form expressions.

With each of these models, we include an external shear profile parameterized with the shear magnitude $\gamma_{\rm ext}$ and shear angle $\varphi_{\rm ext}$.

\subsection{Central deflector's light profile}

We use the S\'ersic profile and the Chameleon profile to model different light components of the lens system. 

\subsubsection{S\'ersic profile}
The S\'ersic profile is given by
\begin{linenomath}\begin{equation}
	I_{\text{S\'ersic}} (\theta_1, \theta_2) = I_{\rm eff} \exp \left[-b_n \left\{\left(\frac{\sqrt{\theta_1^2 + \theta_2^2/q_{\rm L}^2}}{\theta_{\rm eff}} \right)^{1/n_{\rm s}} - 1 \right\} \right],
\end{equation}\end{linenomath}
where $\theta_{\rm eff}$ is the effective radius, $I_{\rm eff}$ is the amplitude at $\theta_{\rm eff}$, and $n_{\rm s}$ is the S\'ersic index \citep{Sersic68}. The factor $b_n$ normalizes the profile such that half of the total luminosity is contained within $\theta_{\rm eff}$.

\subsubsection{Chameleon profile}
We use the same Chameleon profile from equation (\ref{eq:chameleon}) to fit the central deflector's light profile by replacing the convergence amplitude $\kappa_0$ with flux amplitude $I_0$.

\subsection{Quasar host galaxy's light profile}
We choose an elliptical S\'ersic profile to model the smooth component of the quasar host galaxy's light distribution. Additionally, we use a basis set of shapelets to reconstruct the non-smooth features in the extended source light distribution \citep{Refregier03, Birrer15}. The set of shapelets is characterized with a scale size $\varsigma$ and maximum polynomial order $n_{\rm max}$. The order $n_{\rm max}$ determines the total number of shapelet components $n_{\rm shapelet} = (n_{\rm max} + 1)(n_{\rm max} + 2)/2$.

We model the quasar images as point-sources on the image plane convolved with the reconstructed point spread function (PSF). \editnn{We let the amplitudes of each quasar image free.}

\section{Lens model setups} \label{sec:model_choices}
In this section, we present the specific model choices for \lensname. Extending on the baseline models, we choose different options\editnn{ -- that we consider equally viable -- }for some particular components of the model. A combination of these options then make up our model settings $S$ for each mass profile family $M$. \editnn{To be specific, the model settings $S$ include the model components describing the source and the line-of-sight galaxies, and the model settings $M$ include the mass and light profiles of the central deflector galaxy G1.} We marginalize over these model settings $S$ to \editnn{account for} any possible source of systematics that may be introduced \editnn{from adopting only} one specific choice. We first state the baseline models in Section \ref{sec:baseline_models}. Then, we elaborate on the different additional model choices in Sections \ref{sec:g1_mass_light}--\ref{sec:extinction_mask}, and we summarize the set of model settings $S$ that we marginalize over in Section \ref{sec:choice_comb}. \editn{\editref{A summary of the adopted models and the} parameter priors are tabulated in Appendix \ref{app:param_prior}.}

\subsection{Baseline models} \label{sec:baseline_models}
The specifics of the baseline models agreed by the participating independent modelling teams are:
\begin{enumerate}[leftmargin=13pt]
	\item Central deflector G1's mass profile: power-law profile and composite profile (elliptical NFW potential for the dark component, double Chameleon convergence for luminous component),
	\item Central deflectors G1's light profile: 
	\begin{enumerate}[leftmargin=20pt, label=(\arabic*)]
		\item \textit{For models with power-law mass profile:} double S\'ersic profile in all three bands,
		\item \textit{For models with composite mass profile:} double Chameleon light profile in the F160W band linked with the double Chameleon mass profile, double S\'ersic profiles in UVIS bands,
	\end{enumerate}
	\item Satellite G2's mass profile: singular isothermal sphere (SIS) placed on G1's lens plane,
	\item External shear,
	\item Explicit modelling of the line-of-sight galaxies G3--G6, multilens-plane treatment for G3,
	\item Multisource-plane treatment for quasar host S1 and additional source component S2.
\end{enumerate}
In the next sections, we explain these model settings \editnn{and further extend on some of these settings as we see fit.}

\subsection{Central deflector G1's mass and light profiles} \label{sec:g1_mass_light}

We choose two sets of mass profile for the central deflector G1: power-law mass profile and composite mass profile.

For the corresponding light profile distribution of G1 with the power-law mass profile, we adopt \editnn{a double S\'ersic profile in the IR band, and a single S\'ersic profile for each of the UVIS bands.} \editnn{Here, we deviated from the baseline model of double S\'ersic profile for the UVIS bands, as we find a single S\'ersic profile for each of the UVIS bands is sufficient and the posteriors of the lens model parameters are almost identical between the double S\'ersic and single S\'ersic profiles for the UVIS bands. Therefore, we adopt the single S\'ersic profile for the UVIS bands to increase numerical efficiency by simplifying our model. However,} \editth{we still use the double S\'ersic profile in the IR band where the signal-to-noise ratio of the galaxy light is higher and thus more flexibility is needed to render it within the noise.} The centroids are joint \editnn{for all the S\'ersic profiles} between the bands. The axis ratio $q_{\rm L}$, position angle ${\varphi}_{\rm L}$, and the S\'ersic index $n_{\rm s}$ are also joint between the UVIS bands. We let effective radius $\theta_{\rm eff}$ and amplitude $I_{\rm eff}$ as free parameters \editnn{independently} for all bands to allow a color gradient.

For the composite mass profile, we model the dark matter distribution with \editsx{a} NFW profile with ellipticity \editnn{parameterized} in the potential. For the baryonic matter distribution, we adopt two concentric Chameleon profiles to model both the luminous mass distribution and the light distribution in the F160W band. We join the scaling and ellipticity parameters of each pairing of the Chameleon profiles between the baryonic mass distribution and the F160W light distribution. We do not fix the amplitude ratio between the two Chameleon profiles and this ratio is sampled as a non-linear parameter in our model. For \editnn{each of the two UVIS bands, we adopt a single S\'ersic profile}. Similar to power-law profile, the S\'ersic profile parameters except $\theta_{\rm eff}$ and $I_{\rm eff}$ are joint between the UVIS bands and the centroids of the all the deflector light profiles are joint together. The amplitudes of the mass and light profiles are independent of each other, thus we allow the mass-to-light ratio ($M/L$) to be free. \editn{We adopt a Gaussian prior equivalent to $12.74\pm1.71$ arcsec for the NFW scale radius $r_{\rm s}$ based on the results of \citet{Gavazzi07} for the Sloan Lens ACS (SLACS) survey lenses \citep{Bolton06}. \editref{G1's velocity dispersion and redshift are within the range of those from the SLACS lenses, thus SLACS is a representative sample of elliptical galaxies such as G1 \citep{Treu06}.} Similar priors were adopted in previous H0LiCOW analyses of time-delay lenses \citep[e.g.,][]{Wong17, Rusu19}.}

We find the half-light radius $\theta_{\rm eff}$ of the S\'ersic profiles to be degenerate with the S\'ersic index $n_{\rm s}$ in our models and the models tend to optimize towards large values of $\theta_{\rm eff}$ that is inconsistent with our observational prior. To prevent $\theta_{\rm eff}$ from converging towards abnormally large values, we impose an empirical prior on $\theta_{\rm eff}$. We derive a scaling relation from the distribution of the central velocity dispersion $\sigma_{\rm e/2}$ measured within half of effective radius and $R_{\rm eff}$ in physical unit for the lenses in the SLACS sample \citep{Auger10b}. We account for intrinsic scatter in the derived scaling relation as we are ignoring the average surface brightness $\overline{I}(R_{\rm eff})$ in the relation between the three quantities along the fundamental plane. Then, we derive a distribution for $R_{\rm eff}$ for \lensname\ for the given central velocity dispersion measurements from Table \ref{tab:vel_dis}. In practice, we simultaneously sample $R^{\rm J0408}_{\rm eff}$ and the parameters $\{m, b, \mathcal{S}\}$ -- for the scaling relation 
\begin{linenomath}
	\begin{equation}
		\log_{\rm 10} (\sigma_{\rm los}/{\rm km\ s^{-1}}) = m \log_{\rm 10} (R_{\rm eff} / {\rm kpc}) + b
	\end{equation}	
\end{linenomath}
 with scatter $\mathcal{S}$ -- from a joint likelihood for the SLACS sample data and the measured velocity dispersions of \lensname. For each sampled $R_{\rm eff}^{\rm J0408}$, we transform the measured central velocity dispersions within each aperture into $\sigma_{\rm e/2}$ using the aperture correction formulae given by \citet{Jorgensen95}. We include the intrinsic scatter in the likelihood term for \lensname's velocity dispersions, thus the scatter in the scaling relation propagates into the $R_{\rm eff}^{\rm J0408}$ distribution. We estimate the scaling relation parameters as $m=0.18^{+0.05}_{-0.04}$, $b=2.2\pm0.04$, $\mathcal{S}=1.53\pm0.05$. We convert the $R_{\rm eff}^{\rm J0408}$ distribution  in physical unit into $\theta_{\rm eff}$ distribution in angular unit using the angular diameter distance to \lensname\ for our fiducial cosmology, however we add 10 per cent uncertainty to the distribution to remove any strong dependence on the choice of cosmology. \editn{We take a Gaussian prior with the same mean and standard deviation of the resultant $\theta_{\rm eff}$ distribution from the SLACS lenses (Table \ref{tab:param_prior}). We adopt this prior only to prevent $\theta_{\rm eff}$ from veering off to very large values. The adopted prior is broad enough not to bias the $\theta_{\rm eff}$ posterior within the plausible range of values, including for the double S\'ersic profile.} 

\subsection{Satellite G2's mass and light profile} \label{sec:g2_mass_light}
In addition to the power-law or composite mass profile for the central deflector, we add a singular isothermal sphere (SIS) profile for G2's mass distribution and a circular S\'ersic profile for its light distribution. The S\'ersic profile parameters except $\theta_{\rm eff}$ are joint between all bands. We join the centroid between the SIS and S\'ersic profiles. \editref{Although a deviation from the isothermal profile in G2's mass can potentially change the deflection potential at image C, such a change will be negligible in our inference of \Ho\ as time-delays with respect to image C are not used in our inference. The SIS profile is sufficient to capture the astrometric position of the image C in our modelling.}

\subsection{Nearby line-of-sight galaxies}
We explicitly model the mass distributions of line-of-sight galaxies G3--6 to fully capture their higher than second-order lensing effects that cannot be accounted  for by the external convergence and the external shear profiles. First in Section \ref{sec:selction_los_galaxy}, we describe our selection criterion for the line-of-sight galaxies to explicitly include in our lens model. Then in Section \ref{sec:multi_pllane_los_galaxies}, we explain the mutli-lens-plane treatment of the line-of-sight galaxies. Lastly in Sections \ref{sec:perturber_sis} and \ref{sec:perturber_nfw}, we describe the mass profiles we adopt to model these line-of-sight galaxies.

\subsubsection{Selection criterion of the line-of-sight galaxies for explicit modelling} \label{sec:selction_los_galaxy}
To select the line-of-sight galaxies for explicit modelling, we first estimate the contribution in time-delays between the images from higher than second-order derivatives of the deflection potential of these galaxies. To quantify this effect, we set \editsx{a} SIS profile for each perturber with its Einstein radius corresponding to the estimated central velocity dispersion. We infer the velocity dispersion for all the line-of-sight galaxies from their stellar masses using two scaling relations -- one from \editsx{\citet{Auger10b}} and the other from \citet{Zahid16}. To be conservative, we choose the upper limit of the 1$\sigma$ confidence interval of the estimated stellar mass and choose the larger value of the velocity dispersions estimated from the two scaling relations \citep{BuckleyGeer20}. We select the line-of-sight galaxies that may cause more than 1 per cent shift in the measured Hubble constant if higher than second-order derivatives of their deflection potential are ignored. The shift in the Hubble constant can be related to the relative astrometric shift $\delta \btheta_{\rm AB}$ between image A and B as 
\begin{linenomath}\begin{equation}
	\frac{\delta {H_0}}{H_0} \lesssim \frac{D_{\Delta t}}{c \Delta t_{\rm AB}} (\btheta_{\rm A} - \btheta_{\rm B}) \cdot \delta {\btheta_{\rm AB}}
\end{equation}\end{linenomath}
\citep{Birrer19b}. We take the relative astrometric shift $\delta \btheta_{\rm AB} = \balpha_{\rm A}^{(3)} - \balpha_{\rm B}^{(3)}$, where the term on the right-hand side is the relative deflection angle for third and higher order lensing effects from the SIS profile corresponding to each line-of-sight galaxy. Thus, we set the selection criterion
\begin{linenomath}\begin{equation}
	\frac{D_{\Delta t}^{\rm fiducial}} {c \Delta t_{\rm AB}} (\btheta_{\rm A} - \btheta_{\rm B} ) \cdot \left( \balpha_{\rm A}^{(3)} - \balpha_{\rm B}^{(3)} \right) \geq 0.01.
\end{equation}\end{linenomath}
This criterion selects G3--G6 for explicit modelling. \editfr{Note that the perturber selection criterion based on the ``flexion shift'' $\Delta_3{x} > 10^{-4}$ also selects G3--G6 for explicit modelling \citep[][]{McCully17, Sluse19, BuckleyGeer20}.} 


\subsubsection{Multilens-plane modelling of the line-of-sight galaxies}
\label{sec:multi_pllane_los_galaxies}
We model this lens system with a multilens-plane treatment by setting G3's lens plane at its own redshift $z_{\rm G3}=0.769$, as G3 is close enough to G0 to cause more than 1 per cent deviation in the computed time-delays if we place it on G0's lens plane. We place G4--G6 on G0's lens plane as we assume that the deviation in computed time-delays due to this assumption is negligible given the combinations of their stellar masses and distances from G0.

\begin{figure}
	\includegraphics[width=\columnwidth]{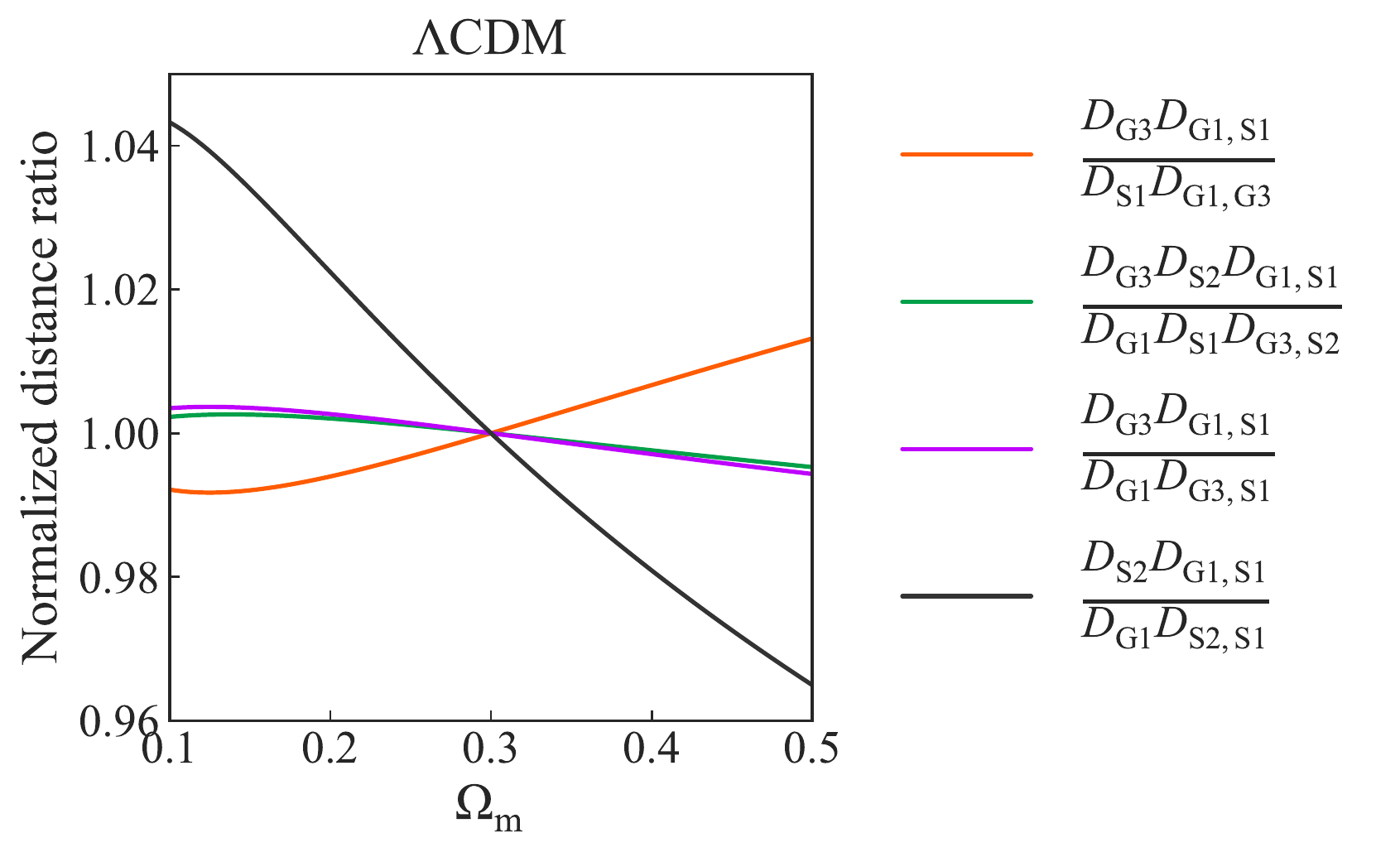}
	\caption{\label{fig:distance_ratio}
	Impact of varying $\Omega_{\rm m}$ in the \lcdm\ cosmology on the angular diameter distance ratios between the lens and source planes. All the distance ratios except for the black line changes less than 1 per cent for a wide range of $\Omega_{\rm m}$. The black line corresponds to the distance ratio involving S2's lens plane. As the S2's Einstein radius is small ($\sim$0.002 arcsec), the change in the black line only has a small effect on the effective Fermat potential [cf. equation (\ref{eq:double_plane_fermat_potential})]. Therefore, fixing the distance ratios for the fiducial cosmology with $\Omega_{\rm m} = 0.3$ is not a strong assumption in our analysis. See Appendix \ref{app:effect_cosmology} for tests validating this point.
	}
\end{figure}

Additionally, we model the mass profile of S2 at its redshift $z_{\rm S2}=2.228$. Therefore, we have three lens-planes in our model. We can express the effective Fermat potential for the triple-lens-plane case from equation (\ref{eq:multi_fermat_potential}) as
\begin{linenomath}\begin{equation} \label{eq:double_plane_fermat_potential}
	\begin{split}
		\phi^{\rm eff} (\btheta) =& \left[ \frac{D_{\rm G3} D_{\rm G1,S1}}{D_{\rm S1}D_{\rm G1,G3} } \frac{(\btheta_{\rm G1} - \btheta_{\rm G3})^2}{2} - \psi_{\rm G1}(\btheta_{\rm G1})  \right] \\
		&+ \frac{1+z_{\rm G3}}{1+z_{\rm G1}}\left[ \frac{D_{\rm G3} D_{\rm S2} D_{\rm G1,S1}}{D_{\rm G1} D_{\rm S1} D_{\rm G3, S2}} \frac{(\btheta_{\rm G3} - \btheta_{\rm S2})^2}{2} \right. \\
		& \qquad\qquad\quad \left. - \frac{D_{\rm G3} D_{\rm G1,S1}}{D_{\rm G1} D_{\rm G3,S1} }  \psi_{\rm G3}(\btheta_{\rm G3})  \right] \\
		&+ \frac{1+z_{\rm S2}}{1+z_{\rm G1}} \frac{D_{\rm S2} D_{\rm G1,S1}}{D_{\rm G1} D_{\rm S2,S1} } \left[ \frac{(\btheta_{\rm S2} - \bbeta)^2}{2} - \psi_{\rm S2}(\btheta_{\rm S2})  \right].
	\end{split}
\end{equation}\end{linenomath}
Here, $\btheta_{{\rm G}}$ is the quasar's image position on G's plane with G $\in$ \{G1, G3, S2\}, $\psi_{{\rm G}}$ is the deflection potential of G, and $\bbeta$ is the quasar's position on the source-plane. We fix the distance ratios in the above equation in our modelling. We adopt the \lcdm\ cosmology with the cosmological density parameters $\Omega_{\rm m} = 0.3$, $\Omega_{\Lambda}=0.7$ to obtain these distance ratios. The relevant distance ratios change by less than 1 per cent within $0.25 \lesssim \Omega_{\rm m} \lesssim 0.35$ and $-1.1 \lesssim w \lesssim -0.9$  (Fig. \ref{fig:distance_ratio}). Therefore, adopting this fiducial cosmology is only a weak assumption in our analysis. Fixing these distance ratios does not linearly affect our inference of \Ho, as the ratios do not depend on \Ho. However, there can potentially be a small non-linear shift in the inferred \Ho\ from our analysis had we adopted a different set of values for $\Omega_{\rm m}$ and $\Omega_{\rm \Lambda}$. The non-linear effect on \Ho\ from fixing density parameters in the multilens-plane treatment was demonstrated to be less than 1 per cent for two previously analysed lens systems, HE 0435--1223 and WFI 2033--4723 \citep{Wong17, Rusu19}. In Appendix \ref{app:effect_cosmology}, we show that \Ho\ shifts by less than 1 per cent if we change the matter density parameter \editnn{to $\Omega_{\rm m}=0.1$ and to $\Omega_{\rm }=0.45$ \editnn{within the \lcdm\ cosmology}. This range in $\Omega_{\rm m}$ covers nearly the full range of our prior $\Omega_{\rm m} \in [0.05, 0.5]$ for inferring \Ho\ for the \lcdm\ cosmology. As a shift less than 1 per cent in \Ho\ is much smaller than the typical precision on \Ho\ ($\sim$5--8 per cent) allowed by the current data quality, we consider that the impact of fixing the distance ratios using a fiducial \lcdm\ cosmology has negligible impact in our analysis. However, we find that our inference of \Ho\ is sensitive to the dark energy equation of state parameter $w$ in the $w$CDM cosmology. As we adopt a double source plane model -- as described in Section \ref{sec:souce_components} -- the distance ratios or the $\zeta$ terms in equation (\ref{eq:multi_time_delay}) become sensitive to $w$ \citep{Gavazzi08, Collett12, Collett14}. Therefore, the distance posteriors from this analysis should not be used to infer \Ho\ in extended cosmologies other than the \lcdm\ model.} \editth{We postpone the derivation of a distance posterior in more general cosmologies to future work.}

\begin{figure}
	\includegraphics[width=\columnwidth]{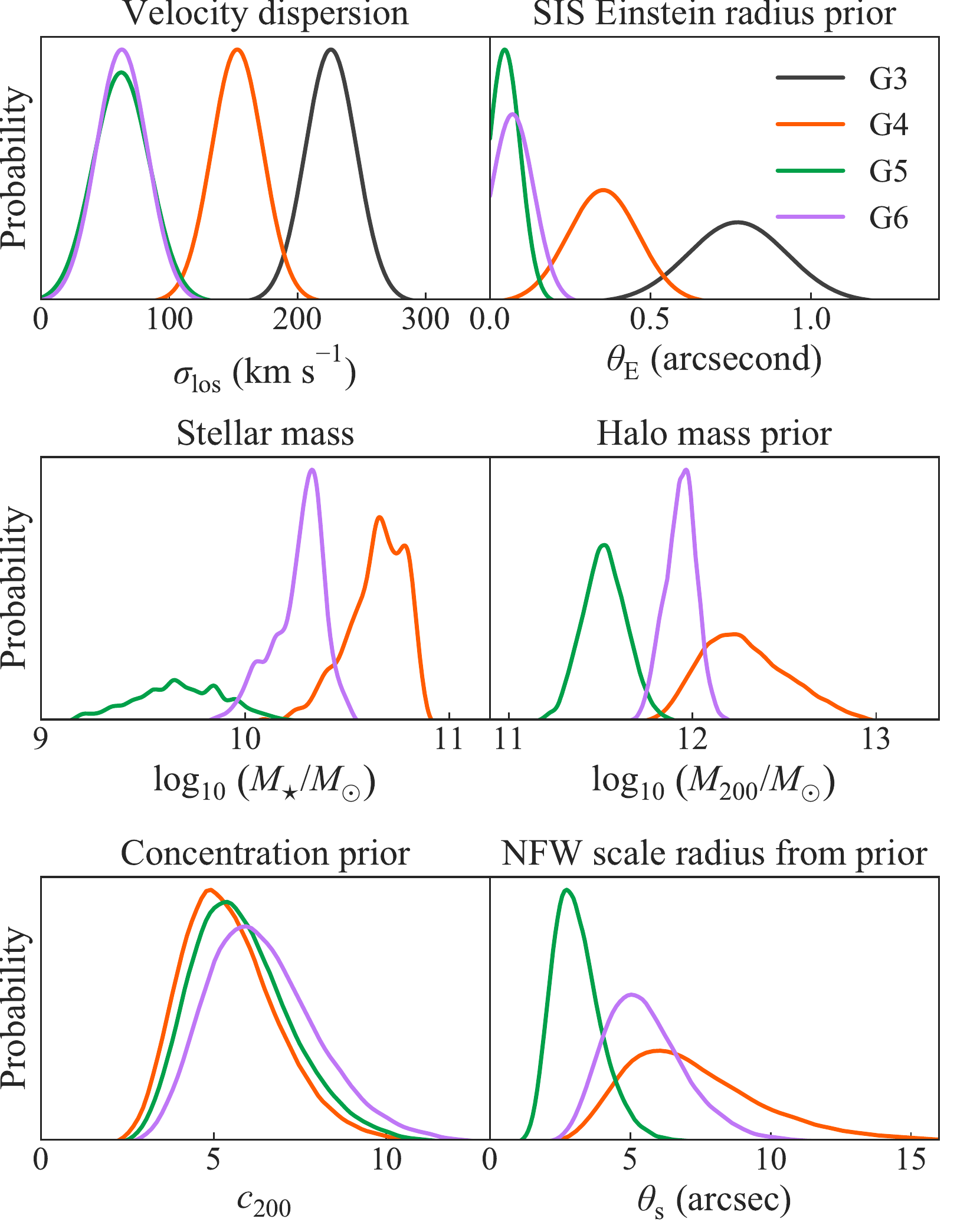}
	\caption{\label{fig:perturber_prior}
	Observed and estimated properties of the line-of-sight galaxies G3--G6. \textbf{Top left:} velocity dispersions derived from the MUSE integral field spectra. \textbf{Top right:} SIS Einstein radius distributions obtained from the observed velocity dispersions. \textbf{Middle left:} estimated stellar masses from \citet{BuckleyGeer20}. \textbf{Middle right:} halo mass $M_{200}$ inferred from the estimated stellar mass using the stellar mass--halo mass relation from \citet{Behroozi19}. \textbf{Bottom left:} halo concentration parameter $c_{200}$ obtained using a halo mass--concentration relation for our fiducial cosmology \citep{Diemer19}. \textbf{Bottom right:} scale radius of the NFW profile in angular unit for our fiducial cosmology from the $M_{200}$ and $c_{200}$ priors. The intrinsic scatter and uncertainties of the adopted scaling relations are accounted for at each conversion step. We use the SIS Einstein radius distributions as priors for the SIS model and the $M_{200}$ and $c_{200}$ distributions as priors for the NFW model for G4--G6.
	} 
\end{figure}

We model G3 and S2 with SIS profiles. We place G3 at it's ``true'' position on its own lens plane by tracing back from its observed position accounting for the foreground deflectors. As we also model the flux distribution from S2, we join the centroid of S2's mass profile with its light centroid on its plane.


For G4--G6, we fix their centroids at their observed position on the lens plane of G1. For the mass profiles of G4--G6, we adopt two choices: the SIS profile and the spherical NFW profile. We choose the additional NFW model for G4--G6 as the NFW scale radius estimated from each of their stellar masses is smaller than the distance between the galaxy and G0 (Fig. \ref{fig:perturber_prior}, Section \ref{sec:perturber_nfw}). Thus, their mass profile slopes can potentially be different from the isothermal profile at the centre of G0. In Sections \ref{sec:perturber_sis} and \ref{sec:perturber_nfw}, we describe the priors for the SIS and NFW profile parameters, respectively, of the line-of-sight galaxies.

\subsubsection{SIS profile for the line-of-sight galaxies} \label{sec:perturber_sis}

We estimate the SIS Einstein radius distribution from each galaxy's SIS velocity dispersion \editth{$\sigma_{\rm SIS}$} using the relation
\begin{linenomath}
	\begin{equation} \label{eq:sis_einstein_radius}
		\theta_{\rm E, SIS} = 4 \uppi \left( \frac{\sigma_{\rm SIS}}{c} \right)^2 \frac{D_{{\rm G},{\rm S1}}}{D_{\rm S1}},
	\end{equation}
\end{linenomath}
where $\theta_{\rm E, SIS}$ is the Einstein radius for an SIS profile, and $D_{{\rm G},{\rm S1}}$ is the angular diameter distance between a line-of-sight galaxy G $\in$ \{\editn{G3,} G4, G5, G6, S2\} and S1. We calculate the distance ratio in the above equation using our fiducial cosmology. \editth{We do not \editsx{need} to add uncertainty to the fiducial cosmology used here as the distance ratios are independent of \Ho\, and a large shift (e.g., by 0.1) in $\Omega_{\rm m}$ changes \editsx{them} by negligible amount relative to the 20 km s\textsuperscript{$-$1} uncertainty we adopted for the velocity dispersions.}

G4 and G6's observed morphologies indicate that they are elliptical galaxies. Therefore, we \editth{take} \editsx{their} observed \editth{stellar} velocity dispersions \editth{$\sigma_{\rm ap}$} as $\sigma_{\rm SIS}$ in equation (\ref{eq:sis_einstein_radius}) to obtain these galaxies' Einstein radius prior distributions \editth{\citep{Treu06, Auger10b}}. In contrast, G5's spectra contains bright [O \textsc{ii}] emission lines indicative of a star-forming galaxy. We also take S2 as a star-forming galaxy due to its blue color in the \textit{HST} three-band imaging (Fig. \ref{fig:cutouts}). Therefore, we estimate the rotational velocities $v_{\rm c}$ of G5 and S2 from their observed `aperture-averaged' velocity dispersions $\sigma_{\rm ap}$ using the scaling relation between $v_{\rm c}^2/\sigma_{\rm ap}^2$ and S\'ersic index $n_{\rm s}$ from \citet{Agnello14}. We obtain the S\'ersic index of G5 $n_{\rm s} = 4$ by fitting a S\'ersic profile to its light distribution in the F814W band. From a preliminary lens model, we adopt S2's S\'ersic index as $n_{\rm s} = 1.5$. For these S\'ersic indices, the $v_{\rm c}^2/\sigma_{\rm ap}^2$ ratios are approximately 2.5 and 2.2, respectively, for G5 and S2. We adopt a Gaussian uncertainty with standard deviation 0.2 for these ratios to account for the scatter observed in the $v_{\rm c}^2/\sigma_{\rm ap}^2$--$n_{\rm s}$ distribution (cf. fig. 6 of \citealt{Agnello14}). Then, to convert the estimated rotational velocity $v_{\rm c}$ into the corresponding SIS velocity dispersion $\sigma_{\rm SIS}$, we use the relation
\begin{linenomath}
	\begin{equation}
		\sigma_{\rm SIS}^2 = \frac{G M(R)}{2 R} = \frac{v_{\rm c}(R)^2}{2},
	\end{equation}	
\end{linenomath}
where $M(R)$ is the enclosed 3D mass within a radius $R$. The estimated SIS velocity dispersions are $\sigma_{\rm SIS}^{\rm G5} =  62 \pm 22$ km s\textsuperscript{$-$1} and $\sigma_{\rm SIS}^{\rm S2} = 48 \pm 11$ km s\textsuperscript{$-$1}. We parameterize the SIS Einstein radius distributions derived from the velocity dispersions as Gaussian priors for the SIS mass profiles of G3--G6 and S2.

%

%
%
 
\subsubsection{NFW profile for the line-of-sight galaxies} \label{sec:perturber_nfw}
We parameterize the NFW profiles for G4--G6 with the halo mass $M_{200}$ and concentration $c_{200}$. We obtain the priors on the NFW profile parameters from the estimated stellar masses of G4--G6 \citep{BuckleyGeer20}. We derive the halo mass distribution from the stellar mass distribution using the stellar mass--halo mass relation from \citet{Behroozi19} for the respective redshift of the line-of-sight galaxy. We weight the halo-mass distribution with the halo mass function for our fiducial cosmology and the relevant redshift from \citet{Tinker08}. We obtain the concentration distribution from the halo mass distribution for our fiducial cosmology using the $M$--$c$ relation from \citet{Diemer19}. We propagate the uncertainties and scatters in these relations when deriving one quantity from another. The $M_{200}$ priors and $c_{200}$ priors for G4--G6 are shown in Fig. \ref{fig:perturber_prior}. \edit{We can also derive the NFW scale radius $r_{\rm s} = R_{200}/c_{200}$ in physical unit, and convert it to the scale radius $\theta_{\rm s}$ in angular unit \editsx{given} our fiducial cosmology (Fig. \ref{fig:perturber_prior}). We do not use these scale radii as prior, we only show the distributions to motivate our choice of the NFW profile for the galaxies G4--G6.}

\subsection{Galaxy group containing G1} \label{sec:group_g1}

\editfv{We do not explicitly model the galaxy group that contains the central deflector G1 [Group 5 in \citet{BuckleyGeer20}]. The estimated flexion shift \editsx{$\log_{10} \Delta x_3 = -3.86^{+0.97}_{-0.72}$} for this group is marginally above the conservative threshold $\Delta x_3 > 10^{-4}$ \citep{BuckleyGeer20}. However, the larger end of the flexion shift is provided by the case where the group is centred near to the central deflector. In that case, the group's halo coincides with the deflector's halo, which is already accounted for in our lens models. However, if the group's centroid is offset from the central deflector, then the flexion shift becomes smaller. Then, the group's contribution can be considered only in the approximated convergence, as the external shear profile already captures the shear contribution from the group. In Appendix \ref{app:impact_group}, we show that the impact of the group's convergence, if explicitly accounted for, only shifts \Ho\ by 0.4 per cent. This small shift justifies our choice of not including the group in our lens model.}

\subsection{Source component light profiles} \label{sec:souce_components}
We use an elliptical S\'ersic light profile and a set of shapelets to reconstruct the quasar host S1's light profile. We join all the S\'ersic profile parameters across the three bands. We join the shapelet scale size $\varsigma$ across the UVIS bands and leave $\varsigma$ in the IR band as a free parameter.

To reconstruct S2's light profile, we take a basis set of one elliptical S\'ersic profile and multiple shapelets. We join the S\'ersic profile parameters and the shapelet scale size $\varsigma$ across bands. For S3's light profile, we adopt only an elliptical S\'ersic profile. All the profile parameters for this profile except the amplitude $I_{\rm eff}$ are joint across the three bands.

For each model setup, we choose three fixed values of $n_{\rm max}$: S1's $n_{\rm max}$ in the IR band, S1's $n_{\rm max}$ in the UVIS band, and S2's $n_{\rm max}$ common across the three bands. We adopt three different sets of $\{n^{\rm S1,\ IR}_{\rm max}, n^{\rm S1,\ UVIS}_{\rm max}, n^{\rm S2}_{\rm max}\}$: $\{6, 3, 2\}, \ \{7, 4, 2\}, \ \{8, 5, 3\}$. \edit{A minimum number of shapelets is necessary to sufficiently capture the complex structures in the lensed arcs, however \editsx{we would start to fit the noise in the data by adopting too many more shapelets than necessary}. We choose these values for $n_{\rm max}$ so that we hit a balance between these two scenarios. As we show in Table \ref{tab:model_logzs}, the model evidence \editsx{peaks} around these values \editref{leading to} our choice of these $n_{\rm max}$ values.} \editnn{We check if the inferred \Ho\ value from our analysis depends on the particular range of $n_{\rm max}$ values adopted above. We find that a set of larger $n_{\rm max}$ values $\{12, 9, 9, 4 \}$ and a set of smaller $n_{\rm max}$ values $\{2, 2, 2, 2\}$ both infer \Ho\ within the range spanned by the models with our adopted $n_{\rm max}$ values. Thus, our inference of \Ho\ is robust against the particular range of adopted $n_{\rm max}$ values.} 

We place the additional source component S2 at the source plane with redshift $z_{\rm S2}=2.228$. As we do not know the redshift of S3, we adopt two choices for its redshift: $z_{\rm QSO}$ and $z_{\rm S2}$. 

For the model where we place S3 on S2's plane, we ignore the mass distribution of S3. From our lens model, we find that S3 is approximately twice further away from the quasar position on S2's plane than S2. We run a lens model ignoring S2's mass profile as well and find that the time-delay distance shifts by 0.94 per cent. The total flux from the reconstructed source light distribution of S2 and S3 are comparable after accounting for lensing magnification. If S2 and S3 are at a similar redshift, then they have similar total mass. If we assume SIS profile for S2 and S3, then the convergence of S3 at the quasar position would be approximately half of that from S2. We estimate that the time-delay distance will shift by $\lesssim0.5$ per cent due to ignoring S3's mass distribution, if it indeed lies at a similar redshift of S2. This shift is negligible compared to the typical uncertainty (5--8 per cent) on the estimated time-delay distance given the \editsx{quality of the current data}. Therefore, we do not include its mass distribution in our model as we do not know the true redshift of S3.

\subsection{Potential additional image C2 split from the image C}
A faint blob is noticeable on a few pixels toward North-East from the position of G2 in the F160W band. This blob can potentially be another quasar image C2 split off from the image C by the nearby satellite G2. This potential additional image is not noticeable in the UVIS bands, but this non-detection in the UVIS bands can be caused by differential extinction through G2. If such an additional image is predicted by the model, we allow the model to assign point-source-like flux at the position of the predicted additional image. Note that we do not impose the existence of this additional image in the model.

\subsection{\textit{HST} image region for likelihood computation}
To compute the imaging likelihood, we choose a large enough circular region of the \textit{HST} image centred on the deflector galaxy in each band so that it contains most of the flux from the lens system in that particular band. The radii of these regions are 4.3, 3.3, and 3.3 arcsec in the F160W, F814W, and F475X bands, respectively. We mask out some of the pixels around the faint blob visible between G1 and G3 to block its light that would otherwise be within the chosen apertures. See the ``Normalized Residual'' plots of Figs \ref{fig:0408_pl_model} or \ref{fig:0408_comp_model} for the shape of the likelihood computation regions. In Appendix \ref{app:effect_mask}, we \editsx{show} that this particular choice of likelihood computation regions is not a source of bias in our analysis.

\subsection{Dust extinction by the satellite G2} \label{sec:extinction_mask}
The satellite G2 may cause differential dust extinction to the lensed light distribution near image C. Ignoring this differential extinction may produce poor fitting around image C in the modelling. To account for this effect, we multiply a differential extinction factor $\exp[-\tau_{\lambda}(\theta_1, \theta_2)]$ to the lensed light distribution from the quasar host galaxy in all three bands. Here, $\tau_{\lambda}$ is equivalent to an optical depth parameter. We set the differential extinction profile proportional to G2's IR surface brightness with a wavelength-dependent normalization. Therefore, we take  $\tau_{\lambda}(\theta_1, \theta_2) = \tau^0_{\lambda} I_{\rm G2} (x, y)$, where $I_{\rm G2}(\theta_1, \theta_1)$ is G2's light distribution parameterized with a S\'ersic profile as described in Section \ref{sec:g2_mass_light}. Thus, we are only modelling the differential extinction effect by G2 and this extinction goes to zero far away from G2. We do not model the differential extinction effect for the central deflector G1 as elliptical galaxies like G1 are typically dust-poor. We connect the proportionality constant $\tau_0^{\lambda}$ for each band using the differential extinction law of \citet{Cardelli89} with $R_{V}=3.1$. As a result, we only have $\tau^0_{\rm F814W}$ as a non-linear parameter in our model. As a check, we run a lens model with the proportionality constant $\tau^0_{\lambda}$ in each band independent of each other and we find that the three constrained $\tau^0_{\lambda}$ parameters follow the extinction law from \citet{Cardelli89} for $R_V \sim $3--5. The amplitudes of the quasar images are free parameters, therefore any possible differential extinction effect in the quasar image flux is already accounted for.


\subsection{Requirement for astrometric precision} \label{sec:astrometric_precision}

\edit{
For the lens system \lensname, a precision of 6 mas is required in the estimated source position to match the precision of the most precise time delay, $\Delta t_{\rm AB}$ \citep{Birrer19b}. Given the magnification and the multiplicity of the images, this precision in the source position translates to an astrometric precision of approximately 40 mas for each image position on the image plane under a fixed lens model. As we can constrain the image positions in our models within 10 mas, \editfv{we meet the requirement for astrometric precision}. We expect any non-accounted astrometric uncertainty on the level of 10 mas or below to be subdominant in the error budget and the systematic impact.
}

\subsection{Model choice combinations} \label{sec:choice_comb}

Assembling the different choices described above for various components in our models, we have the following options that we vary:
\begin{enumerate}[leftmargin=15pt]
	\item Central deflector G1's mass profile: power-law, composite,
	\item Source $n_{\rm max}$:  $\{6, 3, 2\}, \ \{7, 4, 2\}, \ \{8, 5, 3\}$,
	\item S3 redshift: $z_{\rm QSO}=2.375$, $z_{\rm S2}=2.28$, and
	\item G4--G6 mass profile: SIS, NFW.
\end{enumerate}
Taking \editsx{all possible combinations} of these choices, we have 24 different models in total -- 12 for the power-law and \editsx{12 for} the composite mass profiles. All the light profiles for lens and source light distribution form a linear basis set, thus all the amplitude parameters are linear \citep{Birrer15, Birrer18}. We have \edit{85--137} linear parameters and 57--62 non-linear parameters \edit{in the 24 model setups with either power-law or composite mass profiles}.

We can compare the number of chosen models in this study with the 128 model runs performed in the cosmographic analysis of SDSS 1206+4332 \citep{Birrer19}. \editsx{Since} \citet{Birrer19} performed two separate sampling runs for the same model, these authors adopt 64 different models in practice combining the power-law and composite mass profiles. As \citet{Birrer19} find that explicitly accounting for the \editref{non-linear components of the} foreground shear has negligible impact in the cosmographic analysis, we choose not to include it in our analysis. \editref{Note, the linear components of the foreground shear is already accounted by the adopted external shear profile.} Moreover, \citet{Birrer19} incorporate two different likelihood-computation region sizes in their model choices, whereas we do not vary \editsx{it} in our analysis as we show that our analysis is stable against different choices of the likelihood-computation region size (Appendix \ref{app:effect_mask}). As a result, the comparable number of models in \citet{Birrer19} is 16 to contrast with our adopted model number of 24. These numbers, \editsx{although not identical,} are comparable and difference between the exact number of chosen models to check systematics can arise naturally \editsx{due to different complexity in different lens systems}.


\section{Lens modelling and cosmographic inference} \label{sec:results}

In this section, we first present the lens modelling results (Section \ref{sec:modelling_workflow}), combine the time-delay and kinematics likelihoods with the lens model posterior to produce the cosmological distance posterior (Section \ref{sec:combine_td_likelihood}), and infer \Ho\ from the distance posterior (Section \ref{sec:h0_inference}).

\subsection{Modelling workflow and results} \label{sec:modelling_workflow}

We simultaneously model the images from all three \textit{HST} bands. For each model choice from Section \ref{sec:choice_comb}, we reconstruct the PSF for each \textit{HST} band. Thus, a set of three reconstructed PSFs is part of the model choice $S$ that we marginalize (cf. equation \ref{eq:modelling_posterior}). To initiate the PSF reconstruction, we take an initial PSF estimate by taking the median of a few ($\sim$4--6) stars from each \textit{HST} image and then re-centring the median PSF. At each iteration of \editsx{the} PSF reconstruction process, we first realign the IR band's coordinate system with the UVIS bands' coordinate system using the quasar image positions \citep{Shajib19}. Then, we optimize the lens model given the PSF from the initial estimate or the previous iteration \editref{of PSF reconstruction}. Finally, \editref{we subtract the extended host-galaxy and the lens-galaxy light from the image and optimize the PSF using the residual quasar images} \citep[see for details][and for similar procedure \citealt{Chen16}]{Birrer19}. We use \textsc{lenstronomy} for lens modelling and PSF reconstruction and the particle swarm optimization (PSO) routine of \textsc{cosmohammer} for optimizing the model \editfv{\citep{Kennedy95, Akeret13, Birrer18}}. We repeat the set of the following three steps five times in total to reconstruct the PSF:
\begin{enumerate}[label={\arabic*.}, leftmargin=15pt]
	\item IR band image re-alignment,
	\item lens model optimization using PSO,
	\item PSF reconstruction.
\end{enumerate}
\editref{We check that the reconstructed PSF stabilizes after five such iterations as the PSFs from additional iterations do not lead to higher imaging likelihood.}
	
After the PSF reconstruction, we simultaneously sample from the lens model posterior and compute the model evidence $\mathcal{Z}$ using the dynamic nested sampling algorithm \citep{Skilling04, Higson18}. We use the nested sampling software \textsc{dypolychord} \citep{Handley15, Higson19}. In Appendix \ref{app:dypolychord_setup}, we \editfv{describe} the sampler settings, assess the numerical performance, and conclude that the \editfv{choosen} settings allow for sufficient exploration of the posterior space.

We perform our analysis while blinding \Ho\ and \editref{other model parameters and lensing quantities} directly related to \Ho, i.e. the model-predicted time delays. We also blind the mass profile slope $\gamma$ of the power-law model after the initial exploration stage to find a stable preliminary lens model. \editref{In practice, the mean of the blinded quantities are always subtracted from the distribution within the analysis software, so that the printed values or plotted distributions are centred at zero. After all the co-authors had agreed during a teleconference on 2019 September 25 that sufficient amount of checks for modelling systematics were carried out, the analysis was frozen and the actual posterior distribution of \Ho\ was revealed for the first time. We report this \Ho\ posterior in this paper without any further alteration.}

\begin{figure*}
	\center
	\includegraphics[width=1.25\textwidth]{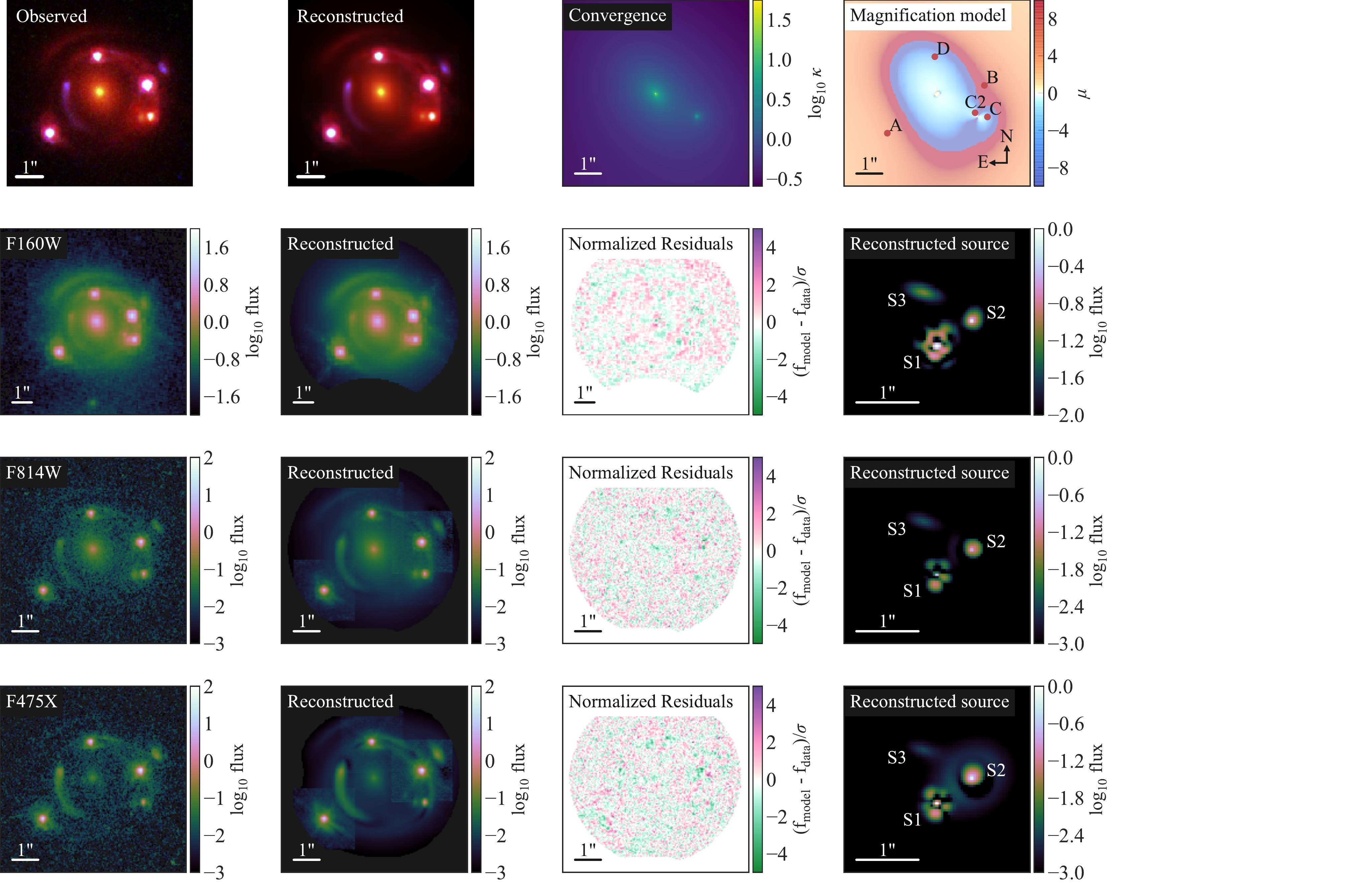} 
	\caption{
	\label{fig:0408_pl_model}
	The most likely lens model and reconstructed image of \lensname\ using the power-law model. The top row shows the observed RGB image, reconstructed RGB image, the convergence profile, and the magnification model in order from the left-hand side to the right-hand side. The next three rows show the observed image, the reconstructed image, the residual, the reconstructed source in order from the left-hand side to the right-hand side for each of the \textit{HST} filters. The three filters are F160W (second row), F814W (third row), and F475X (fourth row). All the scale bars in each plot correspond to 1 arcsec. \editref{The patchy or ring-like artefacts in the source reconstruction translate to lensed features below the noise level in the image, thus they do not affect our lens model.}
	}
\end{figure*}

\begin{figure*}
	\center
	\includegraphics[width=1.3\textwidth]{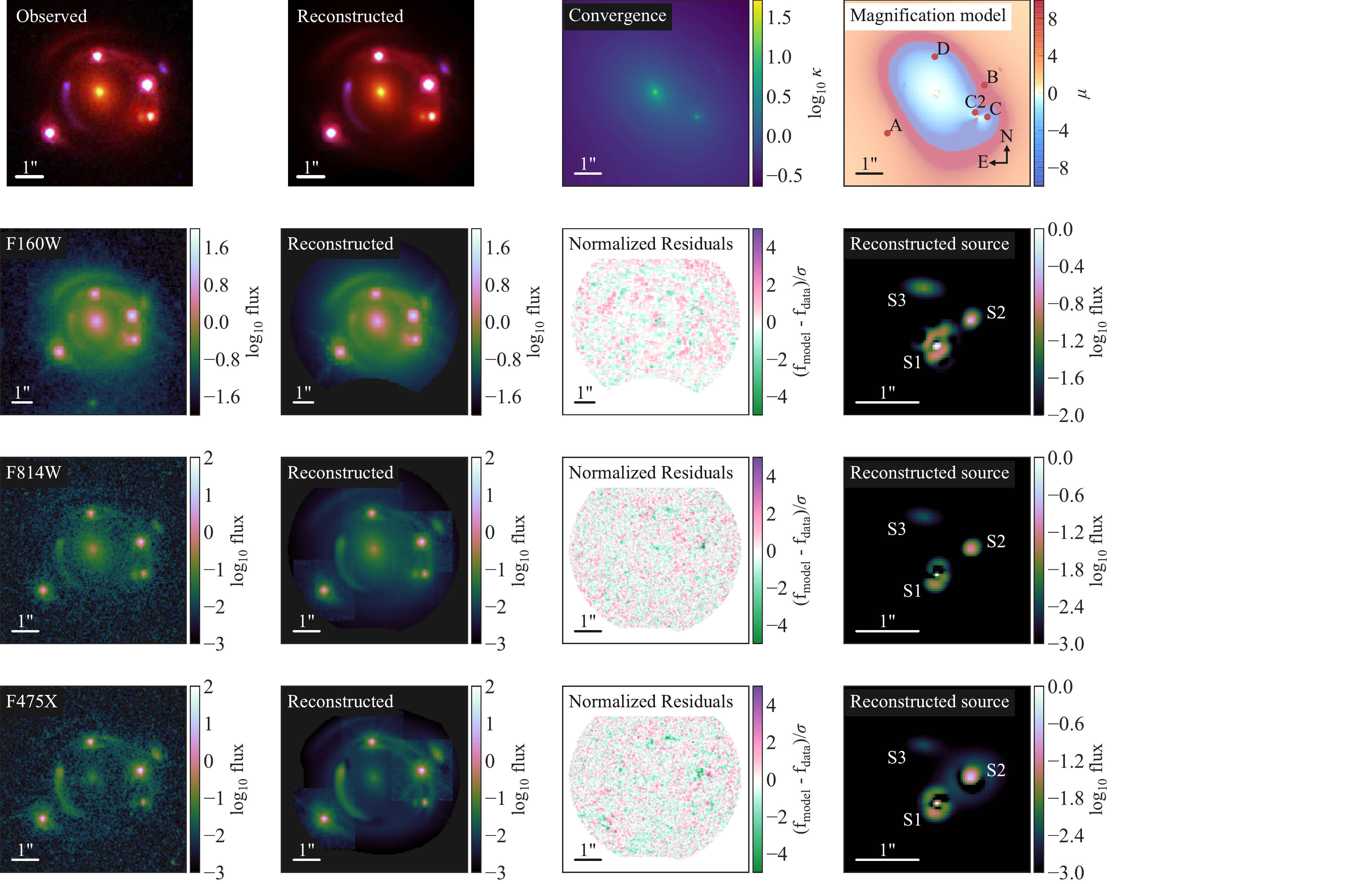} 
	\caption{
	\label{fig:0408_comp_model}
	The most likely lens model and reconstructed image of \lensname\ using the composite model. The top row shows the observed RGB image, reconstructed RGB image, the convergence profile, and the magnification model in order from the left-hand side to the right-hand side. The next three rows show the observed image, the reconstructed image, the residual, the reconstructed source in order from the left-hand side to the right-hand side for each of the \textit{HST} filters. The three filters are F160W (second row), F814W (third row), and F475X (fourth row). All the scale bars in each plot correspond to 1 arcsec. \editref{The patchy or ring-like artefacts in the source reconstruction translate to lensed features below the noise level in the image, thus they do not affect our lens model.}
	}
\end{figure*}

Figs. \ref{fig:0408_pl_model} and \ref{fig:0408_comp_model} display the most likely models for the power-law and composite profiles, respectively. In addition to the lensed complex structures in the Einstein ring from the extended quasar host galaxy, the lensed arcs S2 and S3 are also reproduced very accurately. Moreover, the models reproduce the additional split image C2 on the other side of G2 from image C.

Table \ref{tab:model_logzs} tabulates the evidences for different model choices. We combine the model posteriors weighted by the evidence ratios within each lens model family -- power-law or composite -- to marginalize over the model choices. Previous studies -- e.g. \citet{Birrer19, Rusu19, Chen19} -- use the Bayesian information criterion (BIC) as an estimate of the evidence for Bayesian model averaging \citep[BMA; e.g.][]{Madigan94, Hoeting99}. Whereas BIC estimates the model evidence based on the maxima of the likelihood function under certain assumptions, nested sampling directly computes the model evidence by integrating over the whole prior space. Hence, the evidence obtained from nested sampling is more robust. 

\edit{We account for sparse sampling from the model space by down-weighting the evidence ratios between the models. Effectively, we want to estimate the factor $\Delta S_n$ in equation (\ref{eq:discrete_model_posterior}) to account for sparse sampling. We estimate the sparsity of the sampled models by taking the variance of $\Delta \log \mathcal{Z}$ between ``neighbouring'' model pairs that differ in only one model setting. In this way, we are being conservative by accepting more variance in our lens model posterior to avoid any bias due to sparse sampling from the model space. For 12 models within each mass profile family, we then have 20 such ``neighbouring'' models. We obtain $\sigma_{\Delta \log \mathcal{Z}}^{\rm model}$ = \editfr{436} for the power-law models and $\sigma_{\Delta \log \mathcal{Z}}^{\rm model}$ = \editfr{1210} for the composite models. We follow \citet{Birrer19} to \editfr{adjust} the relative weights of the model by convolving the evidence ratios with a Gaussian kernel with standard deviation $\sigma_{\Delta \log \mathcal{Z}} = ({{\sigma_{\Delta \log \mathcal{Z}}^{\rm model}}^2 + {\sigma^{\rm numeric}_{\Delta \log \mathcal{Z}}}^2})^{1/2}$. Here, we take ${\sigma^{\rm numeric}_{\Delta \log \mathcal{Z}}}=34$ as explained in Appendix \ref{app:dypolychord_setup}.} \editref{Following \citet{Birrer19}, we first calculate the absolute weight $W_{n, \text{abs}}$ of the $n$\textsuperscript{th} model by convolving the evidence uncertainty with the evidence ratio function $f(x)$ as
\begin{linenomath}\begin{equation} \label{eq:abs_model_weight}
	W_{n, \text{abs}} = \frac{1}{\sqrt{2 \pi} \sigma_{\Delta \log \mathcal{Z}}} \int_{-\infty}^{\infty} f(x) \exp \left[-\frac{(\log \mathcal{Z}_n - x)^2}{2 \sigma_{\Delta \log \mathcal{Z}}^2} \right] \mathrm{d} x,
\end{equation}\end{linenomath}
where we define the evidence ratio function $f(x)$ as
\begin{linenomath}\begin{equation}
	f(x) \equiv 
     \begin{cases}
       1 &\quad x \geq \mathcal{\log Z}_{\rm max}, \\\
       \exp\left( x - \mathcal{\log Z_{\
       \rm max}} \right) &\quad x < \mathcal{\log Z}_{\rm max}. \
     \end{cases}
\end{equation}\end{linenomath}
We then obtain the relative weight $W_{n, \text{rel}}$ by simply normalizing the absolute weights by the maximum absolute weight as
\begin{linenomath}\begin{equation}
	W_{n, \text{rel}} = \frac{W_{n, \text{abs}}}{\max \left(\left\{ W_{n, \text{abs} } \right\} \right)}.
\end{equation}\end{linenomath}
In the limit of $n \to \infty$, we would have a perfect sampling of models from the model space. In that case, we have $\sigma^{\rm model}_{\Delta \log \mathcal{Z}} \to 0$ and $\sigma_{\Delta \log \mathcal{Z}} \to \sigma_{\Delta \log \mathcal{Z}}^{\rm numeric}$. Furthermore, if the evidence value is perfectly computed with $\sigma_{\Delta \log \mathcal{Z}}^{\rm numeric}=0$, then the exponential function inside equation (\ref{eq:abs_model_weight}) becomes a Dirac delta function. In that limit, the relative weight $W_{n, \text{rel}}$ approaches the evidence ratio as
\begin{equation}
	\lim_{\sigma_{\Delta \log \mathcal{Z}} \to 0} W_{n, \text{rel}} = \frac{\mathcal{Z}_{n}}{\mathcal{Z_{\rm max}}}.
\end{equation}
}

\begin{figure*}
	\includegraphics[width=\textwidth]{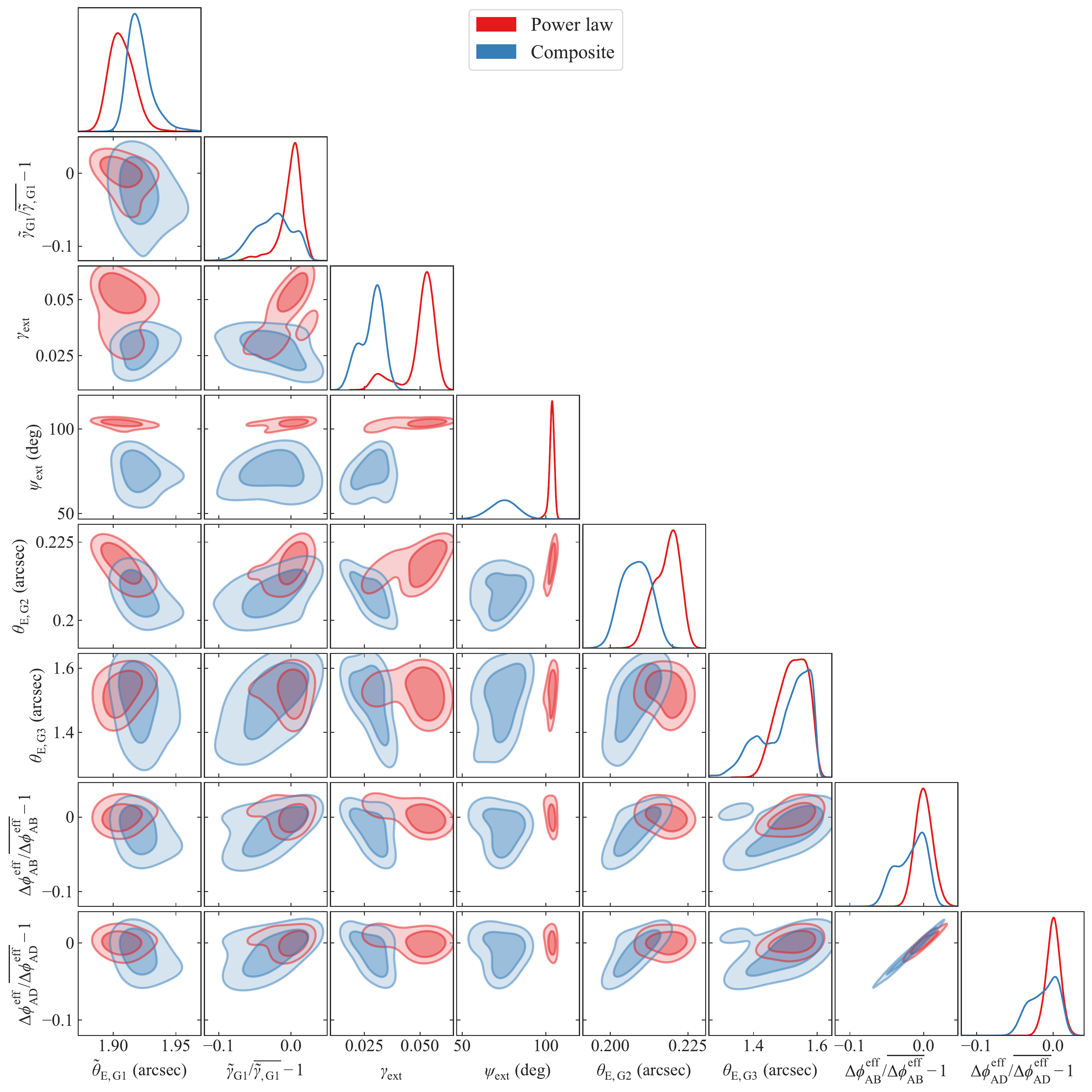}
	\caption{\label{fig:compare_mass_model}
	Comparison of the lensing properties between the power-law and composite mass models. \editth{The posteriors are weighted combinations of 12 models for each mass model family.} Here, $\tilde{\theta}_{\rm E}$ is the Einstein radius defined to contain mean convergence of 1, and the profile slope $\tilde{\gamma}$ is defined as the derivative of the convergence profile at $\tilde{\theta}_{\rm E}$. We blind the profile-slope and the effective Fermat potential differences by subtracting the mean and then normalize it with the mean to show relative offsets in percentage.}
\end{figure*}

In Fig. \ref{fig:compare_mass_model}, we compare the posteriors of important lens parameters between power-law and composite mass profiles after \editfv{marginalizing over the model space using the adjusted evidence ratios as described above}.

\renewcommand{\arraystretch}{1.2}
\begin{table*}
\caption{\label{tab:model_logzs}
	Evidence for different lens model setups. \editfr{The evidence ratio $\Delta \log \mathcal{Z}$ is calculated only within the particular mass profile family -- power law or composite. \editsx{The model setups are ordered from higher to lower evidence within each mass profile family}. The relative weights for each model are obtained from the evidence ratios adjusted for sparse sampling from the model space as described in Section \ref{sec:modelling_workflow}}.
	}
\begin{tabular}{lcccccc}
\hline
Mass profile & Source $n_{\rm max}$ & $z_{\rm S3}$ & G4--G6 mass profile & $\log \mathcal{Z}$ & $\Delta \log \mathcal{Z}$ & Relative weight, $W_{\rm rel}$\\
 & & & & ($\pm 24$) & ($\pm 34$) & \\
\hline
Power law & \{8, 5, 3\} & 2.375 & SIS & $-$25087 & 0 & 1.00 \\
Power law & \{6, 3, 2\} & 2.375 & SIS & $-$25215 & 128 & 0.77 \\
Power law & \{8, 5, 3\} & 2.228 & SIS & $-$25232 & 145 & 0.74 \\
Power law & \{7, 4, 2\} & 2.375 & SIS & $-$25317 & 230 & 0.60 \\
Power law & \{7, 4, 2\} & 2.228 & SIS & $-$25421 & 333 & 0.45 \\
Power law & \{6, 3, 2\} & 2.228 & SIS & $-$25450 & 363 & 0.41 \\
Power law & \{8, 5, 3\} & 2.375 & NFW & $-$25578 & 490 & 0.26 \\
Power law & \{7, 4, 2\} & 2.228 & NFW & $-$25624 & 537 & 0.22 \\
Power law & \{6, 3, 2\} & 2.228 & NFW & $-$25656 & 569 & 0.19 \\
Power law & \{6, 3, 2\} & 2.375 & NFW & $-$26432 & 1345 & 0.00 \\
Power law & \{8, 5, 3\} & 2.228 & NFW & $-$26469 & 1382 & 0.00 \\
Power law & \{7, 4, 2\} & 2.375 & NFW & $-$26551 & 1464 & 0.00 \\
Composite & \{7, 4, 2\} & 2.228 & SIS & $-$25055 & 0 & 1.00 \\
Composite & \{8, 5, 3\} & 2.228 & SIS & $-$25121 & 66 & 0.96 \\
Composite & \{8, 5, 3\} & 2.375 & SIS & $-$25147 & 92 & 0.94 \\
Composite & \{6, 3, 2\} & 2.228 & SIS & $-$25155 & 100 & 0.94 \\
Composite & \{7, 4, 2\} & 2.375 & SIS & $-$25155 & 100 & 0.93 \\
Composite & \{6, 3, 2\} & 2.375 & NFW & $-$25252 & 197 & 0.87 \\
Composite & \{6, 3, 2\} & 2.375 & SIS & $-$25292 & 237 & 0.85 \\
Composite & \{7, 4, 2\} & 2.375 & NFW & $-$25482 & 427 & 0.72 \\
Composite & \{6, 3, 2\} & 2.228 & NFW & $-$25985 & 930 & 0.44 \\
Composite & \{8, 5, 3\} & 2.375 & NFW & $-$26541 & 1486 & 0.22 \\
Composite & \{8, 5, 3\} & 2.228 & NFW & $-$27073 & 2018 & 0.09 \\
Composite & \{7, 4, 2\} & 2.228 & NFW & $-$28979 & 3924 & 0.00 \\
\hline
\end{tabular}
\end{table*}

\subsection{Combining the time delays, kinematics, and external convergence information} \label{sec:combine_td_likelihood}

To combine the time-delay likelihood with the lens imaging likelihood, we importance sample from the lens model posterior weighted by the joint time-delay and kinematics likelihood \citep{Lewis02}. In Section \ref{sec:td_vd_likelihood}, we fold in the time-delay and kinematic likelihoods into the lens model posterior. Then in Section \ref{sec:ext_convergence_sampling}, we add the external convergence distribution to the cosmological distance posteriors. Finally in Section \ref{sec:microlensing}, we check the impact of microlensing time-delay on our inference of the effective time-delay distance. 

\subsubsection{Combining time-delay and kinematics likelihoods}\label{sec:td_vd_likelihood}
The posterior samples from nested sampling carry weights proportional \editsx{to} their contribution to the posterior mass. We obtain 10000 equally weighted posterior samples through weighted random sampling from the nested sampling chain for each lens model setup. \editfr{To combine distance posteriors from different lens model setups, we randomly sample a number of points from the lens model posterior for each setup within a mass-model family, where the sampled number is proportional to the relative weight computed from the adjusted evidence ratio (Table \ref{tab:model_logzs}).} We then uniformly sample \editfr{4000} points of $(D^{\rm eff}_{\Delta t},\ D_{\rm s}/D_{\rm ds})$ for each lens model sample from [0, 2.15]$D_{\Delta t}^{\rm eff,fiducial}\ \times$ [0.35, 1.35]$(D_{\rm s}/D_{\rm ds})^{\rm fiducial}$. The chosen boundaries fully contain (>5$\sigma$) the distance posteriors and they also encompass the range allowed by the priors $H_0 \in [0, 150]$ \Hunit\ and $\Omega_{\rm m} \in [0.05, 0.5]$, \editsx{given} our fiducial cosmology. This procedure effectively \editfr{gives us 4000$\times N_{\rm  sample}$ points from the joint space combining the lens model parameters and $(D^{\rm eff}_{\Delta t},\ D_{\rm s}/D_{\rm ds})$, where $N_{\rm sample}$ is the number of lens model samples.} We then importance sample \editfr{from these 4000$\times N_{\rm  sample}$} points weighted by the joint time-delay and kinematic likelihood to obtain the marginalized posterior distribution of $(D^{\rm eff}_{\Delta t},\ D_{\rm s}/D_{\rm ds})$. We only consider $\Delta t_{\rm AB}$ and $\Delta t_{\rm AD}$ in the time-delay likelihood as $\Delta t_{\rm BD}$ is not independent of the others. We then transform the $(D^{\rm eff}_{\rm \Delta t},\ D_{\rm s}/D_{\rm ds})$ distribution into the $(D^{\rm eff}_{\rm \Delta t}, \ D_{\rm d})$ distribution.

\editsx{Since} there are four observing setup for G1's central \editth{line-of-sight} velocity dispersion, we \editth{compute four line-of-sight velocity dispersions for each sample from the lens model posterior. We account for covariance between these four measurements in the kinematic likelihood (see Section \ref{sec:velocity_dispersion_data} for covariance matrix definition).} We choose a uniform prior for the anisotropy scale radius as $r_{\rm ani} \sim U(0.5\theta_{\rm hl}, 5\theta_{\rm hl})$, where $\theta_{\rm hl}$ is the half-light radius in the F160W band. As we model the deflector light distribution using a double S\'ersic profile, we numerically compute the radius of the circular aperture that contains half of the total flux from the double S\'ersic profile.


\subsubsection{Adding the external convergence distribution into the cosmological distance posterior} \label{sec:ext_convergence_sampling}

\begin{figure}
	\includegraphics[width=\columnwidth]{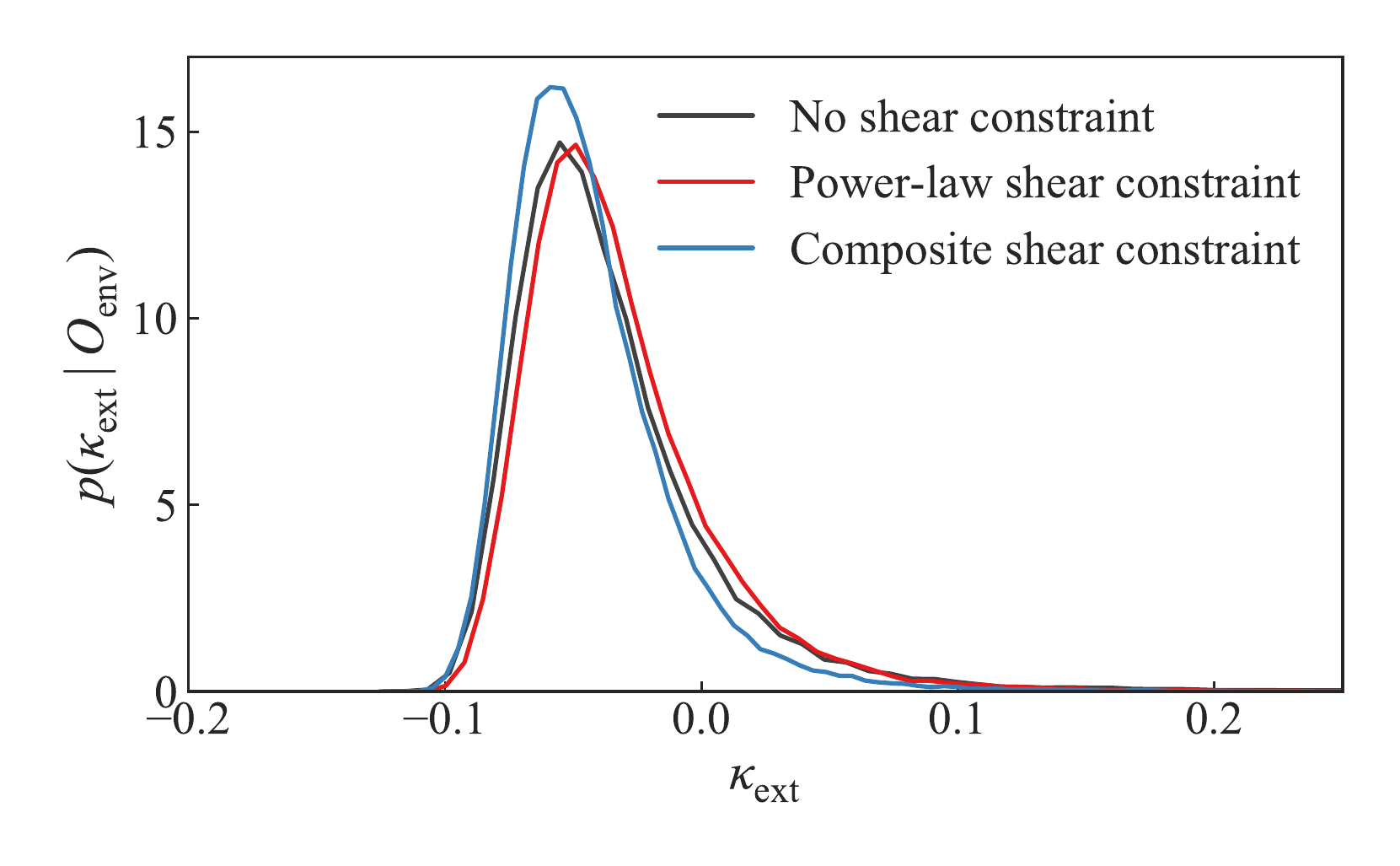}
	\caption{\label{fig:k_ext}
	\editfr{}Distribution of external convergence $\kappa_{\rm ext}$. \editsx{The black line shows the $\kappa_{\rm ext}$ distribution from \citet{BuckleyGeer20} without imposing any shear constraint. The red and blue lines show the distributions with shear constraints from the power-law and the composite mass models, respectively.} 
	}
\end{figure}

\editfr{We apply a selection criterion on the $P(\kappa_{\rm ext})$ estimated in \citet{BuckleyGeer20} by requiring that the selected lines of sight also correspond to the external shear values \editsx{predicted} by our lens models. In Fig. \ref{fig:k_ext}, we show the probability distribution of $\kappa_{\rm ext}$ for the fiducial choice of constraints explored in \citet{BuckleyGeer20}, and the two $\kappa_{\rm ext}$ distributions consistent with the external shear values for the power-law and composite mass profiles [see \citet{BuckleyGeer20} for further details].}

%
%

We simple-sample from the external convergence distribution corresponding to each mass profile.\editfr{We correct the distance posterior using the sampled external convergence} according to equation (\ref{eq:corrected_time_delay_distance}).

\begin{figure*}
	\includegraphics[width=0.33\textwidth]{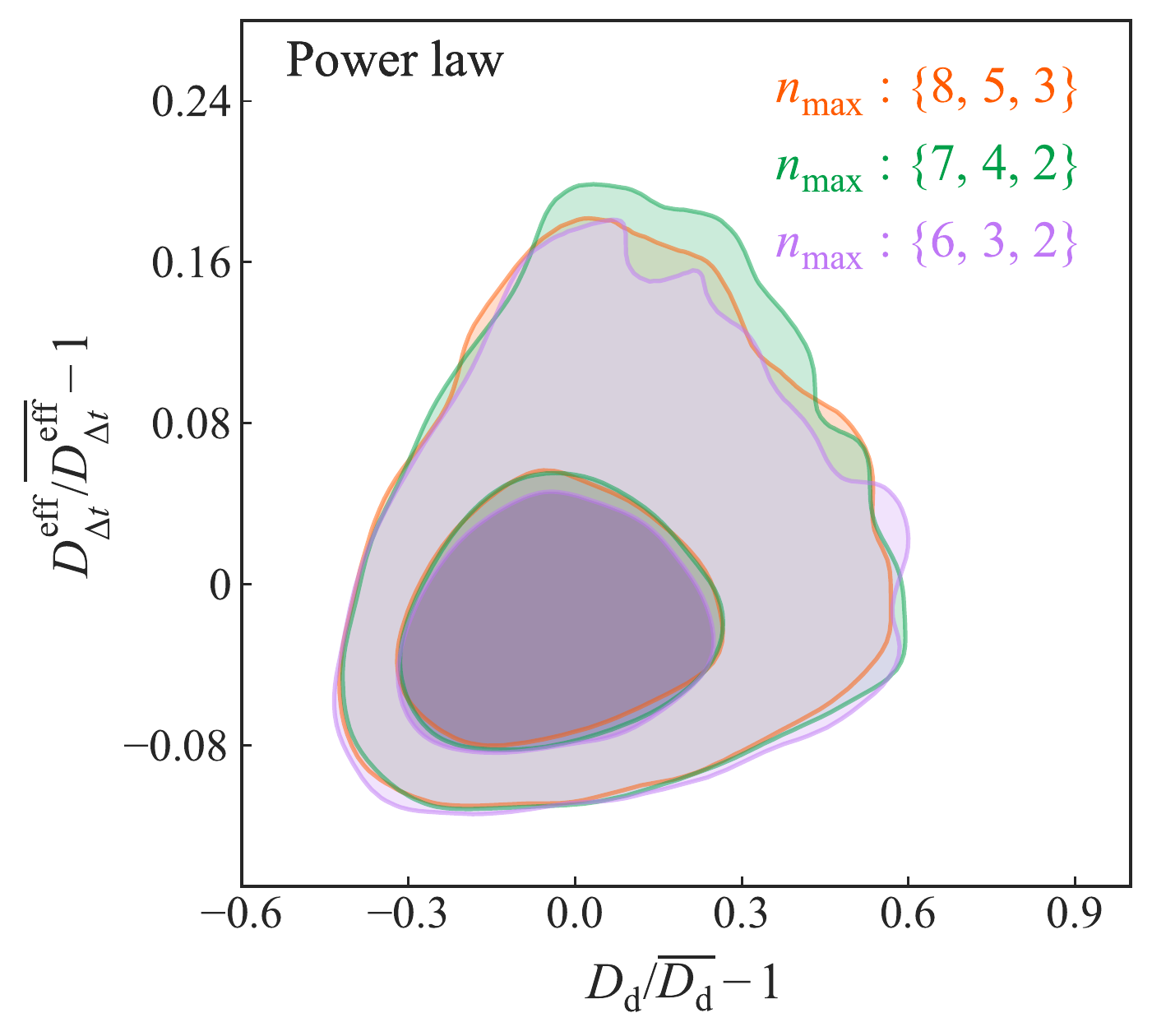}
	\includegraphics[width=0.33\textwidth]{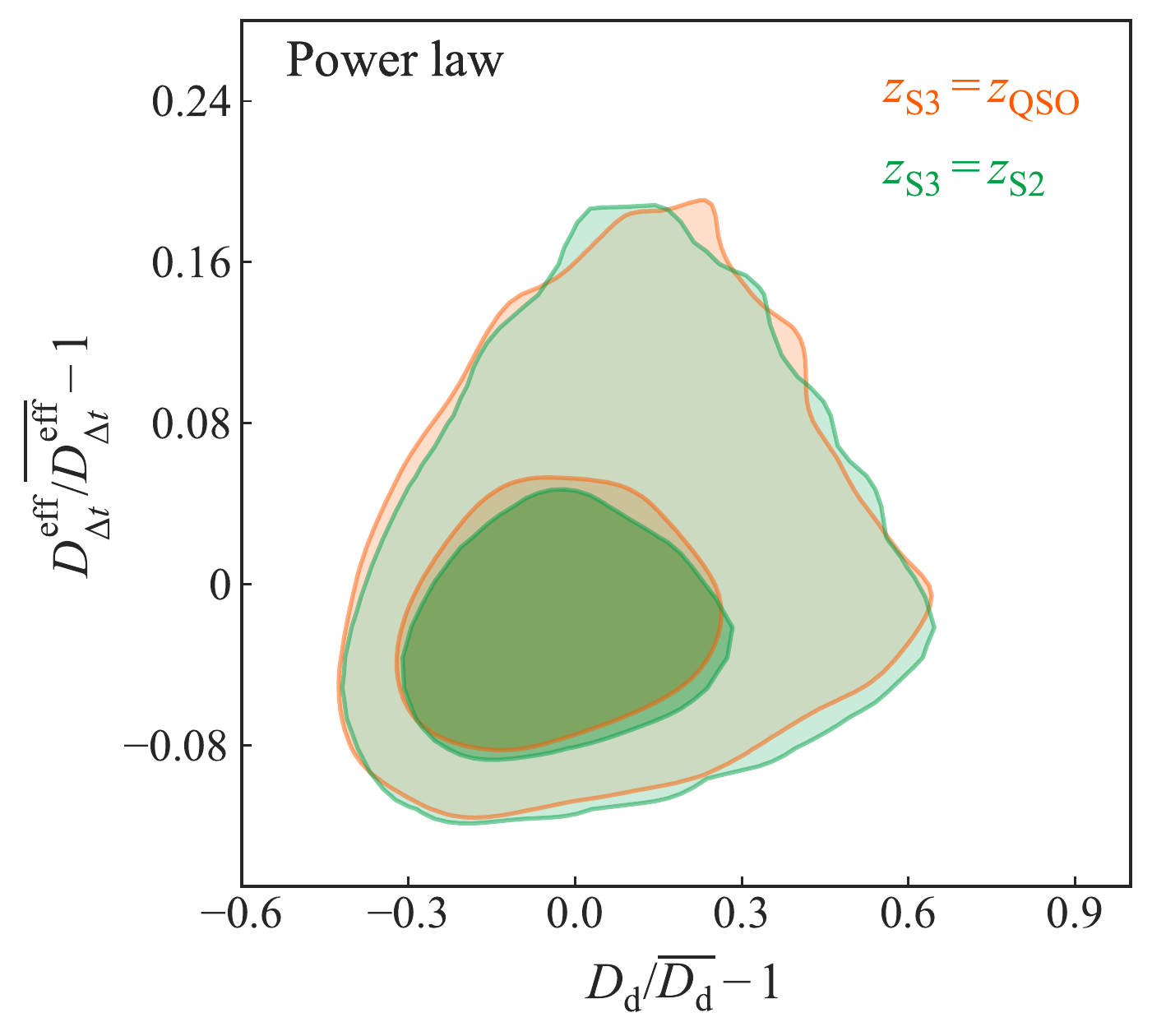}
	\includegraphics[width=0.33\textwidth]{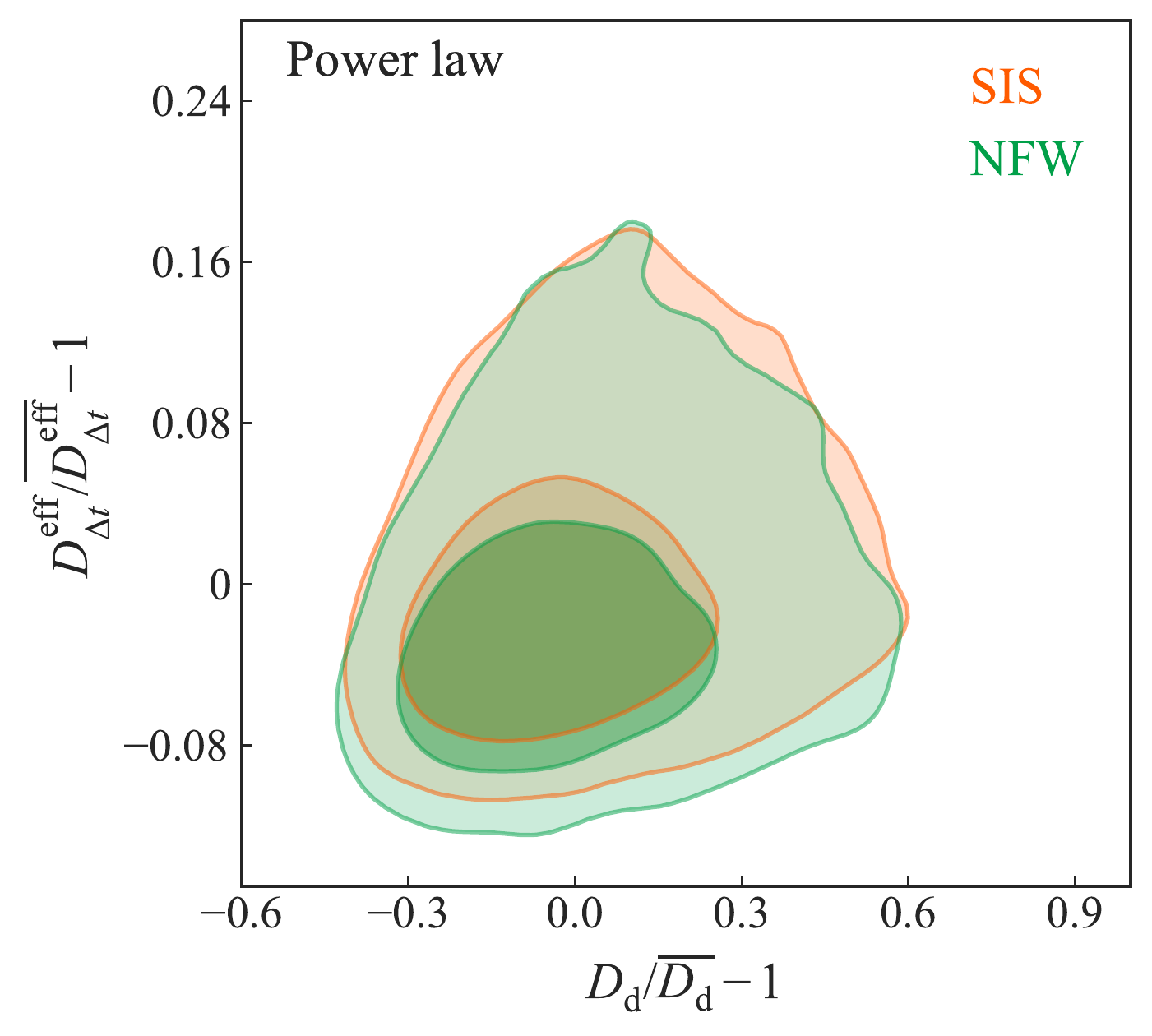} \\
	\includegraphics[width=0.33\textwidth]{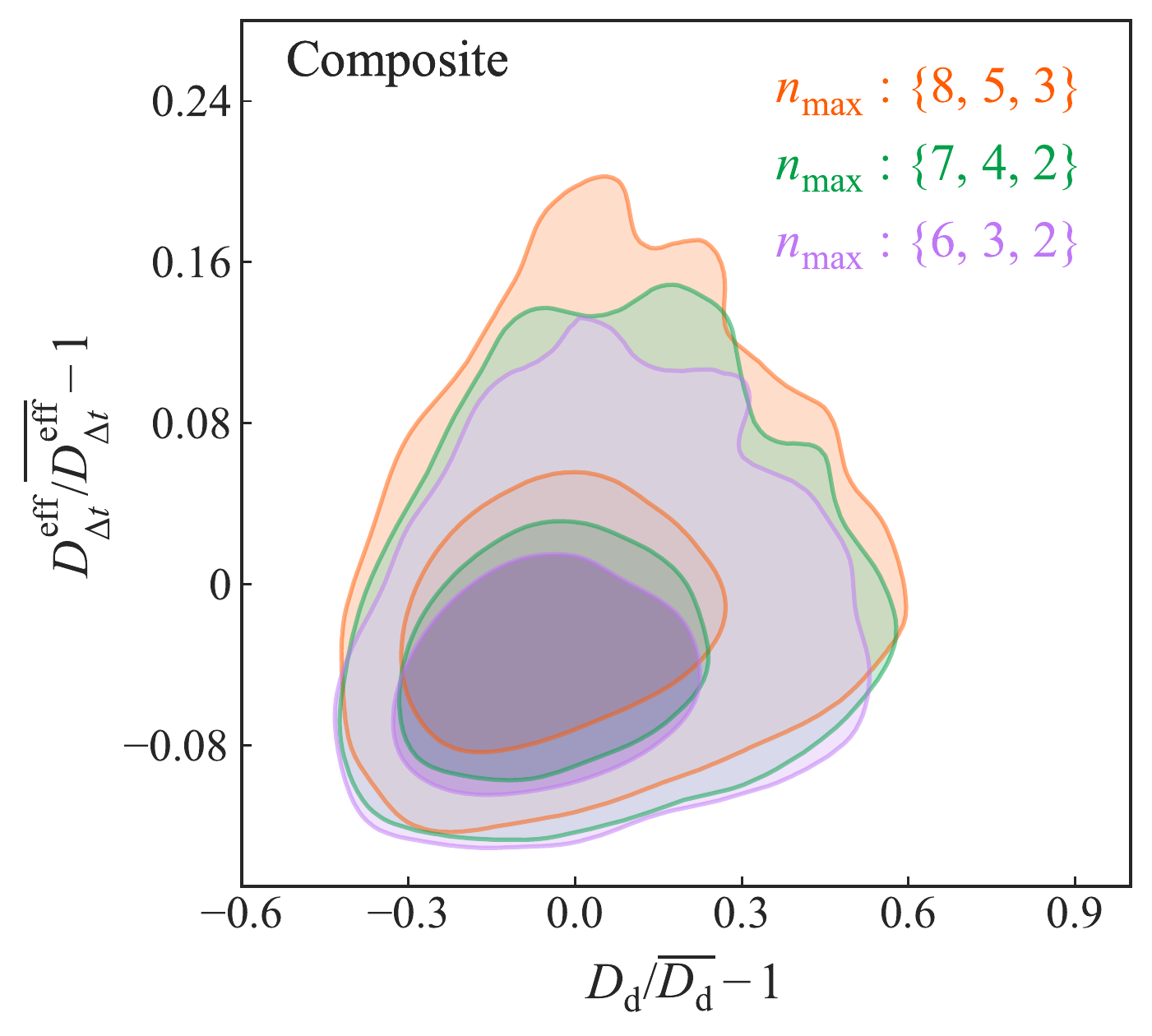}
	\includegraphics[width=0.33\textwidth]{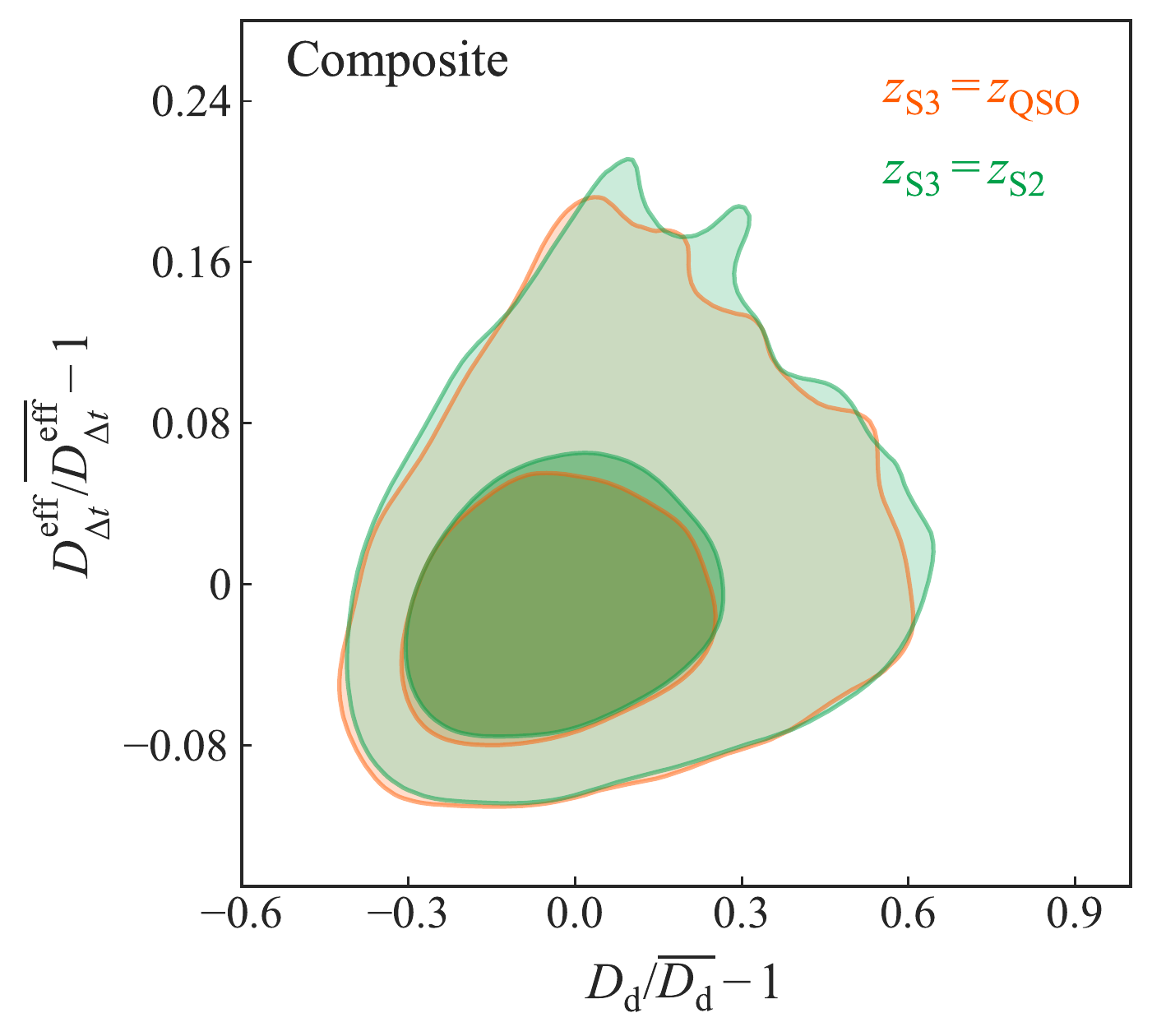}
	\includegraphics[width=0.33\textwidth]{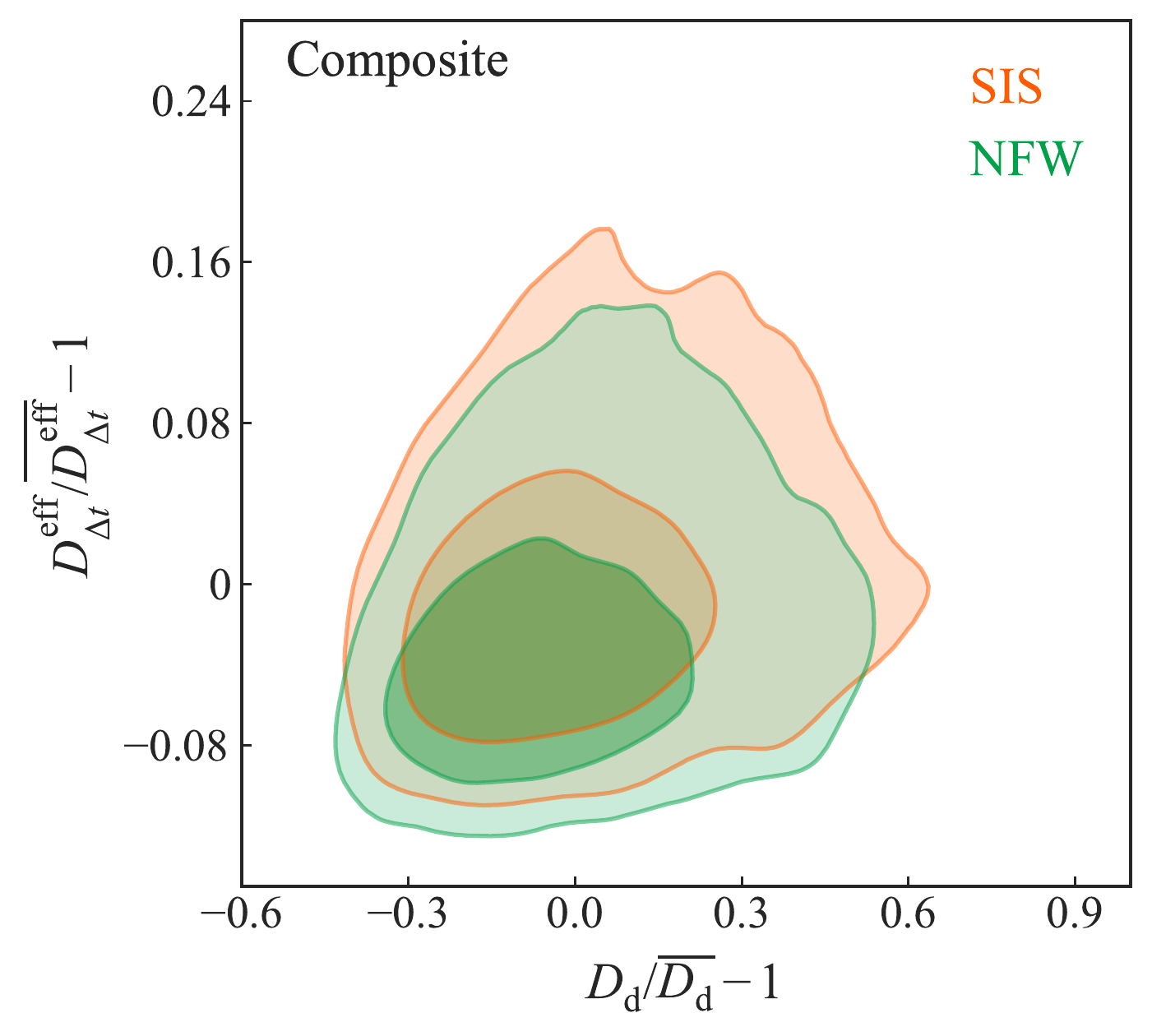} \\
	\caption{\label{fig:model_compare_distance}
	Comparison of the distance posteriors between choices of the lens model setups for the power-law models (top row) and the composite models (bottom row). \editth{The distance posteriors are weighted combinations of different runs with one common model setting as specified.} One of the posteriors mean is subtracted from all them and then they are normalized by the mean to get the relative shifts in percentage. \textbf{Left-hand panel:} the distance posteriors for different settings of the source components' $n_{\rm max}$. \textbf{Centre:} the distance posteriors for different redshifts of S3. \editth{\textbf{Right-hand panel:} the distance posteriors for SIS and NFW mass profiles for the line-of-sight galaxies. The distance posteriors from different models are consistent.}
		}
\end{figure*}

\begin{figure}
	\includegraphics[width=\columnwidth]{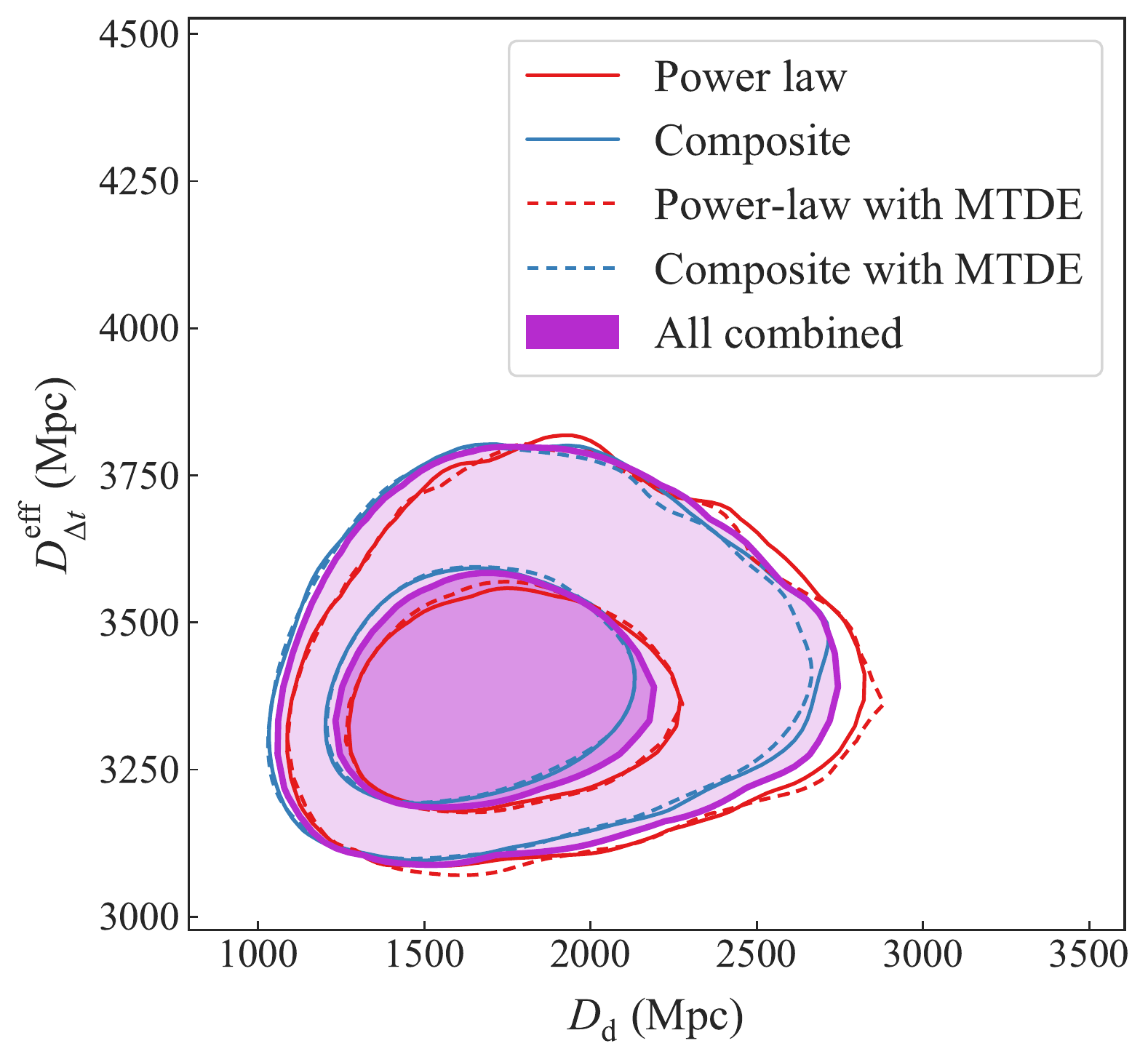}
	\caption{\label{fig:final_combined_distance_posterior}
		\editfr{Cosmological distance posteriors for power-law (red) and composite mass-profile families (blue), and for all models combined (purple). All the models are combined within each categories weighted by their adjusted evidence ratios (Table \ref{tab:model_logzs}). The solid contours for the power-law and composite mass profiles are without accounting for the microlensing time-delay effect (MTDE), and the corresponding dashed contours show the ones with the MTDE. The MTDE is negligible in our analysis and we do not incorporate this effect in the final distance posterior (purple contours) combining all the models.}
	}	
\end{figure}

Fig. \ref{fig:model_compare_distance} shows the comparison of distance posteriors between the mass model families and between different settings within a mass model family. \editfr{The distance posteriors are consistent between different model setups.} From the combined distance posterior from all the models, we obtain the 1D marginalized constraints for the effective time delay distance $D_{\Delta t}^{\rm eff}=$ \Ddtval\ Mpc and the angular diameter distance $D_{\rm d}=$ \Ddval\ Mpc (Fig. \ref{fig:final_combined_distance_posterior}).

\subsubsection{Microlensing time-delay effect} \label{sec:microlensing}
We check the impact of microlensing by the deflector galaxy's stars on the measured time-delays in our analysis \citep{Tie18}. \editref{Note, this microlensing time-delay from \citet{Tie18} is based on the lamp-post model for the AGN accretion disc \citep{Shakura73}. This effect will not necessarily exist for other disc models.} 

We generate the microlensing time-delay maps following \citet{Bonvin18} and \citet{Chen18}. The estimated microlensing time-delay depends on the magnification of the lens model\editsx{, on} the stellar contribution at the image position, and \editsx{on} the properties of the black hole's accretion disc. 

\begin{figure}
	\includegraphics[width=\columnwidth]{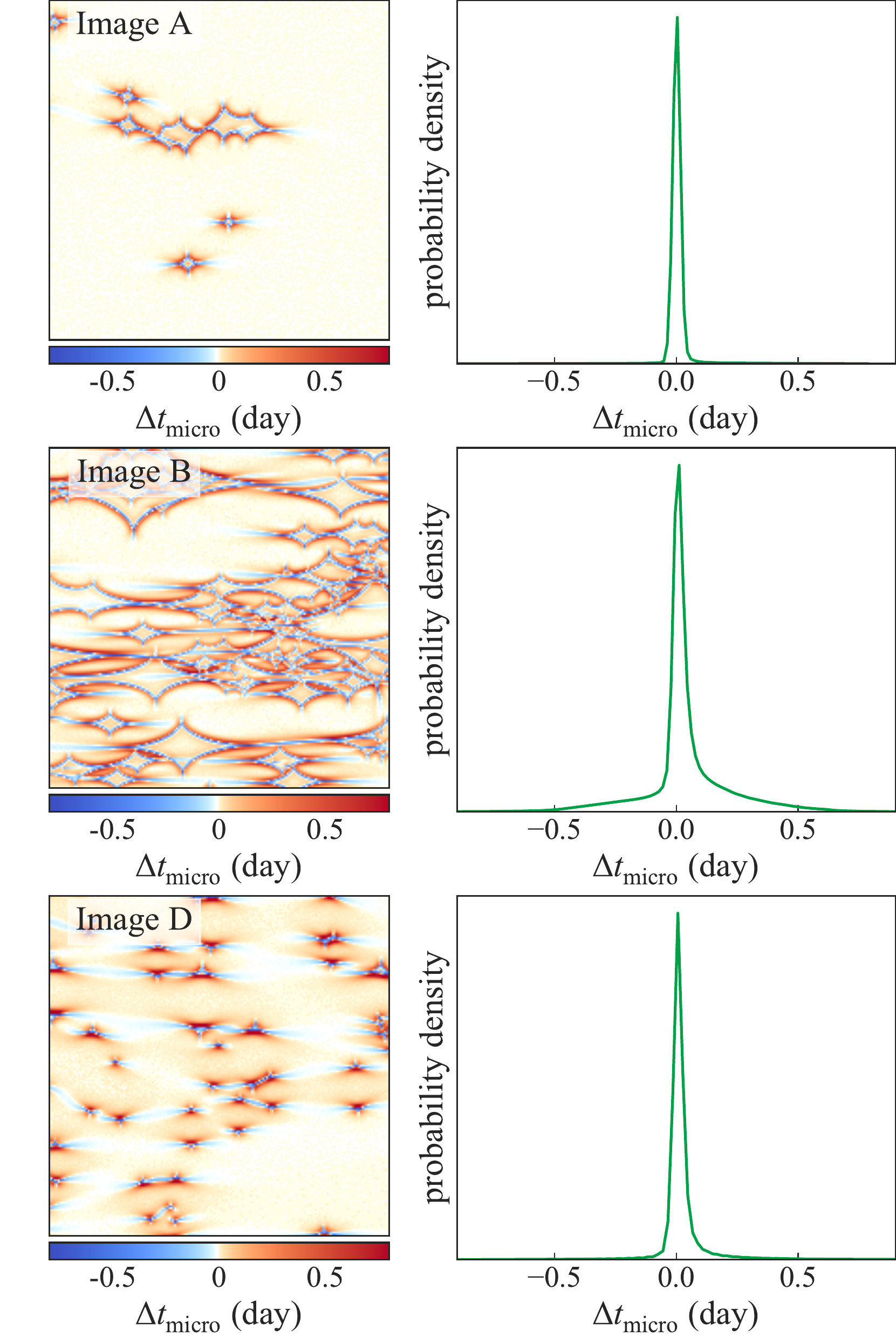}
	\caption{\label{fig:microlensing}
	Microlensing time-delay maps and the probability of microlensing time-delays for images A, B, and D. Each microlensing time-delay map on the left-hand column is created from the relevant image magnification, stellar contribution to the convergence at the image position, and the accretion disc properties of the black hole. The probability density function of the microlensing time-delay at each image position is shown in the right-hand column. The expected fluctuation in the measured time-delays is small relative to the measurement uncertainty.
	}
\end{figure}

We estimate the black hole mass using the scaling relation
\begin{linenomath}
	\begin{equation}
	\begin{split}
		\log_{10} \left( \frac{M_{\rm BH,Mg\ \textsc{ii}}}{M_{\odot}} \right) &= b + m \log_{10} \left( \frac{{\rm EW_{\rm Mg \ \textsc{ii}}}}{\si{\angstrom}}\right)  \\
		& \qquad \qquad + 2 \log_{10} \left( \frac{{\rm FWHM_{Mg\ \textsc{ii}}}}{{\rm km \ s^{-1}}} \right)
		\end{split}
	\end{equation}	
\end{linenomath}
between the black hole mass $M_{\rm BH, vir}$, and the rest-frame full width at half-maximum (FWHM) and equivalent width (EW) of the Mg \textsc{ii} broad line. \edit{This equation is equivalent to the $M_{\rm BH} \propto R_{\rm BLR} \sigma_{\rm BLR}^2$, where $R_{\rm BLR}$ is the radius of the broad-line region and $\sigma_{\rm BLR}$ is the velocity dispersion of the broad-line region. Here, we used the EW as a proxy for $R_{\rm BLR}$ and \editsx{the} FWHM as a proxy for $\sigma_{\rm BLR}$ [cf. equation (2) of \citet{Shen11}].} We estimate the parameters of this scaling relation using the SDSS quasars from the catalogue provided by \citet{Shen11} as $b=2.71$ and $m=-0.61$. We only take the quasars with non-zero entries for $M_{\rm BH, vir}$, FWHM${}_{\rm Mg\ \textsc{ii}}$, and EW${}_{\rm Mg\ \textsc{ii}}$. Moreover, we only select the quasars within 1300 km s\textsuperscript{$-$1} < FWHM${}_{\rm Mg\ \textsc{ii}}$ < 30000 km s\textsuperscript{$-$1} to remove the quasars creating stripe-like features at the boundaries of the $M_{\rm BH, vir}$--FWHM${}_{\rm Mg\ \textsc{ii}}$ scatter plot. As a result, we have 85038 selected quasars to fit the above scaling relation. We obtain the rest-frame FWHM and EW of the Mg \textsc{ii} line from the quasar spectra at image B and image D from \citet{Agnello17} as FWHM${}_{\rm Mg\ \textsc{ii}}^{\rm B}$ = 3413 km s\textsuperscript{$-$1}, EW${}_{\rm Mg\ \textsc{ii}}^{\rm D}$ = 37.3 \si{\angstrom}, FWHM${}_{\rm Mg\ \textsc{ii}}^{\rm D}$ = 2952 km s\textsuperscript{$-$1}, EW${}_{\rm Mg\ \textsc{ii}}^{\rm B}$ = 30.5 \si{\angstrom}. We apply a magnification correction to the estimated black hole mass from each image as
\begin{linenomath}
	\begin{equation}
		\log_{10} \left( \frac{M_{\rm BH}}{M_{\odot}} \right) = \log_{10} \left( \frac{M_{\rm BH, Mg \ \textsc{ii}}}{M_{\odot}} \right) - g \log_{\rm 10} \mu, 
	\end{equation}	
\end{linenomath}
where we take the calibration factor $g=0.5$ \citep{Vestergaard06}. We also add 0.25 dex uncertainty to the estimated black hole mass to account for the limitation of using Mg \textsc{ii} to measure \editsx{it} \citep{Woo18}. Averaging over the estimates from image B and D, we obtain the black hole mass of the quasar as $\log_{10} \left(M_{\rm BH,vir}^{\rm J0408}/M_{\odot}\right) = 8.41 \pm 0.27$. We also estimate the Eddington ratio using the scaling relation
\begin{linenomath}
	\begin{equation}
		\log_{\rm 10} \left( \frac{L_{\rm bol}^\prime}{L_{\rm Edd}^\prime} \right) = b + m\log_{10} \left( \frac{M_{\rm BH, Mg\ \textsc{ii}}}{M_{\odot}} \right).
	\end{equation}	
\end{linenomath}
We estimate $m=-0.33$, $b=2.2$ with an intrinsic scatter of 0.64 \editth{dex} by fitting the relation to the same objects selected from \citet{Shen11}'s catalogue. We also apply a magnification correction on the Eddington ratio obtained for each image as
\begin{linenomath}
	\begin{equation}
		\log_{\rm 10} \left( \frac{L_{\rm bol}}{L_{\rm Edd}} \right) = \log_{\rm 10} \left( \frac{L_{\rm bol}^\prime}{L_{\rm Edd}^\prime} \right) + (g-1) \log_{10} \mu
	\end{equation}
\end{linenomath}
\citep{Birrer19}. As a result, we obtain $\log_{10}\left( L_{\rm bol} / L_{\rm Edd}\right) = -1.48 \pm 0.27$ after averaging over the estimates from images B and D. \editfr{The accretion disc size $R_0$ is determined assuming a standard accretion disc model \citep{Shakura73}.} In Tables \ref{tab:bh_mass_lens_parameters} and \ref{tab:bh_mass_disk_parameters}, we tabulate the values used to create the microlensing time delay maps shown in Fig. \ref{fig:microlensing}. \editref{We assumed Salpeter initial mass function (IMF) and ignored the uncertainty on the convergence and shear parameters. Shifting the stellar convergence $\kappa_{\star}$ by the typical uncertainty of 10 per cent or changing the IMF has negligible impact on the estimated microlensing time-delay distribution in Fig. \ref{fig:microlensing}.}

\begin{table}
\caption{\label{tab:bh_mass_lens_parameters}
	Lensing quantities at the image positions used to create the microlensing time-delay map. \editref{The total convergence $\kappa$, the stellar convergence $\kappa_{\star}$, the shear $\gamma_{\rm shear}$, and the magnification $\mu$ are obtained from the best-fitting composite model. As the local slope of the composite model at the image positions can deviate from the 3D slope $\gamma=2$ of an isothermal profile, $\kappa$ and $\gamma_{\rm shear}$ are not necessarily identical.}
	}
\begin{tabular}{lcccc}
\hline
TImage & $\kappa$ & $\kappa_{\star}$ & $\gamma_{\rm shear}$ & $\mu$ \\ 
\hline
A & 0.46 & 0.03 & 0.19 & \phantom{0}3.9 \\
B & 0.59 & 0.06 & 0.32 & 15.5 \\
D & 0.70 & 0.13 & 0.69 & -2.6 \\
\hline
\end{tabular}
\end{table}

\begin{table}
\caption{\label{tab:bh_mass_disk_parameters}
	Properties of the quasar accretion disc used to compute the microlensing time-delay maps. 
	}
\begin{tabular}{lr}
\hline
Quantity & Value\\ 
\hline
Black hole mass, $\log_{10} (M_{\rm BH}/M_{\odot})$ & 8.41 $\pm$ 0.27 \\
Eddington ratio, $\log_{10} (L_{\rm bol}/L_{\rm Edd})$ &  -1.48 $\pm$ 0.27 \\
Accretion disc size, $R_0$ (cm)& 3.125$\times 10^{14}$\\
Accretion efficiency, $\eta$ & 0.1 \\
Central wavelength for light curve observation, $\lambda$ ($\mu$m) & 0.668 \\
Average foreground stellar mass, $\langle M_{\star} / M_{\odot}\rangle$ & 0.3 \\
\hline
\end{tabular}
\end{table}

We account for the microlensing time-delay effect in the measured time delay by sampling from the microlensing time-delay distribution and adjusting the measured time delay as
\begin{linenomath}
	\begin{equation}
		\Delta t_{\rm XY, adjusted} = \Delta t_{\rm XY, measured} + t_{\rm X, micro} - t_{\rm Y, micro}.
	\end{equation}	
\end{linenomath}
The microlensing time-delay effects is small compared to the uncertainty on the measured time delays. Thus, accounting for this microlensing time-delay effect does not shift the effective time-delay distance \editsx{by more than \editfv{0.1} per cent} \editfr{(Fig. \ref{fig:final_combined_distance_posterior})}. We only perform this step as a check and we do not include this effect in our inference of \Ho.

\subsection{Inference of \Ho} \label{sec:h0_inference}
\begin{figure}
	\includegraphics[width=\columnwidth]{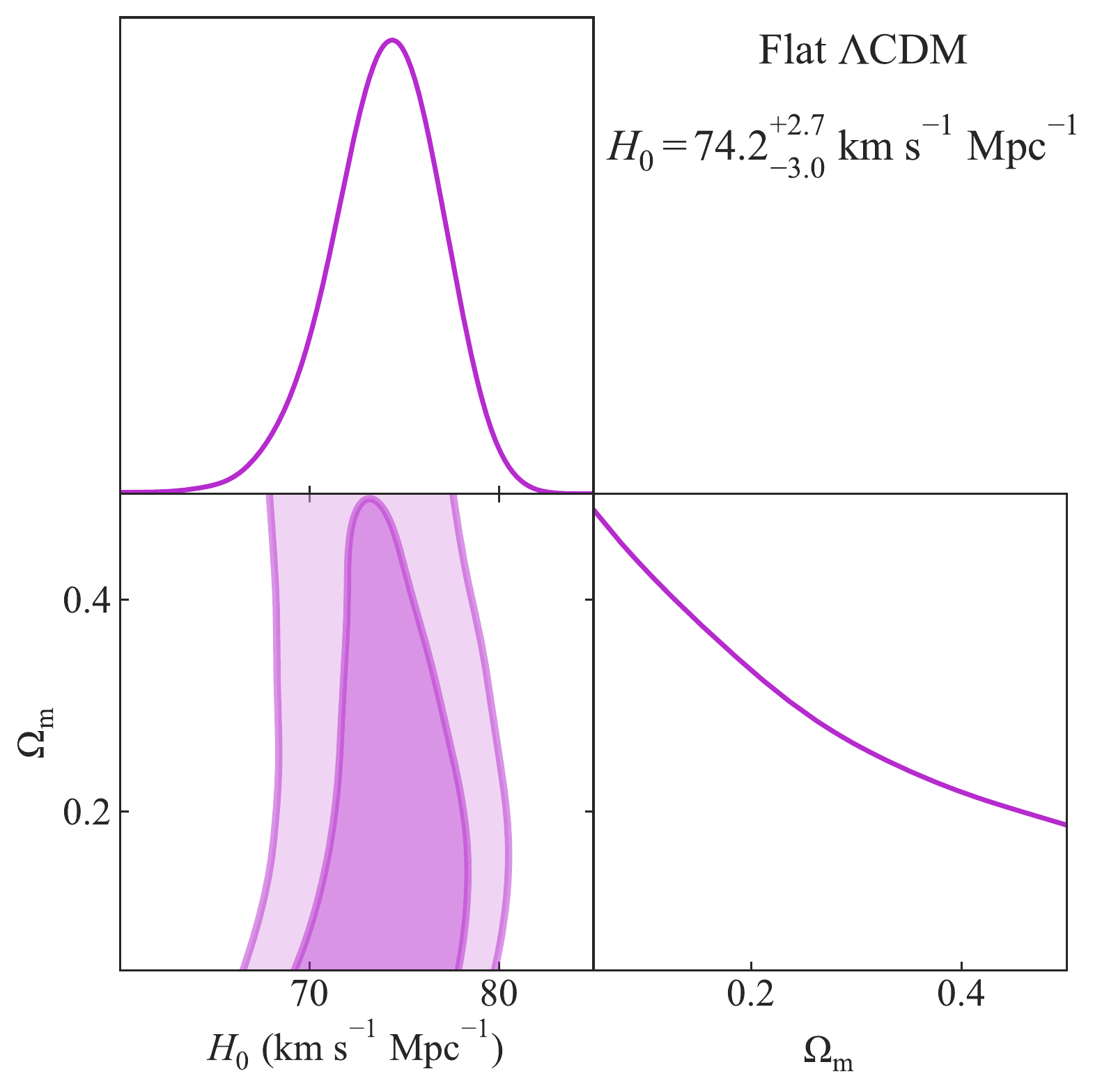}
	\caption{\label{fig:final_h0_posterior}
		\editfr{\editfv{Posterior probability distribution functions} of \Ho\ and $\Omega_{\rm m}$ in the \lcdm\ cosmology for the 2D distance posterior in Fig. \ref{fig:final_combined_distance_posterior}. The inferred Hubble constant is \Ho\ = \Hval\ \Hunit.}
	}	
\end{figure}

The cosmological distance posterior contains all the cosmographic information. We infer \Ho\ from this distance posterior for a flat \lcdm\ cosmology with priors $H_0 \in$ [0, 150] \Hunit\ and $\Omega_{\rm m}\in$ [0.05, 0.5]. \editth{We take these priors for consistency with previous H0LiCOW analyses \editref{\citep[][]{Birrer19, Rusu19, Chen19}}. The $\Omega_{\rm m}$ prior is based on our knowledge from various observations that the Universe is neither empty nor closed.} We take a kernel density estimate of the distance posterior as the likelihood function for cosmological parameters to retain the full covariance between $D_{\Delta t}^{\rm eff}$ and $D_{\rm d}$. \editth{Similar to \citet{Birrer19}, we take the bandwidth for the kernel density estimation to be sufficiently narrow so as to not affect the resultant posteriors of the cosmological parameters.} We infer \Ho\ = \Hval\ \Hunit\ \editth{in the \lcdm\ cosmology, which is a 3.9 per cent measurement} (Fig. \ref{fig:final_h0_posterior}). \editsx{In this 3.9 per cent uncertainty, we estimate that the time-delay measurement contributes 1.8 per cent, the external convergence contributes 3.3 per cent, and the lens modelling and other sources contribute the remaining 1 per cent uncertainty.} \editref{As a systematic check on our model weighting scheme, we infer \Ho\ by combining all the models with equal weight as done in the first few analyses of the H0LiCOW lenses \citep[e.g.,][]{Wong17}. For this most conservative weighting scheme, we find $H_0 = 74.8^{+2.7}_{-3.0}$ km s\textsuperscript{$-$1} Mpc\textsuperscript{$-$1}, which is a 0.8 per cent deviation from our quoted \Ho\ above.} \editref{We summarize the various systematic checks performed in this paper and their corresponding impacts on the inferred \Ho\ in Appendix \ref{app:systematic_summary}.}

\section{Discussion and summary} \label{sec:discussion}
In this paper, we analyse the lens system \lensname\ to blindly infer the effective time-delay distance from the observed time delays. We \editsx{model} the mass profile of the lens system using high-resolution \textit{HST} imaging from three bands. We combine the time-delay and kinematic likelihoods with the lens model posterior, \editsx{and factor in the statistically inferred external convergence} to obtain the cosmological distance posteriors in the $D_{\Delta t}^{\rm eff}$--$D_{\rm d}$ plane. We perform a thorough check for systematic effects \editth{arising from model choices}, and \editth{we marginalize over them to account for this source of systematic uncertainties in our analysis.} As a result, we \editnn{constrain the 2D joint posterior of the effective time delay distance $\Ddt^{\rm eff}$ and the angular diameter distance to the deflector $\Dd$ that fully incorporates their covariance. The marginalized estimates for these two distances are $D_{\Delta t}^{\rm eff}=$ \Ddtval\ with 3.9 per cent uncertainty, and $\Dd=$ \Ddval\ with 19.2 per cent uncertainty.} These constraints translate into \Ho = \Hval\ \Hunit\ with a precision of 3.9 per cent. This estimated value of \Ho\ is consistent with from the previously analysed sample of six lenses by the H0LiCOW collaboration, $H_0 = 73.3^{+1.7}_{-1.8}$ \Hunit\ \citep{Wong19}. \edit{It is also consistent with measurements of \Ho\ based on the local distance ladder \citep{Riess19,Freedman19}, reinforcing the tension \citep{Verde19} with the inference from early Universe probes  \citep{PlanckCollaboration18, Abbott18b}.}

\editth{The one presented in \editsx{this} paper is} the first of two independent cosmographic analysis of the lens system \lensname, \editth{which is based on the lens-modelling software \textsc{lenstronomy}}. A second independent and blind analysis using the lens-modelling software \textsc{glee} will be presented in a future work (Y{\i}ld{\i}r{\i}m et al., in preparation). In this future \editsx{paper}, we will compare the results from the two modelling efforts and \editth{quantitatively evaluate the systematic uncertainty} that may arise due to using different softwares \editth{and adopting different modelling choices} by different investigators. \editth{The posterior probability distribution function of \Ho\ from \lensname\ will be combined with previous measurements by the H0LiCOW team after the second analysis is complete, so as to include this modelling systematic uncertainty in the combination.}

The analysis presented in this paper can be improved in the future. Due to a multitude complexities required in the lens model of \lensname, we fix the distance ratios between the multiple lens and source planes in our analysis using a fiducial \editth{\lcdm} cosmology to make our analysis computationally feasible. We show that the choice of fiducial cosmology has negligible impact on our inference \editth{within the \lcdm\ model}. \editth{However,} it would be ideal to treat the distance ratios as independent non-linear parameters in the model. We leave this improvement to be implemented in future works, \editth{where more general cosmological models will be considered}. Furthermore, the precision on the inferred \Ho\  can be improved in the future with the help of spatially resolved kinematics \citep{Shajib18, Yildirim19}. 

\editref{In the modelling, more general mass profiles can be used in the composite model -- e.g., the generalized NFW profile, or a mass-to-light ratio gradient in the stellar component \citep[e.g.,][]{Zhao96, Sonnenfeld18b}. However, these types of generalized mass profile had been computationally intractable for lens modelling in the elliptical case until only recently \citep{Shajib19b}. Given the computational cost of this study ($\sim10^6$ CPU h) already pushing far beyond the typical case of modelling endeavours for a single lens system, we leave the explorations of more general mass models and estimation of the corresponding impact in the inference of \Ho\ for future studies.}

\editth{Improving the precision \editsx{on} \Ho\ measurement from each \editsx{single} lens system, increasing the number of systems to $\sim$40, and investigating the presence of yet unknown systematic errors are all necessary steps towards reaching 1 per cent precision from time-delay cosmography \citep{Shajib18}. The analysis presented in this paper took one step in each direction.}

\section*{Acknowledgements}
We thank the anonymous referee for helpful comments that improved this manuscript. AJS, SB, TT, GCFC, and CDF were supported by the National Aeronautics and Space Administration (NASA) through the Space Telescope Science Institute (STScI) grant HST-GO-15320. AJS was also supported by the Dissertation Year Fellowship from the University of California, Los Angeles (UCLA) graduate division. TT acknowledges support by the Packard Foundation through a Packard Research fellowship and by the National Science Foundation through NSF grants AST-1714953 and AST-1906976. This project is partly funded by the Danish council for independent research under the project ``Fundamentals of Dark Matter Structures'', DFF - 6108-00470. AA was supported by a grant from VILLUM FONDEN (project number 16599). JHHC, MM, CL, DS, FC, and AG acknowledge support from the Swiss National Science Foundation (SNSF) and by the European Research Council (ERC) under the European Union's Horizon 2020 research and innovation program (COSMICLENS: grant agreement No 787866). CS has received funding from the European Union's Horizon 2020 research and innovation programme under the Marie Sk\l odowska-Curie actions grant agreement No 664931. GCFC acknowledges support from the Ministry of Education in Taiwan via Government Scholarship to Study Abroad (GSSA). CDF and GCFC acknowledge support for this work from the National Science Foundation under Grant No AST-1715611. TA acknowledges support from Proyecto FONDECYT N$\degr$ 1190335. KCW was supported by World Premier International Research Center Initiative (WPI Initiative), MEXT, Japan.

This work used computational and storage services associated with the Hoffman2 Shared Cluster provided by UCLA Institute for Digital Research and Education’s Research Technology Group. This work used the Extreme Science and Engineering Discovery Environment (XSEDE) through allocation TG-AST190038, which is supported by National Science Foundation grant number ACI-1548562 \citep{Towns14}. Specifically, this work used the Comet and Oasis system at the San Diego Supercomputer Center (SDSC), and the Bridges system, which is supported by NSF award number ACI-1445606, at the Pittsburgh Supercomputing Center \citep[PSC,][]{Nystrom15}. AJS thanks Smadar Naoz for providing access to additional computing nodes on the Hoffman2 Shared Cluster and Khalid Jawed for assisting with additional computational resources from the Structures--Computer Interaction Laboratory at UCLA.

AJS acknowledges the hospitality of the Aspen Center of Physics (ACP) and the Munich Institute for Astro- and Particle Physics (MIAPP) of the Excellence Cluster "Universe", where part of this research was completed. ACP is supported by National Science Foundation grant PHY-1607611.

Funding for the DES Projects has been provided by the U.S. Department of Energy, the U.S. National Science Foundation, the Ministry of Science and Education of Spain, 
the Science and Technology Facilities Council of the United Kingdom, the Higher Education Funding Council for England, the National Center for Supercomputing 
Applications at the University of Illinois at Urbana-Champaign, the Kavli Institute of Cosmological Physics at the University of Chicago, 
the Center for Cosmology and Astro-Particle Physics at the Ohio State University,
the Mitchell Institute for Fundamental Physics and Astronomy at Texas A\&M University, Financiadora de Estudos e Projetos, 
Funda{\c c}{\~a}o Carlos Chagas Filho de Amparo {\`a} Pesquisa do Estado do Rio de Janeiro, Conselho Nacional de Desenvolvimento Cient{\'i}fico e Tecnol{\'o}gico and 
the Minist{\'e}rio da Ci{\^e}ncia, Tecnologia e Inova{\c c}{\~a}o, the Deutsche Forschungsgemeinschaft and the Collaborating Institutions in the Dark Energy Survey. 

The Collaborating Institutions are Argonne National Laboratory, the University of California at Santa Cruz, the University of Cambridge, Centro de Investigaciones Energ{\'e}ticas, 
Medioambientales y Tecnol{\'o}gicas-Madrid, the University of Chicago, University College London, the DES-Brazil Consortium, the University of Edinburgh, 
the Eidgen{\"o}ssische Technische Hochschule (ETH) Z{\"u}rich, 
Fermi National Accelerator Laboratory, the University of Illinois at Urbana-Champaign, the Institut de Ci{\`e}ncies de l'Espai (IEEC/CSIC), 
the Institut de F{\'i}sica d'Altes Energies, Lawrence Berkeley National Laboratory, the Ludwig-Maximilians Universit{\"a}t M{\"u}nchen and the associated Excellence Cluster Universe, 
the University of Michigan, the National Optical Astronomy Observatory, the University of Nottingham, The Ohio State University, the University of Pennsylvania, the University of Portsmouth, 
SLAC National Accelerator Laboratory, Stanford University, the University of Sussex, Texas A\&M University, and the OzDES Membership Consortium.

Based in part on observations at Cerro Tololo Inter-American Observatory, National Optical Astronomy Observatory, which is operated by the Association of 
Universities for Research in Astronomy (AURA) under a cooperative agreement with the National Science Foundation.

The DES data management system is supported by the National Science Foundation under Grant Numbers AST-1138766 and AST-1536171.
The DES participants from Spanish institutions are partially supported by MINECO under grants AYA2015-71825, ESP2015-66861, FPA2015-68048, SEV-2016-0588, SEV-2016-0597, and MDM-2015-0509, 
some of which include ERDF funds from the European Union. IFAE is partially funded by the CERCA program of the Generalitat de Catalunya.
Research leading to these results has received funding from the European Research
Council under the European Union's Seventh Framework Program (FP7/2007-2013) including ERC grant agreements 240672, 291329, and 306478.
We  acknowledge support from the Australian Research Council Centre of Excellence for All-sky Astrophysics (CAASTRO), through project number CE110001020, and the Brazilian Instituto Nacional de Ci\^encia
e Tecnologia (INCT) e-Universe (CNPq grant 465376/2014-2).

This manuscript has been authored by Fermi Research Alliance, LLC under Contract No. DE-AC02-07CH11359 with the U.S. Department of Energy, Office of Science, Office of High Energy Physics. The United States Government retains and the publisher, by accepting the article for publication, acknowledges that the United States Government retains a non-exclusive, paid-up, irrevocable, world-wide license to publish or reproduce the published form of this manuscript, or allow others to do so, for United States Government purposes.

This research made use of \textsc{lenstronomy} \editnn{\citep{Birrer15, Birrer18}}, \textsc{dypolychord} \citep{Higson19}, \textsc{polychord} \citep{Handley15}, \textsc{cosmohammer} \citep{Akeret13}, \editnn{\textsc{fastell} \citep{Barkana99}}, \textsc{numpy} \citep{Oliphant15}, \textsc{scipy} \citep{Jones01}, \textsc{astropy} \citep{AstropyCollaboration13, AstropyCollaboration18}, \textsc{jupyter} \citep{Kluyver16}, \textsc{matplotlib} \citep{Hunter07}, \textsc{seaborn} \citep{Waskom14}, \textsc{nestcheck} \citep{Higson18b}, \textsc{sextractor} \citep{Bertin96}, \textsc{emcee} \citep{Foreman-Mackey13}, \textsc{colossus} \citep{Diemer18}, and \textsc{getdist} (\url{https://github.com/cmbant/getdist}).






\bibliographystyle{mnras}
\bibliography{/Users/ajshajib/mylib/STRIDES/Papers/Anowar/ajshajib}

\begin{thebibliography}{}
\makeatletter
\relax
\def\mn@urlcharsother{\let\do\@makeother \do\$\do\&\do\#\do\^\do\_\do\%\do\~}
\def\mn@doi{\begingroup\mn@urlcharsother \@ifnextchar [ {\mn@doi@}
  {\mn@doi@[]}}
\def\mn@doi@[#1]#2{\def\@tempa{#1}\ifx\@tempa\@empty \href
  {http://dx.doi.org/#2} {doi:#2}\else \href {http://dx.doi.org/#2} {#1}\fi
  \endgroup}
\def\mn@eprint#1#2{\mn@eprint@#1:#2::\@nil}
\def\mn@eprint@arXiv#1{\href {http://arxiv.org/abs/#1} {{\tt arXiv:#1}}}
\def\mn@eprint@dblp#1{\href {http://dblp.uni-trier.de/rec/bibtex/#1.xml}
  {dblp:#1}}
\def\mn@eprint@#1:#2:#3:#4\@nil{\def\@tempa {#1}\def\@tempb {#2}\def\@tempc
  {#3}\ifx \@tempc \@empty \let \@tempc \@tempb \let \@tempb \@tempa \fi \ifx
  \@tempb \@empty \def\@tempb {arXiv}\fi \@ifundefined
  {mn@eprint@\@tempb}{\@tempb:\@tempc}{\expandafter \expandafter \csname
  mn@eprint@\@tempb\endcsname \expandafter{\@tempc}}}

\bibitem[\protect\citeauthoryear{{Abbott} et~al.,}{{Abbott}
  et~al.}{2018}]{Abbott18b}
{Abbott} T.~M.~C.,  et~al., 2018, \mn@doi [\prd] {10.1103/PhysRevD.98.043526},
  \href {https://ui.adsabs.harvard.edu/abs/2018PhRvD..98d3526A} {98, 043526}

\bibitem[\protect\citeauthoryear{{Agnello}, {Evans}  \& {Romanowsky}}{{Agnello}
  et~al.}{2014}]{Agnello14}
{Agnello} A.,  {Evans} N.~W.,   {Romanowsky} A.~J.,  2014, \mn@doi [\mnras]
  {10.1093/mnras/stu959}, \href
  {http://adsabs.harvard.edu/abs/2014MNRAS.442.3284A} {442, 3284}

\bibitem[\protect\citeauthoryear{{Agnello}, {Kelly}, {Treu}  \&
  {Marshall}}{{Agnello} et~al.}{2015a}]{Agnello15}
{Agnello} A.,  {Kelly} B.~C.,  {Treu} T.,   {Marshall} P.~J.,  2015a, \mn@doi
  [\mnras] {10.1093/mnras/stv037}, \href
  {http://adsabs.harvard.edu/abs/2015MNRAS.448.1446A} {448, 1446}

\bibitem[\protect\citeauthoryear{{Agnello} et~al.,}{{Agnello}
  et~al.}{2015b}]{Agnello15b}
{Agnello} A.,  et~al., 2015b, \mn@doi [\mnras] {10.1093/mnras/stv2171}, \href
  {http://adsabs.harvard.edu/abs/2015MNRAS.454.1260A} {454, 1260}

\bibitem[\protect\citeauthoryear{{Agnello} et~al.,}{{Agnello}
  et~al.}{2017}]{Agnello17}
{Agnello} A.,  et~al., 2017, \mn@doi [\mnras] {10.1093/mnras/stx2242}, \href
  {http://adsabs.harvard.edu/abs/2017MNRAS.472.4038A} {472, 4038}

\bibitem[\protect\citeauthoryear{{Agnello} et~al.,}{{Agnello}
  et~al.}{2018a}]{Agnello18b}
{Agnello} A.,  et~al., 2018a, \mn@doi [\mnras] {10.1093/mnras/stx3226}, \href
  {http://adsabs.harvard.edu/abs/2018MNRAS.475.2086A} {475, 2086}

\bibitem[\protect\citeauthoryear{{Agnello} et~al.,}{{Agnello}
  et~al.}{2018b}]{Agnello18d}
{Agnello} A.,  et~al., 2018b, \mn@doi [\mnras] {10.1093/mnras/sty1419}, \href
  {https://ui.adsabs.harvard.edu/abs/2018MNRAS.479.4345A} {479, 4345}

\bibitem[\protect\citeauthoryear{Akeret, Seehars, Amara, Refregier  \&
  Csillaghy}{Akeret et~al.}{2013}]{Akeret13}
Akeret J.,  Seehars S.,  Amara A.,  Refregier A.,   Csillaghy A.,  2013,
  \mn@doi [Astronomy and Computing] {10.1016/j.ascom.2013.06.003}, 2, 27

\bibitem[\protect\citeauthoryear{{Alam} et~al.,}{{Alam} et~al.}{2017}]{Alam17}
{Alam} S.,  et~al., 2017, \mn@doi [\mnras] {10.1093/mnras/stx721}, \href
  {http://adsabs.harvard.edu/abs/2016arXiv160703155A} {470, 2617}

\bibitem[\protect\citeauthoryear{{Anguita} et~al.,}{{Anguita}
  et~al.}{2018}]{Anguita18}
{Anguita} T.,  et~al., 2018, \mn@doi [\mnras] {10.1093/mnras/sty2172}, \href
  {http://adsabs.harvard.edu/abs/2018arXiv180512151A} {480, 5017}

\bibitem[\protect\citeauthoryear{{Astropy Collaboration}}{{Astropy
  Collaboration}}{2013}]{AstropyCollaboration13}
{Astropy Collaboration} 2013, \mn@doi [\aap] {10.1051/0004-6361/201322068},
  \href {http://adsabs.harvard.edu/abs/2013A%26A...558A..33A} {558, A33}

\bibitem[\protect\citeauthoryear{{Astropy Collaboration}}{{Astropy
  Collaboration}}{2018}]{AstropyCollaboration18}
{Astropy Collaboration} 2018, \mn@doi [\aj] {10.3847/1538-3881/aabc4f}, \href
  {https://ui.adsabs.harvard.edu/abs/2018AJ....156..123A} {156, 123}

\bibitem[\protect\citeauthoryear{{Auger}, {Treu}, {Bolton}, {Gavazzi},
  {Koopmans}, {Marshall}, {Moustakas}  \& {Burles}}{{Auger}
  et~al.}{2010}]{Auger10b}
{Auger} M.~W.,  {Treu} T.,  {Bolton} A.~S.,  {Gavazzi} R.,  {Koopmans}
  L.~V.~E.,  {Marshall} P.~J.,  {Moustakas} L.~A.,   {Burles} S.,  2010,
  \mn@doi [\apj] {10.1088/0004-637X/724/1/511}, \href
  {http://adsabs.harvard.edu/abs/2010ApJ...724..511A} {724, 511}

\bibitem[\protect\citeauthoryear{{Avila}, {Hack}, {Cara}, {Borncamp}, {Mack},
  {Smith}  \& {Ubeda}}{{Avila} et~al.}{2015}]{Avila15}
{Avila} R.~J.,  {Hack} W.,  {Cara} M.,  {Borncamp} D.,  {Mack} J.,  {Smith} L.,
    {Ubeda} L.,  2015, in {Taylor} A.~R.,  {Rosolowsky} E.,  eds,  Astronomical
  Society of the Pacific Conference Series Vol. 495, Astronomical Data Analysis
  Software an Systems XXIV (ADASS XXIV). p.~281 (\mn@eprint {arXiv}
  {1411.5605})

\bibitem[\protect\citeauthoryear{{Barkana}}{{Barkana}}{1998}]{Barkana98}
{Barkana} R.,  1998, \mn@doi [\apj] {10.1086/305950}, \href
  {http://adsabs.harvard.edu/abs/1998ApJ...502..531B} {502, 531}

\bibitem[\protect\citeauthoryear{{Barkana}}{{Barkana}}{1999}]{Barkana99}
{Barkana} R.,  1999, {FASTELL: Fast calculation of a family of elliptical mass
  gravitational lens models}, Astrophysics Source Code Library (\mn@eprint
  {ascl} {9910.003})

\bibitem[\protect\citeauthoryear{{Behroozi}, {Wechsler}, {Hearin}  \&
  {Conroy}}{{Behroozi} et~al.}{2019}]{Behroozi19}
{Behroozi} P.,  {Wechsler} R.~H.,  {Hearin} A.~P.,   {Conroy} C.,  2019,
  \mn@doi [\mnras] {10.1093/mnras/stz1182}, \href
  {https://ui.adsabs.harvard.edu/abs/2019MNRAS.488.3143B} {488, 3143}

\bibitem[\protect\citeauthoryear{{Bendinelli}}{{Bendinelli}}{1991}]{Bendinelli91}
{Bendinelli} O.,  1991, \mn@doi [\apj] {10.1086/169595}, \href
  {http://adsabs.harvard.edu/abs/1991ApJ...366..599B} {366, 599}

\bibitem[\protect\citeauthoryear{{Bertin} \& {Arnouts}}{{Bertin} \&
  {Arnouts}}{1996}]{Bertin96}
{Bertin} E.,  {Arnouts} S.,  1996, \mn@doi [\aaps] {10.1051/aas:1996164}, \href
  {http://adsabs.harvard.edu/abs/1996A%26AS..117..393B} {117, 393}

\bibitem[\protect\citeauthoryear{{Bertin} \& {Lombardi}}{{Bertin} \&
  {Lombardi}}{2006}]{Bertin06}
{Bertin} G.,  {Lombardi} M.,  2006, \mn@doi [\apjl] {10.1086/507298}, \href
  {https://ui.adsabs.harvard.edu/abs/2006ApJ...648L..17B} {648, L17}

\bibitem[\protect\citeauthoryear{{Birrer} \& {Amara}}{{Birrer} \&
  {Amara}}{2018}]{Birrer18}
{Birrer} S.,  {Amara} A.,  2018, \mn@doi [Physics of the Dark Universe]
  {10.1016/j.dark.2018.11.002}, \href
  {http://adsabs.harvard.edu/abs/2018PDU....22..189B} {22, 189}

\bibitem[\protect\citeauthoryear{{Birrer} \& {Treu}}{{Birrer} \&
  {Treu}}{2019}]{Birrer19b}
{Birrer} S.,  {Treu} T.,  2019, \mn@doi [\mnras] {10.1093/mnras/stz2254}, \href
  {https://ui.adsabs.harvard.edu/abs/2019arXiv190410965B} {489, 2097}

\bibitem[\protect\citeauthoryear{{Birrer}, {Amara}  \& {Refregier}}{{Birrer}
  et~al.}{2015}]{Birrer15}
{Birrer} S.,  {Amara} A.,   {Refregier} A.,  2015, \mn@doi [\apj]
  {10.1088/0004-637X/813/2/102}, \href
  {http://adsabs.harvard.edu/abs/2015ApJ...813..102B} {813, 102}

\bibitem[\protect\citeauthoryear{{Birrer}, {Amara}  \& {Refregier}}{{Birrer}
  et~al.}{2016}]{Birrer16}
{Birrer} S.,  {Amara} A.,   {Refregier} A.,  2016, \mn@doi [\jcap]
  {10.1088/1475-7516/2016/08/020}, \href
  {http://adsabs.harvard.edu/abs/2016JCAP...08..020B} {8, 020}

\bibitem[\protect\citeauthoryear{{Birrer} et~al.,}{{Birrer}
  et~al.}{2019}]{Birrer19}
{Birrer} S.,  et~al., 2019, \mn@doi [\mnras] {10.1093/mnras/stz200}, \href
  {http://adsabs.harvard.edu/abs/2018arXiv180901274B} {484, 4726}

\bibitem[\protect\citeauthoryear{{Blandford} \& {Narayan}}{{Blandford} \&
  {Narayan}}{1992}]{Blandford92}
{Blandford} R.~D.,  {Narayan} R.,  1992, \mn@doi [\araa]
  {10.1146/annurev.astro.30.1.311}, \href
  {http://adsabs.harvard.edu/abs/1992ARA%26A..30..311B} {30, 311}

\bibitem[\protect\citeauthoryear{{Bolton}, {Burles}, {Koopmans}, {Treu}  \&
  {Moustakas}}{{Bolton} et~al.}{2006}]{Bolton06}
{Bolton} A.~S.,  {Burles} S.,  {Koopmans} L.~V.~E.,  {Treu} T.,   {Moustakas}
  L.~A.,  2006, \mn@doi [\apj] {10.1086/498884}, \href
  {https://ui.adsabs.harvard.edu/abs/2006ApJ...638..703B} {638, 703}

\bibitem[\protect\citeauthoryear{{Bonvin} et~al.,}{{Bonvin}
  et~al.}{2017}]{Bonvin17}
{Bonvin} V.,  et~al., 2017, \mn@doi [\mnras] {10.1093/mnras/stw3006}, \href
  {http://adsabs.harvard.edu/abs/2017MNRAS.465.4914B} {465, 4914}

\bibitem[\protect\citeauthoryear{{Bonvin} et~al.,}{{Bonvin}
  et~al.}{2018}]{Bonvin18}
{Bonvin} V.,  et~al., 2018, \mn@doi [\aap] {10.1051/0004-6361/201833287}, \href
  {https://ui.adsabs.harvard.edu/abs/2018A%26A...616A.183B} {616, A183}

\bibitem[\protect\citeauthoryear{{Buckley-Geer} et~al.,}{{Buckley-Geer}
  et~al.}{2020}]{BuckleyGeer20}
{Buckley-Geer} E.~J.,  et~al., 2020, arXiv e-prints, \href
  {https://ui.adsabs.harvard.edu/abs/2020arXiv200312117B} {p. arXiv:2003.12117}

\bibitem[\protect\citeauthoryear{{Cappellari}}{{Cappellari}}{2002}]{Cappellari02}
{Cappellari} M.,  2002, \mn@doi [\mnras] {10.1046/j.1365-8711.2002.05412.x},
  \href {http://adsabs.harvard.edu/abs/2002MNRAS.333..400C} {333, 400}

\bibitem[\protect\citeauthoryear{{Cardelli}, {Clayton}  \& {Mathis}}{{Cardelli}
  et~al.}{1989}]{Cardelli89}
{Cardelli} J.~A.,  {Clayton} G.~C.,   {Mathis} J.~S.,  1989, \mn@doi [\apj]
  {10.1086/167900}, \href
  {https://ui.adsabs.harvard.edu/abs/1989ApJ...345..245C} {345, 245}

\bibitem[\protect\citeauthoryear{{Chen} et~al.,}{{Chen} et~al.}{2016}]{Chen16}
{Chen} G.~C.-F.,  et~al., 2016, \mn@doi [\mnras] {10.1093/mnras/stw991}, \href
  {http://adsabs.harvard.edu/abs/2016MNRAS.462.3457C} {462, 3457}

\bibitem[\protect\citeauthoryear{{Chen} et~al.,}{{Chen} et~al.}{2018}]{Chen18}
{Chen} G.~C.-F.,  et~al., 2018, \mn@doi [\mnras] {10.1093/mnras/sty2350}, \href
  {https://ui.adsabs.harvard.edu/abs/2018MNRAS.481.1115C} {481, 1115}

\bibitem[\protect\citeauthoryear{{Chen} et~al.,}{{Chen} et~al.}{2019}]{Chen19}
{Chen} G.~C.-F.,  et~al., 2019, \mn@doi [\mnras] {10.1093/mnras/stz2547}, \href
  {https://ui.adsabs.harvard.edu/abs/2019arXiv190702533C} {}

\bibitem[\protect\citeauthoryear{{Collett} \& {Auger}}{{Collett} \&
  {Auger}}{2014}]{Collett14}
{Collett} T.~E.,  {Auger} M.~W.,  2014, \mn@doi [\mnras]
  {10.1093/mnras/stu1190}, \href
  {https://ui.adsabs.harvard.edu/abs/2014MNRAS.443..969C} {443, 969}

\bibitem[\protect\citeauthoryear{{Collett}, {Auger}, {Belokurov}, {Marshall}
  \& {Hall}}{{Collett} et~al.}{2012}]{Collett12}
{Collett} T.~E.,  {Auger} M.~W.,  {Belokurov} V.,  {Marshall} P.~J.,   {Hall}
  A.~C.,  2012, \mn@doi [\mnras] {10.1111/j.1365-2966.2012.21424.x}, \href
  {https://ui.adsabs.harvard.edu/abs/2012MNRAS.424.2864C} {424, 2864}

\bibitem[\protect\citeauthoryear{{Courbin} et~al.,}{{Courbin}
  et~al.}{2018}]{Courbin18}
{Courbin} F.,  et~al., 2018, \mn@doi [\aap] {10.1051/0004-6361/201731461},
  \href {http://adsabs.harvard.edu/abs/2017arXiv170609424C} {609, A71}

\bibitem[\protect\citeauthoryear{{Diehl} et~al.,}{{Diehl}
  et~al.}{2017}]{Diehl17}
{Diehl} H.~T.,  et~al., 2017, \mn@doi [\apjs] {10.3847/1538-4365/aa8667}, \href
  {http://adsabs.harvard.edu/abs/2017ApJS..232...15D} {232, 15}

\bibitem[\protect\citeauthoryear{{Diemer}}{{Diemer}}{2018}]{Diemer18}
{Diemer} B.,  2018, \mn@doi [\apjs] {10.3847/1538-4365/aaee8c}, \href
  {https://ui.adsabs.harvard.edu/abs/2018ApJS..239...35D} {239, 35}

\bibitem[\protect\citeauthoryear{{Diemer} \& {Joyce}}{{Diemer} \&
  {Joyce}}{2019}]{Diemer19}
{Diemer} B.,  {Joyce} M.,  2019, \mn@doi [\apj] {10.3847/1538-4357/aafad6},
  \href {https://ui.adsabs.harvard.edu/abs/2019ApJ...871..168D} {871, 168}

\bibitem[\protect\citeauthoryear{{Dutton} et~al.,}{{Dutton}
  et~al.}{2011}]{Dutton11}
{Dutton} A.~A.,  et~al., 2011, \mn@doi [\mnras]
  {10.1111/j.1365-2966.2011.18706.x}, \href
  {http://adsabs.harvard.edu/abs/2011MNRAS.417.1621D} {417, 1621}

\bibitem[\protect\citeauthoryear{{Emsellem}, {Monnet}, {Bacon}  \&
  {Nieto}}{{Emsellem} et~al.}{1994}]{Emsellem94}
{Emsellem} E.,  {Monnet} G.,  {Bacon} R.,   {Nieto} J.-L.,  1994, \aap, \href
  {http://adsabs.harvard.edu/abs/1994A%26A...285..739E} {285, 739}

\bibitem[\protect\citeauthoryear{{Falco}, {Gorenstein}  \& {Shapiro}}{{Falco}
  et~al.}{1985}]{Falco85}
{Falco} E.~E.,  {Gorenstein} M.~V.,   {Shapiro} I.~I.,  1985, \mn@doi [\apjl]
  {10.1086/184422}, \href {http://adsabs.harvard.edu/abs/1985ApJ...289L...1F}
  {289, L1}

\bibitem[\protect\citeauthoryear{{Fassnacht}, {Xanthopoulos}, {Koopmans}  \&
  {Rusin}}{{Fassnacht} et~al.}{2002}]{Fassnacht02}
{Fassnacht} C.~D.,  {Xanthopoulos} E.,  {Koopmans} L.~V.~E.,   {Rusin} D.,
  2002, \mn@doi [\apj] {10.1086/344368}, \href
  {https://ui.adsabs.harvard.edu/abs/2002ApJ...581..823F} {581, 823}

\bibitem[\protect\citeauthoryear{{Foreman-Mackey}, {Hogg}, {Lang}  \&
  {Goodman}}{{Foreman-Mackey} et~al.}{2013}]{Foreman-Mackey13}
{Foreman-Mackey} D.,  {Hogg} D.~W.,  {Lang} D.,   {Goodman} J.,  2013, \mn@doi
  [\pasp] {10.1086/670067}, \href
  {http://adsabs.harvard.edu/abs/2013PASP..125..306F} {125, 306}

\bibitem[\protect\citeauthoryear{{Freedman} et~al.,}{{Freedman}
  et~al.}{2019}]{Freedman19}
{Freedman} W.~L.,  et~al., 2019, \mn@doi [\apj] {10.3847/1538-4357/ab2f73},
  \href {https://ui.adsabs.harvard.edu/abs/2019arXiv190705922F} {882, 34}

\bibitem[\protect\citeauthoryear{{Gavazzi}, {Treu}, {Rhodes}, {Koopmans},
  {Bolton}, {Burles}, {Massey}  \& {Moustakas}}{{Gavazzi}
  et~al.}{2007}]{Gavazzi07}
{Gavazzi} R.,  {Treu} T.,  {Rhodes} J.~D.,  {Koopmans} L.~V.~E.,  {Bolton}
  A.~S.,  {Burles} S.,  {Massey} R.~J.,   {Moustakas} L.~A.,  2007, \mn@doi
  [\apj] {10.1086/519237}, \href
  {http://adsabs.harvard.edu/abs/2007ApJ...667..176G} {667, 176}

\bibitem[\protect\citeauthoryear{{Gavazzi}, {Treu}, {Koopmans}, {Bolton},
  {Moustakas}, {Burles}  \& {Marshall}}{{Gavazzi} et~al.}{2008}]{Gavazzi08}
{Gavazzi} R.,  {Treu} T.,  {Koopmans} L.~V.~E.,  {Bolton} A.~S.,  {Moustakas}
  L.~A.,  {Burles} S.,   {Marshall} P.~J.,  2008, \mn@doi [\apj]
  {10.1086/529541}, \href {http://adsabs.harvard.edu/abs/2008ApJ...677.1046G}
  {677, 1046}

\bibitem[\protect\citeauthoryear{{Goobar} et~al.,}{{Goobar}
  et~al.}{2017}]{Goobar17}
{Goobar} A.,  et~al., 2017, \mn@doi [Science] {10.1126/science.aal2729}, \href
  {https://ui.adsabs.harvard.edu/abs/2017Sci...356..291G} {356, 291}

\bibitem[\protect\citeauthoryear{{Greene} et~al.,}{{Greene}
  et~al.}{2013}]{Greene13}
{Greene} Z.~S.,  et~al., 2013, \mn@doi [\apj] {10.1088/0004-637X/768/1/39},
  \href {http://adsabs.harvard.edu/abs/2013ApJ...768...39G} {768, 39}

\bibitem[\protect\citeauthoryear{{Grillo} et~al.,}{{Grillo}
  et~al.}{2018}]{Grillo18}
{Grillo} C.,  et~al., 2018, \mn@doi [\apj] {10.3847/1538-4357/aac2c9}, \href
  {https://ui.adsabs.harvard.edu/abs/2018ApJ...860...94G} {860, 94}

\bibitem[\protect\citeauthoryear{{Gu{\'e}rou} et~al.,}{{Gu{\'e}rou}
  et~al.}{2017}]{Guerou17}
{Gu{\'e}rou} A.,  et~al., 2017, \mn@doi [\aap] {10.1051/0004-6361/201730905},
  \href {https://ui.adsabs.harvard.edu/abs/2017A%26A...608A...5G} {608, A5}

\bibitem[\protect\citeauthoryear{{Handley}, {Hobson}  \& {Lasenby}}{{Handley}
  et~al.}{2015}]{Handley15}
{Handley} W.~J.,  {Hobson} M.~P.,   {Lasenby} A.~N.,  2015, \mn@doi [\mnras]
  {10.1093/mnras/stv1911}, \href
  {https://ui.adsabs.harvard.edu/abs/2015MNRAS.453.4384H} {453, 4384}

\bibitem[\protect\citeauthoryear{Higson}{Higson}{2018}]{Higson18b}
Higson E.,  2018, \mn@doi [Journal of Open Source Software]
  {10.21105/joss.00916}, 3, 916

\bibitem[\protect\citeauthoryear{Higson, Handley, Hobson  \& Lasenby}{Higson
  et~al.}{2018}]{Higson18}
Higson E.,  Handley W.,  Hobson M.,   Lasenby A.,  2018, \mn@doi [Statistics
  and Computing] {10.1007/s11222-018-9844-0}

\bibitem[\protect\citeauthoryear{{Higson}, {Handley}, {Hobson}  \&
  {Lasenby}}{{Higson} et~al.}{2019}]{Higson19}
{Higson} E.,  {Handley} W.,  {Hobson} M.,   {Lasenby} A.,  2019, {dyPolyChord:
  Super fast dynamic nested sampling with PolyChord}, Astrophysics Source Code
  Library (\mn@eprint {ascl} {1902.010})

\bibitem[\protect\citeauthoryear{{Hilbert}, {Hartlap}, {White}  \&
  {Schneider}}{{Hilbert} et~al.}{2009}]{Hilbert09}
{Hilbert} S.,  {Hartlap} J.,  {White} S.~D.~M.,   {Schneider} P.,  2009,
  \mn@doi [\aap] {10.1051/0004-6361/200811054}, \href
  {http://adsabs.harvard.edu/abs/2009A%26A...499...31H} {499, 31}

\bibitem[\protect\citeauthoryear{Hoeting, Madigan, Raftery  \&
  Volinsky}{Hoeting et~al.}{1999}]{Hoeting99}
Hoeting J.~A.,  Madigan D.,  Raftery A.~E.,   Volinsky C.~T.,  1999, \mn@doi
  [Statist. Sci.] {10.1214/ss/1009212519}, 14, 382

\bibitem[\protect\citeauthoryear{Hunter}{Hunter}{2007}]{Hunter07}
Hunter J.~D.,  2007, \mn@doi [Computing in Science and Engineering]
  {10.1109/MCSE.2007.55}, 9, 90

\bibitem[\protect\citeauthoryear{{Jee}, {Komatsu}  \& {Suyu}}{{Jee}
  et~al.}{2015}]{Jee15}
{Jee} I.,  {Komatsu} E.,   {Suyu} S.~H.,  2015, \mn@doi [\jcap]
  {10.1088/1475-7516/2015/11/033}, \href
  {http://adsabs.harvard.edu/abs/2015JCAP...11..033J} {11, 033}

\bibitem[\protect\citeauthoryear{{Jee}, {Komatsu}, {Suyu}  \& {Huterer}}{{Jee}
  et~al.}{2016}]{Jee16}
{Jee} I.,  {Komatsu} E.,  {Suyu} S.~H.,   {Huterer} D.,  2016, \mn@doi [\jcap]
  {10.1088/1475-7516/2016/04/031}, \href
  {http://adsabs.harvard.edu/abs/2016JCAP...04..031J} {4, 031}

\bibitem[\protect\citeauthoryear{Jones, Oliphant, Peterson  \& Others}{Jones
  et~al.}{2001}]{Jones01}
Jones E.,  Oliphant T.,  Peterson P.,   Others 2001, {SciPy}: Open source
  scientific tools for Python, \url {http://www.scipy.org/}

\bibitem[\protect\citeauthoryear{{Jorgensen}, {Franx}  \&
  {Kjaergaard}}{{Jorgensen} et~al.}{1995}]{Jorgensen95}
{Jorgensen} I.,  {Franx} M.,   {Kjaergaard} P.,  1995, \mn@doi [\mnras]
  {10.1093/mnras/276.4.1341}, \href
  {https://ui.adsabs.harvard.edu/abs/1995MNRAS.276.1341J} {276, 1341}

\bibitem[\protect\citeauthoryear{{Kelly} et~al.,}{{Kelly}
  et~al.}{2015}]{Kelly15}
{Kelly} P.~L.,  et~al., 2015, \mn@doi [Science] {10.1126/science.aaa3350},
  \href {https://ui.adsabs.harvard.edu/abs/2015Sci...347.1123K} {347, 1123}

\bibitem[\protect\citeauthoryear{{Kennedy} \& {Eberhart}}{{Kennedy} \&
  {Eberhart}}{1995}]{Kennedy95}
{Kennedy} J.,  {Eberhart} R.,  1995, in Proceedings of
  {ICNN}{\textquotesingle}95 - International Conference on Neural Networks.
  {IEEE}, \mn@doi{10.1109/icnn.1995.488968}, \url
  {https://doi.org/10.1109/icnn.1995.488968}

\bibitem[\protect\citeauthoryear{Kluyver et~al.,}{Kluyver
  et~al.}{2016}]{Kluyver16}
Kluyver T.,  et~al., 2016, in Loizides F.,  Schmidt B.,  eds, Positioning and
  Power in Academic Publishing: Players, Agents and Agendas. IOS Press BV,
  Amsterdam, Netherlands, pp 87 -- 90, \mn@doi{10.3233/978-1-61499-649-1-87}

\bibitem[\protect\citeauthoryear{{Lagattuta}, {Vegetti}, {Fassnacht}, {Auger},
  {Koopmans}  \& {McKean}}{{Lagattuta} et~al.}{2012}]{Lagattuta12}
{Lagattuta} D.~J.,  {Vegetti} S.,  {Fassnacht} C.~D.,  {Auger} M.~W.,
  {Koopmans} L.~V.~E.,   {McKean} J.~P.,  2012, \mn@doi [\mnras]
  {10.1111/j.1365-2966.2012.21406.x}, \href
  {https://ui.adsabs.harvard.edu/abs/2012MNRAS.424.2800L} {424, 2800}

\bibitem[\protect\citeauthoryear{{Lemon} et~al.,}{{Lemon}
  et~al.}{2019}]{Lemon19}
{Lemon} C.,  et~al., 2019, arXiv e-prints, \href
  {https://ui.adsabs.harvard.edu/abs/2019arXiv191209133L} {p. arXiv:1912.09133}

\bibitem[\protect\citeauthoryear{{Lewis} \& {Bridle}}{{Lewis} \&
  {Bridle}}{2002}]{Lewis02}
{Lewis} A.,  {Bridle} S.,  2002, \mn@doi [\prd] {10.1103/PhysRevD.66.103511},
  \href {http://adsabs.harvard.edu/abs/2002PhRvD..66j3511L} {66, 103511}

\bibitem[\protect\citeauthoryear{{Lin} et~al.,}{{Lin} et~al.}{2017}]{Lin17}
{Lin} H.,  et~al., 2017, \mn@doi [\apjl] {10.3847/2041-8213/aa624e}, \href
  {http://adsabs.harvard.edu/abs/2017ApJ...838L..15L} {838, L15}

\bibitem[\protect\citeauthoryear{Madigan \& Raftery}{Madigan \&
  Raftery}{1994}]{Madigan94}
Madigan D.,  Raftery A.~E.,  1994, \mn@doi [Journal of the American Statistical
  Association] {10.1080/01621459.1994.10476894}, 89, 1535

\bibitem[\protect\citeauthoryear{{Mamon} \& {{\L}okas}}{{Mamon} \&
  {{\L}okas}}{2005}]{Mamon05}
{Mamon} G.~A.,  {{\L}okas} E.~L.,  2005, \mn@doi [\mnras]
  {10.1111/j.1365-2966.2005.09400.x}, \href
  {http://adsabs.harvard.edu/abs/2005MNRAS.363..705M} {363, 705}

\bibitem[\protect\citeauthoryear{{McCully}, {Keeton}, {Wong}  \&
  {Zabludoff}}{{McCully} et~al.}{2017}]{McCully17}
{McCully} C.,  {Keeton} C.~R.,  {Wong} K.~C.,   {Zabludoff} A.~I.,  2017,
  \mn@doi [\apj] {10.3847/1538-4357/836/1/141}, \href
  {https://ui.adsabs.harvard.edu/abs/2017ApJ...836..141M} {836, 141}

\bibitem[\protect\citeauthoryear{{Merritt}}{{Merritt}}{1985a}]{Merritt85b}
{Merritt} D.,  1985a, \mn@doi [\aj] {10.1086/113810}, \href
  {http://adsabs.harvard.edu/abs/1985AJ.....90.1027M} {90, 1027}

\bibitem[\protect\citeauthoryear{{Merritt}}{{Merritt}}{1985b}]{Merritt85}
{Merritt} D.,  1985b, \mn@doi [\mnras] {10.1093/mnras/214.1.25P}, \href
  {http://adsabs.harvard.edu/abs/1985MNRAS.214P..25M} {214, 25P}

\bibitem[\protect\citeauthoryear{{Munari}, {Biviano}, {Borgani}, {Murante}  \&
  {Fabjan}}{{Munari} et~al.}{2013}]{Munari13}
{Munari} E.,  {Biviano} A.,  {Borgani} S.,  {Murante} G.,   {Fabjan} D.,  2013,
  \mn@doi [\mnras] {10.1093/mnras/stt049}, \href
  {https://ui.adsabs.harvard.edu/abs/2013MNRAS.430.2638M} {430, 2638}

\bibitem[\protect\citeauthoryear{{Navarro}, {Frenk}  \& {White}}{{Navarro}
  et~al.}{1997}]{Navarro97}
{Navarro} J.~F.,  {Frenk} C.~S.,   {White} S.~D.~M.,  1997, \mn@doi [\apj]
  {10.1086/304888}, \href {http://adsabs.harvard.edu/abs/1997ApJ...490..493N}
  {490, 493}

\bibitem[\protect\citeauthoryear{{Nord} et~al.,}{{Nord} et~al.}{2016}]{Nord16}
{Nord} B.,  et~al., 2016, \mn@doi [\apj] {10.3847/0004-637X/827/1/51}, \href
  {https://ui.adsabs.harvard.edu/abs/2016ApJ...827...51N} {827, 51}

\bibitem[\protect\citeauthoryear{Nystrom, Levine, Roskies  \& Scott}{Nystrom
  et~al.}{2015}]{Nystrom15}
Nystrom N.~A.,  Levine M.~J.,  Roskies R.~Z.,   Scott J.~R.,  2015, in
  Proceedings of the 2015 XSEDE Conference: Scientific Advancements Enabled by
  Enhanced Cyberinfrastructure. XSEDE '15.
ACM, New York, NY, USA, pp 30:1--30:8, \mn@doi{10.1145/2792745.2792775}, \url
  {http://doi.acm.org/10.1145/2792745.2792775}

\bibitem[\protect\citeauthoryear{Oliphant}{Oliphant}{2015}]{Oliphant15}
Oliphant T.~E.,  2015, Guide to NumPy, 2nd edn.
CreateSpace Independent Publishing Platform, USA

\bibitem[\protect\citeauthoryear{{Osipkov}}{{Osipkov}}{1979}]{Osipkov79}
{Osipkov} L.~P.,  1979, Pisma v Astronomicheskii Zhurnal, \href
  {http://adsabs.harvard.edu/abs/1979PAZh....5...77O} {5, 77}

\bibitem[\protect\citeauthoryear{{Ostrovski} et~al.,}{{Ostrovski}
  et~al.}{2017}]{Ostrovski17}
{Ostrovski} F.,  et~al., 2017, \mn@doi [\mnras] {10.1093/mnras/stw2958}, \href
  {http://adsabs.harvard.edu/abs/2017MNRAS.465.4325O} {465, 4325}

\bibitem[\protect\citeauthoryear{{Paraficz} \& {Hjorth}}{{Paraficz} \&
  {Hjorth}}{2009}]{Paraficz09}
{Paraficz} D.,  {Hjorth} J.,  2009, \mn@doi [\aap]
  {10.1051/0004-6361/200913307}, \href
  {http://adsabs.harvard.edu/abs/2009A%26A...507L..49P} {507, L49}

\bibitem[\protect\citeauthoryear{{Perlmutter} et~al.,}{{Perlmutter}
  et~al.}{1999}]{Perlmutter99}
{Perlmutter} S.,  et~al., 1999, \mn@doi [\apj] {10.1086/307221}, \href
  {http://adsabs.harvard.edu/abs/1999ApJ...517..565P} {517, 565}

\bibitem[\protect\citeauthoryear{{Planck Collaboration}}{{Planck
  Collaboration}}{2018}]{PlanckCollaboration18}
{Planck Collaboration} 2018, arXiv e-prints, \href
  {https://ui.adsabs.harvard.edu/abs/2018arXiv180706209P} {p. arXiv:1807.06209}

\bibitem[\protect\citeauthoryear{{Refregier}}{{Refregier}}{2003}]{Refregier03}
{Refregier} A.,  2003, \mn@doi [\mnras] {10.1046/j.1365-8711.2003.05901.x},
  \href {http://adsabs.harvard.edu/abs/2003MNRAS.338...35R} {338, 35}

\bibitem[\protect\citeauthoryear{{Refsdal}}{{Refsdal}}{1964}]{Refsdal64}
{Refsdal} S.,  1964, \mn@doi [\mnras] {10.1093/mnras/128.4.307}, \href
  {http://adsabs.harvard.edu/abs/1964MNRAS.128..307R} {128, 307}

\bibitem[\protect\citeauthoryear{{Riess} et~al.,}{{Riess}
  et~al.}{1998}]{Riess98}
{Riess} A.~G.,  et~al., 1998, \mn@doi [\aj] {10.1086/300499}, \href
  {http://adsabs.harvard.edu/abs/1998AJ....116.1009R} {116, 1009}

\bibitem[\protect\citeauthoryear{{Riess}, {Casertano}, {Yuan}, {Macri}  \&
  {Scolnic}}{{Riess} et~al.}{2019}]{Riess19}
{Riess} A.~G.,  {Casertano} S.,  {Yuan} W.,  {Macri} L.~M.,   {Scolnic} D.,
  2019, \mn@doi [\apj] {10.3847/1538-4357/ab1422}, \href
  {http://adsabs.harvard.edu/abs/2019arXiv190307603R} {876, 85}

\bibitem[\protect\citeauthoryear{{Rusu} et~al.,}{{Rusu} et~al.}{2017}]{Rusu17}
{Rusu} C.~E.,  et~al., 2017, \mn@doi [\mnras] {10.1093/mnras/stx285}, \href
  {http://adsabs.harvard.edu/abs/2017MNRAS.467.4220R} {467, 4220}

\bibitem[\protect\citeauthoryear{{Rusu} et~al.,}{{Rusu} et~al.}{2019}]{Rusu19}
{Rusu} C.~E.,  et~al., 2019, arXiv e-prints, \href
  {https://ui.adsabs.harvard.edu/abs/2019arXiv190509338R} {p. arXiv:1905.09338}

\bibitem[\protect\citeauthoryear{{Schechter} et~al.,}{{Schechter}
  et~al.}{1997}]{Schechter97}
{Schechter} P.~L.,  et~al., 1997, \mn@doi [\apjl] {10.1086/310478}, \href
  {https://ui.adsabs.harvard.edu/abs/1997ApJ...475L..85S} {475, L85}

\bibitem[\protect\citeauthoryear{{Schneider} \& {Sluse}}{{Schneider} \&
  {Sluse}}{2014}]{Schneider14}
{Schneider} P.,  {Sluse} D.,  2014, \mn@doi [\aap]
  {10.1051/0004-6361/201322106}, \href
  {http://adsabs.harvard.edu/abs/2014A%26A...564A.103S} {564, A103}

\bibitem[\protect\citeauthoryear{{Schneider}, {Ehlers}  \& {Falco}}{{Schneider}
  et~al.}{1992}]{Schneider92}
{Schneider} P.,  {Ehlers} J.,   {Falco} E.~E.,  1992, {Gravitational Lenses},
  \mn@doi{10.1007/978-3-662-03758-4.
}

\bibitem[\protect\citeauthoryear{{S\'ersic}}{{S\'ersic}}{1968}]{Sersic68}
{S\'ersic} J.~L.,  1968, {Atlas de Galaxias Australes}.
\url {http://adsabs.harvard.edu/abs/1968adga.book.....S}

\bibitem[\protect\citeauthoryear{{Shajib}}{{Shajib}}{2019}]{Shajib19b}
{Shajib} A.~J.,  2019, \mn@doi [\mnras] {10.1093/mnras/stz1796}, \href
  {https://ui.adsabs.harvard.edu/abs/2019MNRAS.tmp.1753S} {488, 1387}

\bibitem[\protect\citeauthoryear{{Shajib}, {Treu}  \& {Agnello}}{{Shajib}
  et~al.}{2018}]{Shajib18}
{Shajib} A.~J.,  {Treu} T.,   {Agnello} A.,  2018, \mn@doi [\mnras]
  {10.1093/mnras/stx2302}, \href
  {http://adsabs.harvard.edu/abs/2018MNRAS.473..210S} {473, 210}

\bibitem[\protect\citeauthoryear{{Shajib} et~al.,}{{Shajib}
  et~al.}{2019}]{Shajib19}
{Shajib} A.~J.,  et~al., 2019, \mn@doi [\mnras] {10.1093/mnras/sty3397}, \href
  {http://adsabs.harvard.edu/abs/2019MNRAS.483.5649S} {483, 5649}

\bibitem[\protect\citeauthoryear{{Shakura} \& {Sunyaev}}{{Shakura} \&
  {Sunyaev}}{1973}]{Shakura73}
{Shakura} N.~I.,  {Sunyaev} R.~A.,  1973, \aap, \href
  {https://ui.adsabs.harvard.edu/abs/1973A%26A....24..337S} {24, 337}

\bibitem[\protect\citeauthoryear{{Shen} et~al.,}{{Shen} et~al.}{2011}]{Shen11}
{Shen} Y.,  et~al., 2011, \mn@doi [\apjs] {10.1088/0067-0049/194/2/45}, \href
  {https://ui.adsabs.harvard.edu/abs/2011ApJS..194...45S} {194, 45}

\bibitem[\protect\citeauthoryear{{Skilling}}{{Skilling}}{2004}]{Skilling04}
{Skilling} J.,  2004, in {Fischer} R.,  {Preuss} R.,   {Toussaint} U.~V.,  eds,
   American Institute of Physics Conference Series Vol. 735, American Institute
  of Physics Conference Series. pp 395--405, \mn@doi{10.1063/1.1835238}

\bibitem[\protect\citeauthoryear{{Sluse}, {Hutsem{\'e}kers}, {Courbin},
  {Meylan}  \& {Wambsganss}}{{Sluse} et~al.}{2012}]{Sluse12b}
{Sluse} D.,  {Hutsem{\'e}kers} D.,  {Courbin} F.,  {Meylan} G.,   {Wambsganss}
  J.,  2012, \mn@doi [\aap] {10.1051/0004-6361/201219125}, \href
  {https://ui.adsabs.harvard.edu/abs/2012A%26A...544A..62S} {544, A62}

\bibitem[\protect\citeauthoryear{{Sluse} et~al.,}{{Sluse}
  et~al.}{2019}]{Sluse19}
{Sluse} D.,  et~al., 2019, \mn@doi [\mnras] {10.1093/mnras/stz2483}, \href
  {https://ui.adsabs.harvard.edu/abs/2019arXiv190508800S} {}

\bibitem[\protect\citeauthoryear{{Sonnenfeld}, {Treu}, {Gavazzi}, {Marshall},
  {Auger}, {Suyu}, {Koopmans}  \& {Bolton}}{{Sonnenfeld}
  et~al.}{2012}]{Sonnenfeld12}
{Sonnenfeld} A.,  {Treu} T.,  {Gavazzi} R.,  {Marshall} P.~J.,  {Auger} M.~W.,
  {Suyu} S.~H.,  {Koopmans} L.~V.~E.,   {Bolton} A.~S.,  2012, \mn@doi [\apj]
  {10.1088/0004-637X/752/2/163}, \href
  {http://adsabs.harvard.edu/abs/2012ApJ...752..163S} {752, 163}

\bibitem[\protect\citeauthoryear{{Sonnenfeld}, {Leauthaud}, {Auger}, {Gavazzi},
  {Treu}, {More}  \& {Komiyama}}{{Sonnenfeld} et~al.}{2018}]{Sonnenfeld18b}
{Sonnenfeld} A.,  {Leauthaud} A.,  {Auger} M.~W.,  {Gavazzi} R.,  {Treu} T.,
  {More} S.,   {Komiyama} Y.,  2018, \mn@doi [\mnras] {10.1093/mnras/sty2262},
  \href {http://adsabs.harvard.edu/abs/2018MNRAS.481..164S} {481, 164}

\bibitem[\protect\citeauthoryear{{Soto}, {Lilly}, {Bacon}, {Richard}  \&
  {Conseil}}{{Soto} et~al.}{2016}]{Soto16}
{Soto} K.~T.,  {Lilly} S.~J.,  {Bacon} R.,  {Richard} J.,   {Conseil} S.,
  2016, \mn@doi [\mnras] {10.1093/mnras/stw474}, \href
  {https://ui.adsabs.harvard.edu/abs/2016MNRAS.458.3210S} {458, 3210}

\bibitem[\protect\citeauthoryear{{Spiniello} et~al.,}{{Spiniello}
  et~al.}{2018}]{Spiniello18}
{Spiniello} C.,  et~al., 2018, \mn@doi [\mnras] {10.1093/mnras/sty1923}, \href
  {https://ui.adsabs.harvard.edu/abs/2018MNRAS.480.1163S} {480, 1163}

\bibitem[\protect\citeauthoryear{{Springel} et~al.,}{{Springel}
  et~al.}{2005}]{Springel05}
{Springel} V.,  et~al., 2005, \mn@doi [\nat] {10.1038/nature03597}, \href
  {https://ui.adsabs.harvard.edu/abs/2005Natur.435..629S} {435, 629}

\bibitem[\protect\citeauthoryear{{Suyu} \& {Halkola}}{{Suyu} \&
  {Halkola}}{2010}]{Suyu10b}
{Suyu} S.~H.,  {Halkola} A.,  2010, \mn@doi [\aap]
  {10.1051/0004-6361/201015481}, \href
  {https://ui.adsabs.harvard.edu/abs/2010A%26A...524A..94S} {524, A94}

\bibitem[\protect\citeauthoryear{{Suyu}, {Marshall}, {Auger}, {Hilbert},
  {Blandford}, {Koopmans}, {Fassnacht}  \& {Treu}}{{Suyu}
  et~al.}{2010}]{Suyu10}
{Suyu} S.~H.,  {Marshall} P.~J.,  {Auger} M.~W.,  {Hilbert} S.,  {Blandford}
  R.~D.,  {Koopmans} L.~V.~E.,  {Fassnacht} C.~D.,   {Treu} T.,  2010, \mn@doi
  [\apj] {10.1088/0004-637X/711/1/201}, \href
  {http://adsabs.harvard.edu/abs/2010ApJ...711..201S} {711, 201}

\bibitem[\protect\citeauthoryear{{Suyu} et~al.,}{{Suyu} et~al.}{2013}]{Suyu13}
{Suyu} S.~H.,  et~al., 2013, \mn@doi [\apj] {10.1088/0004-637X/766/2/70}, \href
  {http://adsabs.harvard.edu/abs/2013ApJ...766...70S} {766, 70}

\bibitem[\protect\citeauthoryear{{Suyu} et~al.,}{{Suyu} et~al.}{2014}]{Suyu14}
{Suyu} S.~H.,  et~al., 2014, \mn@doi [\apjl] {10.1088/2041-8205/788/2/L35},
  \href {http://adsabs.harvard.edu/abs/2014ApJ...788L..35S} {788, L35}

\bibitem[\protect\citeauthoryear{{Suyu} et~al.,}{{Suyu} et~al.}{2017}]{Suyu17}
{Suyu} S.~H.,  et~al., 2017, \mn@doi [\mnras] {10.1093/mnras/stx483}, \href
  {http://adsabs.harvard.edu/abs/2017MNRAS.468.2590S} {468, 2590}

\bibitem[\protect\citeauthoryear{{Tewes} et~al.,}{{Tewes}
  et~al.}{2013}]{Tewes13}
{Tewes} M.,  et~al., 2013, \mn@doi [\aap] {10.1051/0004-6361/201220352}, \href
  {https://ui.adsabs.harvard.edu/abs/2013A%26A...556A..22T} {556, A22}

\bibitem[\protect\citeauthoryear{{Tie} \& {Kochanek}}{{Tie} \&
  {Kochanek}}{2018}]{Tie18}
{Tie} S.~S.,  {Kochanek} C.~S.,  2018, \mn@doi [\mnras]
  {10.1093/mnras/stx2348}, \href
  {https://ui.adsabs.harvard.edu/abs/2018MNRAS.473...80T} {473, 80}

\bibitem[\protect\citeauthoryear{{Tinker}, {Kravtsov}, {Klypin}, {Abazajian},
  {Warren}, {Yepes}, {Gottl{\"o}ber}  \& {Holz}}{{Tinker}
  et~al.}{2008}]{Tinker08}
{Tinker} J.,  {Kravtsov} A.~V.,  {Klypin} A.,  {Abazajian} K.,  {Warren} M.,
  {Yepes} G.,  {Gottl{\"o}ber} S.,   {Holz} D.~E.,  2008, \mn@doi [\apj]
  {10.1086/591439}, \href
  {https://ui.adsabs.harvard.edu/abs/2008ApJ...688..709T} {688, 709}

\bibitem[\protect\citeauthoryear{Towns et~al.,}{Towns et~al.}{2014}]{Towns14}
Towns J.,  et~al., 2014, \mn@doi [Computing in Science \& Engineering]
  {10.1109/MCSE.2014.80}, 16, 62

\bibitem[\protect\citeauthoryear{{Treu} \& {Koopmans}}{{Treu} \&
  {Koopmans}}{2002}]{Treu02b}
{Treu} T.,  {Koopmans} L.~V.~E.,  2002, \mn@doi [\mnras]
  {10.1046/j.1365-8711.2002.06107.x}, \href
  {http://adsabs.harvard.edu/cgi-bin/nph-bib_query?bibcode=2002MNRAS.337L...6T&db_key=AST}
  {337, L6}

\bibitem[\protect\citeauthoryear{{Treu} \& {Marshall}}{{Treu} \&
  {Marshall}}{2016}]{Treu16}
{Treu} T.,  {Marshall} P.~J.,  2016, \mn@doi [The Astronomy and Astrophysics
  Review] {10.1007/s00159-016-0096-8}, \href
  {http://adsabs.harvard.edu/abs/2016arXiv160505333T} {24, 11}

\bibitem[\protect\citeauthoryear{{Treu}, {Koopmans}, {Bolton}, {Burles}  \&
  {Moustakas}}{{Treu} et~al.}{2006}]{Treu06}
{Treu} T.,  {Koopmans} L.~V.,  {Bolton} A.~S.,  {Burles} S.,   {Moustakas}
  L.~A.,  2006, \mn@doi [\apj] {10.1086/500124}, \href
  {https://ui.adsabs.harvard.edu/abs/2006ApJ...640..662T} {640, 662}

\bibitem[\protect\citeauthoryear{{Treu} et~al.,}{{Treu} et~al.}{2016}]{Treu16b}
{Treu} T.,  et~al., 2016, \mn@doi [\apj] {10.3847/0004-637X/817/1/60}, \href
  {https://ui.adsabs.harvard.edu/abs/2016ApJ...817...60T} {817, 60}

\bibitem[\protect\citeauthoryear{{Treu} et~al.,}{{Treu} et~al.}{2018}]{Treu18}
{Treu} T.,  et~al., 2018, \mn@doi [\mnras] {10.1093/mnras/sty2329}, \href
  {http://adsabs.harvard.edu/abs/2018MNRAS.481.1041T} {481, 1041}

\bibitem[\protect\citeauthoryear{{Verde}, {Treu}  \& {Riess}}{{Verde}
  et~al.}{2019}]{Verde19}
{Verde} L.,  {Treu} T.,   {Riess} A.~G.,  2019, \mn@doi [Nature Astronomy]
  {10.1038/s41550-019-0902-0}, \href
  {https://ui.adsabs.harvard.edu/abs/2019arXiv190710625V} {3, 891}

\bibitem[\protect\citeauthoryear{{Vestergaard} \& {Peterson}}{{Vestergaard} \&
  {Peterson}}{2006}]{Vestergaard06}
{Vestergaard} M.,  {Peterson} B.~M.,  2006, \mn@doi [\apj] {10.1086/500572},
  \href {https://ui.adsabs.harvard.edu/abs/2006ApJ...641..689V} {641, 689}

\bibitem[\protect\citeauthoryear{Waskom et~al.,}{Waskom
  et~al.}{2014}]{Waskom14}
Waskom M.,  et~al., 2014, seaborn: v0.5.0 (November 2014),
  \mn@doi{10.5281/zenodo.12710}, \url {https://doi.org/10.5281/zenodo.12710}

\bibitem[\protect\citeauthoryear{{Wong} et~al.,}{{Wong} et~al.}{2017}]{Wong17}
{Wong} K.~C.,  et~al., 2017, \mn@doi [\mnras] {10.1093/mnras/stw3077}, \href
  {http://adsabs.harvard.edu/abs/2017MNRAS.465.4895W} {465, 4895}

\bibitem[\protect\citeauthoryear{{Wong} et~al.,}{{Wong} et~al.}{2019}]{Wong19}
{Wong} K.~C.,  et~al., 2019, arXiv e-prints, \href
  {https://ui.adsabs.harvard.edu/abs/2019arXiv190704869W} {p. arXiv:1907.04869}

\bibitem[\protect\citeauthoryear{{Woo}, {Le}, {Karouzos}, {Park}, {Park},
  {Malkan}, {Treu}  \& {Bennert}}{{Woo} et~al.}{2018}]{Woo18}
{Woo} J.-H.,  {Le} H.~A.~N.,  {Karouzos} M.,  {Park} D.,  {Park} D.,  {Malkan}
  M.~A.,  {Treu} T.,   {Bennert} V.~N.,  2018, \mn@doi [\apj]
  {10.3847/1538-4357/aabf3e}, \href
  {https://ui.adsabs.harvard.edu/abs/2018ApJ...859..138W} {859, 138}

\bibitem[\protect\citeauthoryear{{Y{\i}ld{\i}r{\i}m}, {Suyu}  \&
  {Halkola}}{{Y{\i}ld{\i}r{\i}m} et~al.}{2019}]{Yildirim19}
{Y{\i}ld{\i}r{\i}m} A.,  {Suyu} S.~H.,   {Halkola} A.,  2019, arXiv e-prints,
  \href {https://ui.adsabs.harvard.edu/abs/2019arXiv190407237Y} {p.
  arXiv:1904.07237}

\bibitem[\protect\citeauthoryear{{Yuan}, {Riess}, {Macri}, {Casertano}  \&
  {Scolnic}}{{Yuan} et~al.}{2019}]{Yuan19}
{Yuan} W.,  {Riess} A.~G.,  {Macri} L.~M.,  {Casertano} S.,   {Scolnic} D.,
  2019, arXiv e-prints, \href
  {https://ui.adsabs.harvard.edu/abs/2019arXiv190800993Y} {p. arXiv:1908.00993}

\bibitem[\protect\citeauthoryear{{Zahid}, {Geller}, {Fabricant}  \&
  {Hwang}}{{Zahid} et~al.}{2016}]{Zahid16}
{Zahid} H.~J.,  {Geller} M.~J.,  {Fabricant} D.~G.,   {Hwang} H.~S.,  2016,
  \mn@doi [\apj] {10.3847/0004-637X/832/2/203}, \href
  {https://ui.adsabs.harvard.edu/abs/2016ApJ...832..203Z} {832, 203}

\bibitem[\protect\citeauthoryear{{Zhao}}{{Zhao}}{1996}]{Zhao96}
{Zhao} H.,  1996, \mn@doi [\mnras] {10.1093/mnras/278.2.488}, \href
  {http://adsabs.harvard.edu/abs/1996MNRAS.278..488Z} {278, 488}

\makeatother
\end{thebibliography}



\appendix

\section{Impact of fiducial cosmology in the lens modelling} \label{app:effect_cosmology}
\begin{figure}
	\includegraphics[width=0.5\columnwidth]{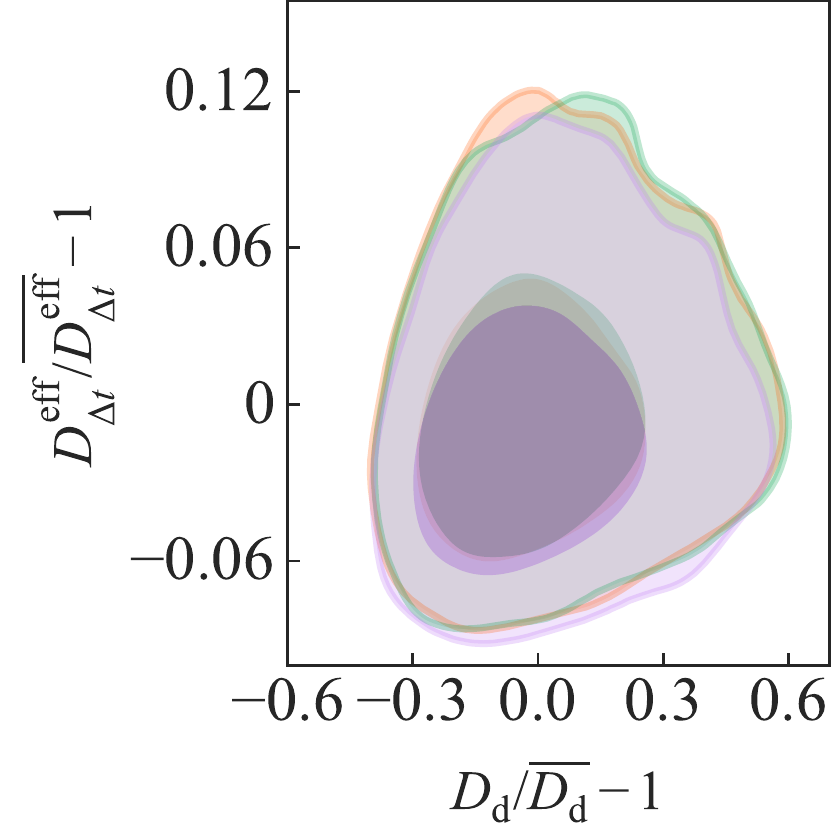}
	\includegraphics[width=0.5\columnwidth]{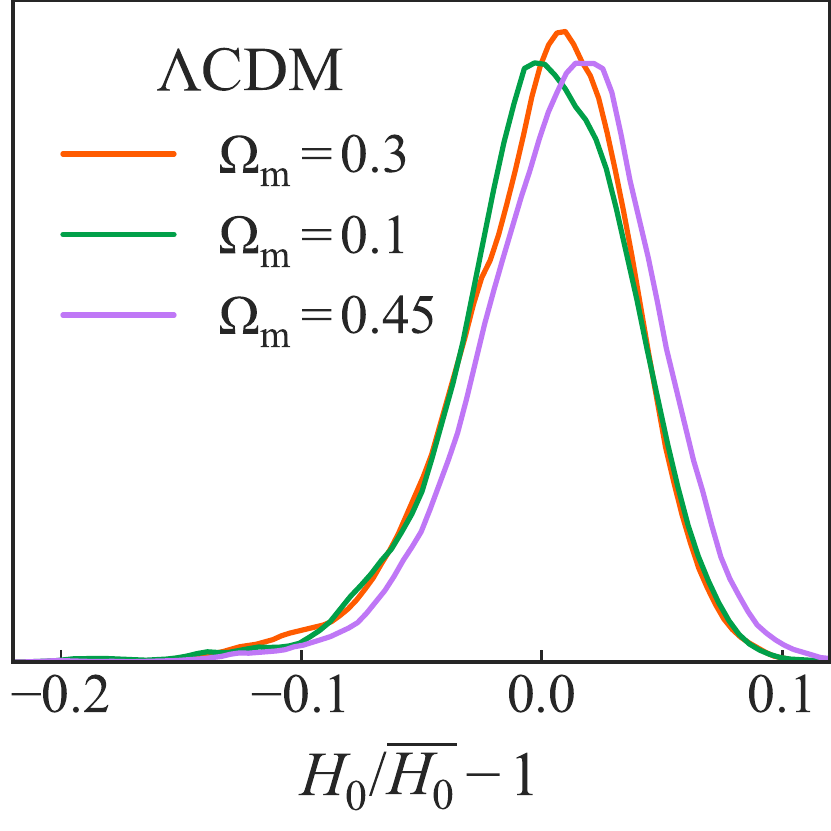}	
	\caption{\label{fig:cosmo_compare_distance}
	Comparison of the distance posteriors and inferred \Ho\ for different fiducial cosmologies within the flat \lcdm\ model. We compare between three set of cosmological parameters: {$\Omega_{\rm m}=0.3$, $\Omega_{\Lambda}=0.7$} (orange), {$\Omega_{\rm m}=0.1$, $\Omega_{\Lambda}=0.9$} (green), and {$\Omega_{\rm m}=0.45$, $\Omega_{\Lambda}=0.55$} (purple). The distance posteriors are from identical lens model setups with the power-law mass profile except for the fiducial cosmology. \Ho\ shifts by less than 1 per cent within these wide range of $\Omega_{\rm m}$ values. This shift is much smaller than the precision on \Ho\ allowed by our current data quality. As a result, we can treat the distance posteriors inferred from our analysis to be independent of cosmological assumptions within the flat \lcdm\ cosmology.
	}
\end{figure}

We check the impact of fixing the distance ratios between the lens and the source planes with a fiducial \lcdm\ cosmology with density parameters $\Omega_{\rm m}=0.3$ and $\Omega_{\Lambda}=0.7$. For \editsx{this purpose}, \editn{we run two separate lens models with the same model setup for the power-law mass profile, but with the cosmological parameters ($\Omega_{\rm m}=0.1$, $\Omega_{\Lambda}=0.9$) and ($\Omega_{\rm m}=0.45$, $\Omega_{\Lambda}=0.55$) to fix the distance ratios. Within this wide-range of $\Omega_{\rm m}$ values within the flat \lcdm\ cosmology, \Ho\ \editsx{only} shifts by $\lesssim$ 1 per cent (Fig. \ref{fig:cosmo_compare_distance}). As this shift is much smaller than the precision of the measured Hubble constant allowed by the quality of our data, we \editfv{conclude} that our inferred cosmological distance posterior on the $\Ddt$--$\Dd$ plane is \editfv{effectively} independent of the choice of cosmological parameters within the flat $\Lambda$CDM cosmology.}

\begin{figure}
	\includegraphics[width=0.5\columnwidth]{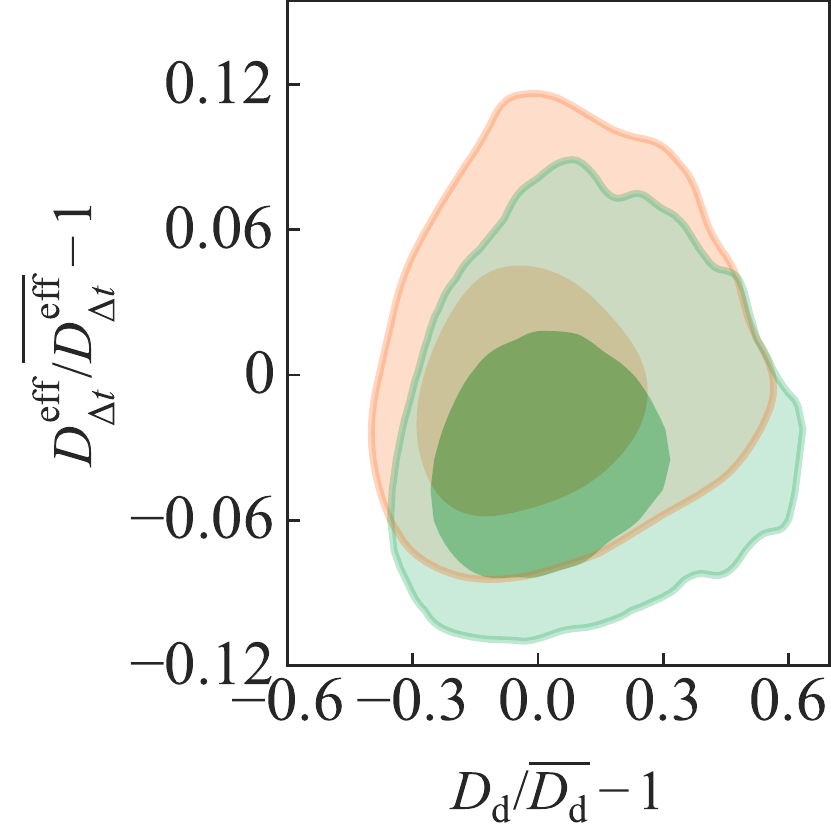}
	\includegraphics[width=0.5\columnwidth]{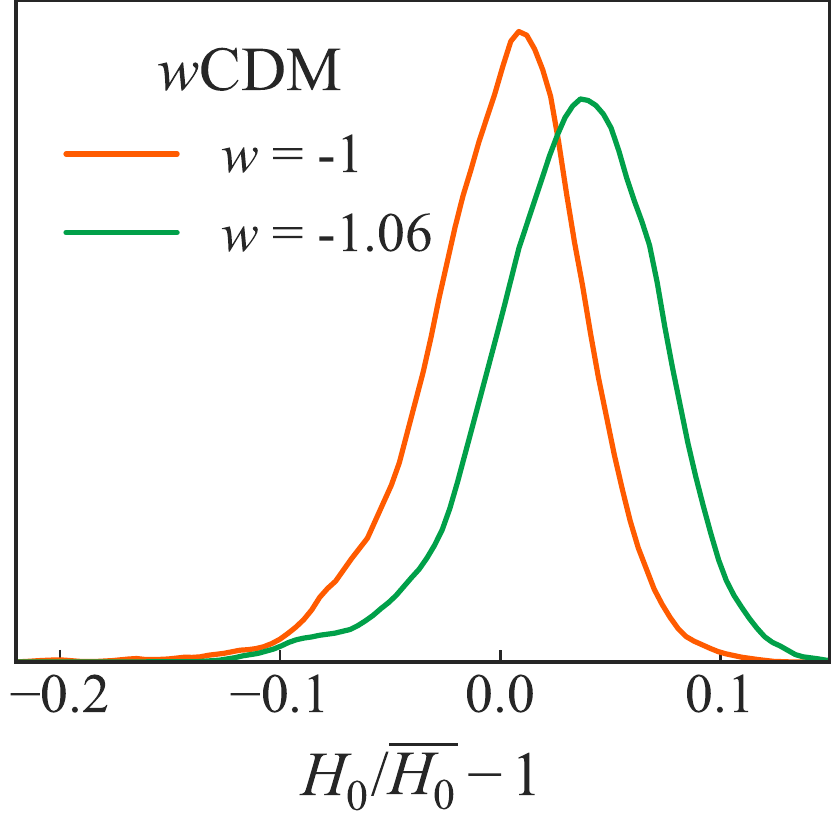}	
	\caption{\label{fig:cosmo_compare_distance_wcdm}
	Comparison of the distance posteriors and inferred \Ho\ for different fiducial cosmologies within the $w$CDM model. We compare between three set of cosmological parameters: {$w=-1$, $\Omega_{\rm m}=0.3$, $\Omega_{\Lambda}=0.7$} (orange), and {$w=-1.06$, $\Omega_{\rm m}=0.1$, $\Omega_{\Lambda}=0.9$} (green). The distance posteriors are from identical lens model setups with the power-law mass profile except for the fiducial cosmology. \Ho\ shifts by  $\sim$3 per cent for a shift $\Delta w=0.06$, which is approximately the joint precision on $w$ from the \textit{Planck} with CMB lensing, SNIae, and the baryon acoustic oscillation measurements \citep{PlanckCollaboration18}. This shift in \Ho\ shows that the double source plane treatment in our analysis is sensitive to the dark energy equation of state parameter $w$ \citep{Gavazzi08, Collett14}. As a result, our distance posterior should not be used to constrain parameters in cosmologies that extend the flat \lcdm\ model.
	}
\end{figure}

\editn{However, we find that the inferred distance posteriors \editsx{depend on} the fiducial cosmology within the $w$CDM model. If we adopt the fiducial $w$CDM cosmology with $w=-1.06$, $\Omega_{\rm m}=0.3$, $\Omega_{\rm de}=0.7$, then the inferred \Ho\ shifts by approximately 3 per cent from that inferred from the cosmology with $w=-1$ (Fig. \ref{fig:cosmo_compare_distance_wcdm}). We adopt the shift $\Delta w=0.06$ for comparison, as this range is in $w$ is the joint precision from the \textit{Planck} with CMB lensing, SNIae, and baryon acoustic oscillation measurements \citep{PlanckCollaboration18}. This significant shift in \Ho\ demonstrate that our double source plane treatment is sensitive to the dark energy equation of state parameter $w$ \citep[e.g.][]{Gavazzi08, Collett12, Collett14}. Therefore, our distance posteriors should not be used to constrain cosmological parameters in extended cosmologies other than the flat \lcdm\ model. \editfv{We leave the computation of a posterior distribution function valid in more general cosmologies for future work.}
}

\section{Impact of likelihood computation region choice} \label{app:effect_mask}

\begin{figure}
	\includegraphics[width=\columnwidth]{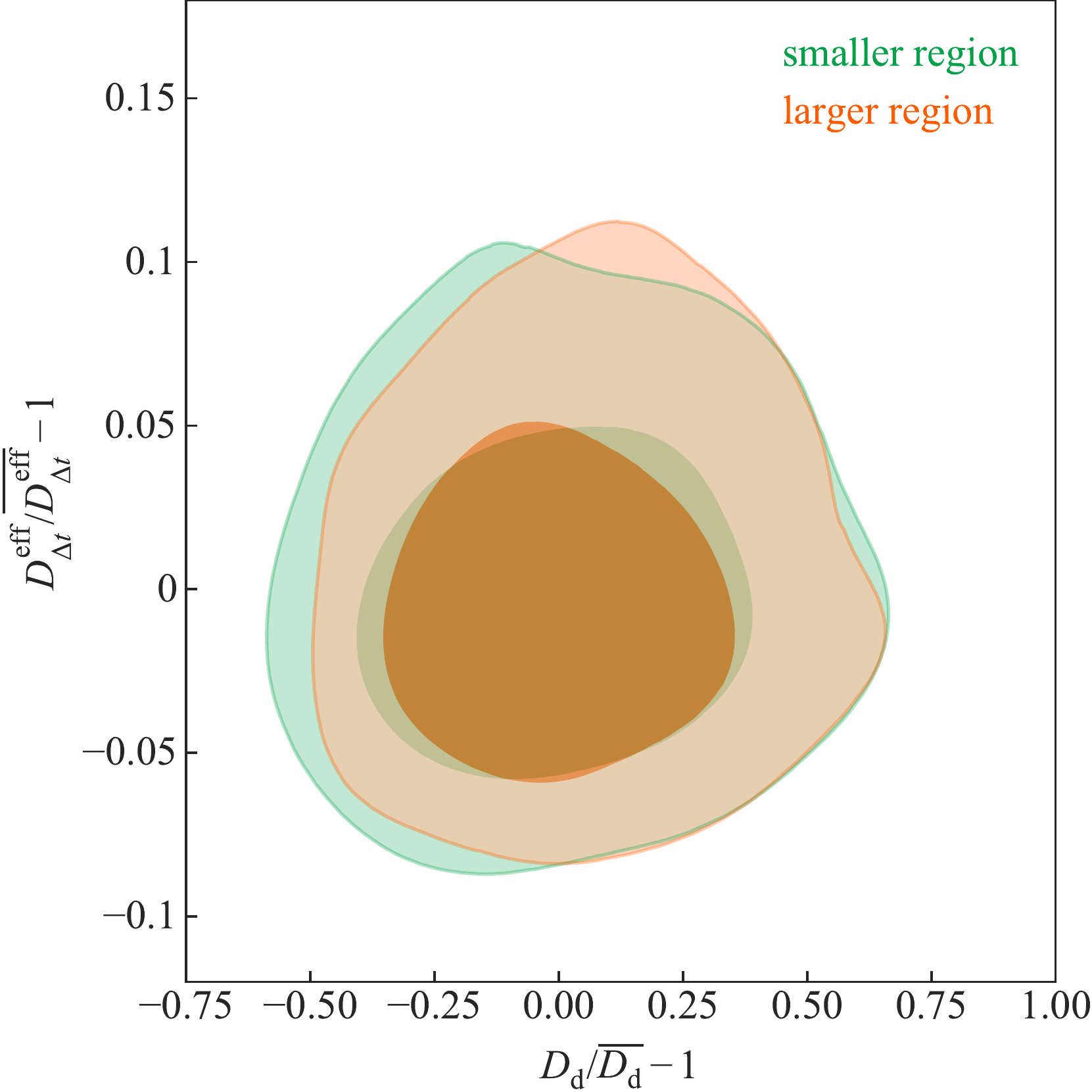}	
	\caption{\label{fig:mask_compare_distance}
	Comparison of the distance posteriors from the power-law lens model for two likelihood computation region sizes. The smaller regions have radii 4.3, 3.3, and 3.3 arcsec in the F160W, F814W, F475X bands, respectively. The larger region sizes are 4.5, 3.5, and 3.5 arcsec in the same order. All the other model setups are same between the two runs. The distance posteriors are almost identical. Therefore, the choice of likelihood computation region has negligible impact in our analysis.
	}
\end{figure}

We check if our adopted region for imaging likelihood computation can be a source of systematic bias in the lens modelling. We perform the modelling procedure for two different region sizes keeping every other settings in the model the same for a power-law model. The regular region sizes are 4.3, 3.3, and 3.3 arcsec in radius for the F160W, F814W, and F475X bands, respectively. The larger region sizes are larger by 0.2 arcsec in each band.  The median of the effective time-delay distance shifts by less than 0.1 per cent and the median of the angular diameter distance shifts by less than 2 per cent \editth{between these two choices of the likelihood computation region} (Fig. \ref{fig:mask_compare_distance}).

\section{Impact of the convergence from the group containing G1} \label{app:impact_group}
\begin{figure}
	\includegraphics[width=\columnwidth]{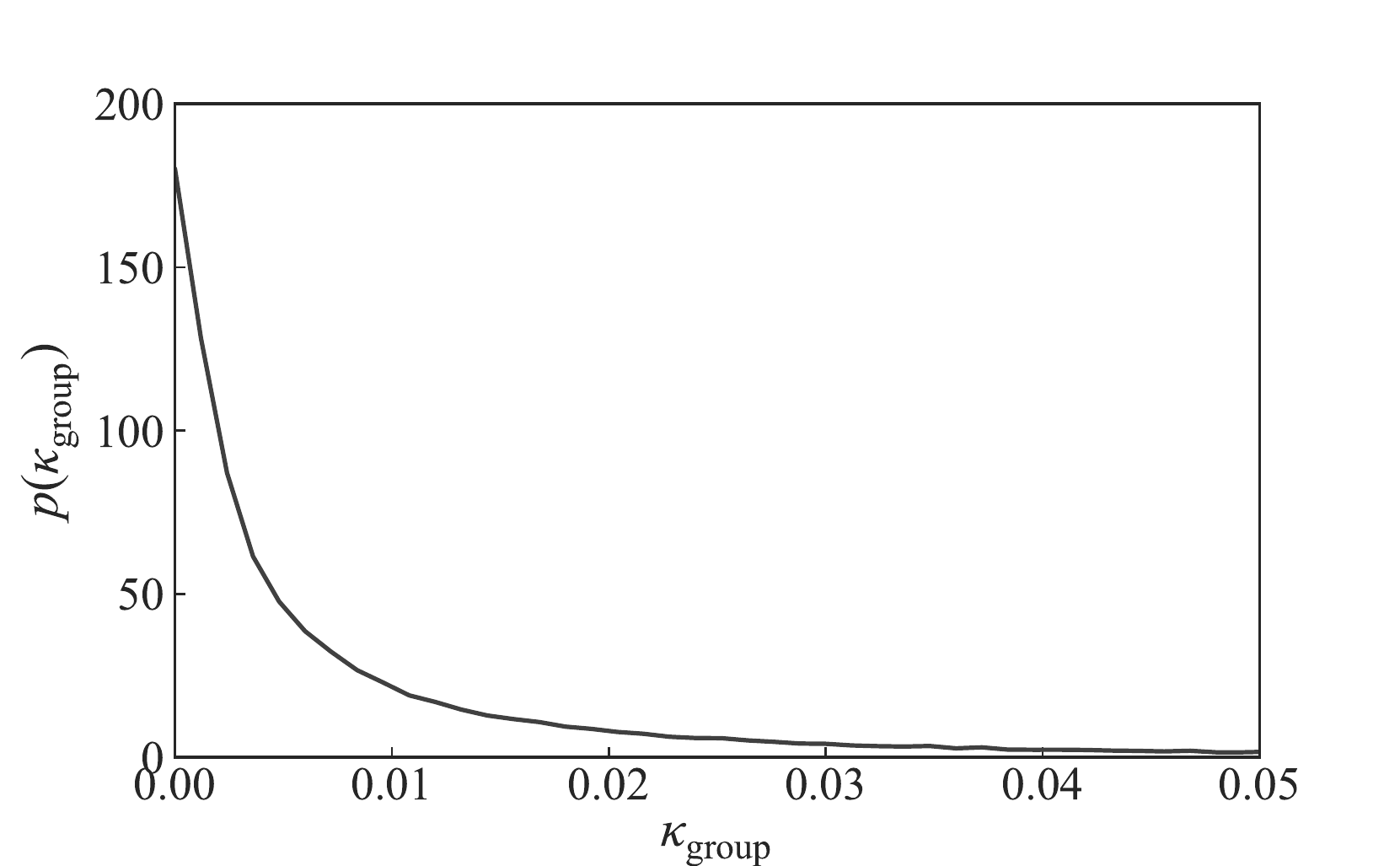}
	\caption{\label{fig:k_group}
	Distribution of the convergence at G1's centre from the galaxy group's halo containing G1. The distribution $p(\kappa_{\rm group})$ is estimated for the estimated centroid and velocity dispersion of this group in \citet{BuckleyGeer20}.
	}
\end{figure}

\editfv{We estimate the convergence at G1's centre from the galaxy group containing G1 [group ID 5 in \citet{BuckleyGeer20}]. We randomly sample halos from the centroid and velocity dispersion distributions \editsx{estimated as described in} \citet{BuckleyGeer20}. \editsx{However, we use a uniform prior to obtain the velocity dispersion for the halo, whereas \citet{BuckleyGeer20} adopt the Jeffrey's prior.} We convert the group's velocity dispersion into halo mass using the scaling relation}
\begin{equation}
	\log_{10} \left[h(z) M_{200}\right] = 13.98 + 2.75 \log_{10} \left(\frac{\sigma_{\rm group}}{500\ \text{km\ s}^{-1}} \right)
\end{equation}
\editfv{\citep{Munari13}. We weight this halo mass distribution using the halo mass function from \citet{Tinker08} corresponding to our fiducial cosmology and the lens redshift. We obtain the concentration parameter distribution using the theoretical $M_{\rm 200}$--$c$ relation from \citet{Diemer19} with 0.16 dex uncertainty. We also apply 10 per cent uncertainty on $M_{200}$ to remove any strong dependency on $H_0$ through the fiducial cosmology. We compute the convergence distribution at G1's centre due \editfv{to} this distribution of the halo masses and we apply a cut in the group's shear distribution $\gamma_{\rm group} < 0.1$ to remove halos that are inconsistent with the model predicted shear (Fig. \ref{fig:k_group}). The median of the group's convergence distribution is 0.004. \editsx{As we are explicitly accounting for the group's convergence here, we re-estimate $\kappa_{\rm ext}$ after removing the galaxies in this group from the number count statistic of \citet{BuckleyGeer20}. The re-estimated $\kappa_{\rm ext}$ decreases by 0.005 for the power-law mass models and by 0.008 for the composite mass models. As a result, explicitly accounting for the group's convergence decreases \Ho\ by approximately 0.4 per cent. This shift is negligible compared to the 3.9 per cent uncertainty in our estimated value of \Ho.}}

\section{Checking for the existence and impact of a dark substructure} \label{app:effect_dark_perturber}

\citet{Agnello17} propose a possible dark subhalo near image D toward the North--East direction. We check the impact of such a dark substructure in our analysis by including a mass profile for the substructure in our lens model. We check with both SIS and spherical NFW profile for the substructure. We take a broad uniform prior of 0.8 arcsec $\times$ 0.8 arcsec for the centroid of the mass profile to encompass the possible position of the substructure given in fig. 9 of \citet{Agnello17}.

\begin{figure}
	\includegraphics[width=\columnwidth]{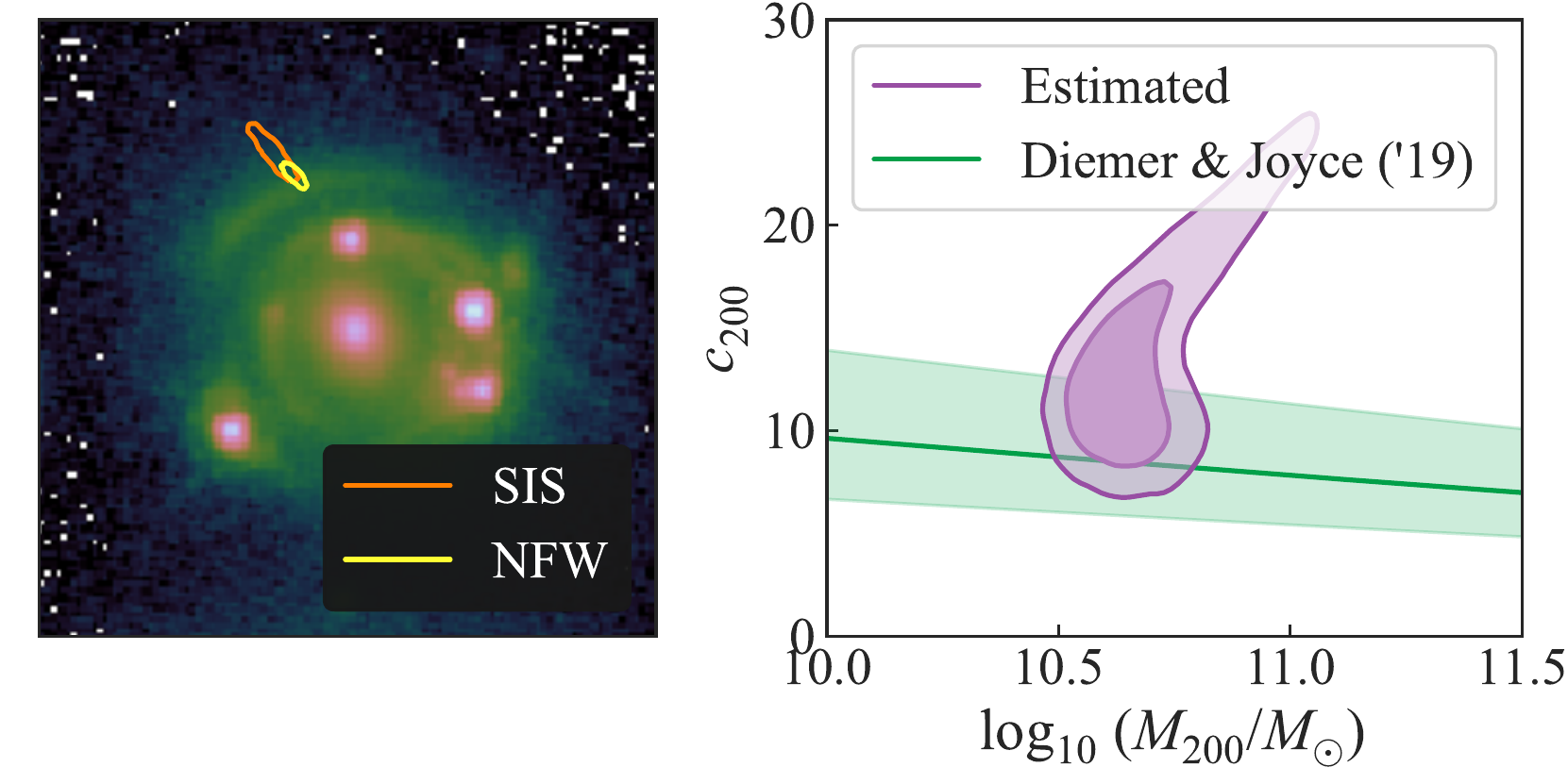}	
	\caption{\label{fig:dark_substructure}
	Constraints from our lens model for the position and NFW halo properties of the dark substructure proposed by \citet{Agnello17}. \textbf{Left-hand panel:} \editnn{2$\sigma$} credible region for the position of the dark substructure in our lens model assuming the SIS profile (the orange contour) and the NFW profile (the yellow contour) for the substructure. \textbf{Right-hand panel:} The constraint on the $M_{\rm 200}$--$c_{200}$ plane assuming the NFW profile for the dark substructure. The purple contours show the 1$\sigma$ and 2$\sigma$ credible regions. We show a comparison with the theoretical $c_{200}$--$M_{200}$ relation for our fiducial cosmology from \citet{Diemer19}.
	}
\end{figure}

For both of the SIS and NFW profiles, our lens model constrains the possible position of the mass profile for the \editth{potential} substructure (Fig. \ref{fig:dark_substructure}). Interestingly, the constrained position is consistent with the proposed position by \citet{Agnello17}, although the model had the freedom to offset the position by $\sim$0.4 arcsec from the constrained position. We estimate the SIS velocity dispersion of this possible dark substructure to be $\sigma_{\rm SIS}=33.7_{-1.3}^{+1.9}$ km s\textsuperscript{$-$1}. From the model with the NFW profile for the dark substructure, we estimate the halo mass $\log_{10} (M_{200}/M_{\odot})=10.65^{+0.10}_{-0.06}$, halo radius $r_{\rm 200}=45.7^{+3.5}_{-2.1}$ kpc, and concentration $c_{200}=12.2^{+4.1}_{-2.2}$.

However, we do not add the \editth{potential} substructure in our final lens model as the addition of the dark substructure shifts the estimated \Ho\ by less than 1 per cent. Moreover, it is not clear if the constrained parameters for the additional mass profile to account for the dark substructure indeed reflect the existence of the substructure. \editsx{A} similar effect can also arise if the additional source component S2 lies at a redshift between the quasar and the central deflector G1. The proximity of the constrained position of the dark substructure and the lensed position of S2 hints this scenario to be a possibility. 

\citet{Agnello17} use the dust-corrected and delay-corrected flux ratios observed in the DES data as a constraint for the lens model and propose that the existence of a dark substructure fits the data better. We check if microlensing can be a possible source for the deviation of the flux ratios from a smooth model observed by \citet{Agnello17}. We derive the amplitude of microlensing in images A, B, and D by comparing their MUSE spectra. Microlensing is stronger in the continuum than in the broad emission lines. Therefore, we can isolate the microlensed fraction of the spectra if we assume that microlensing is more important in one of the lensed images under scrutiny and derive a lower limit on the amplitude of microlensing effect in the continuum emission \citep[e.g.][]{Sluse12b}. This procedure reveals substantial differential microlensing between the continuum and the broad lines when we consider image pairs A--D and B--D, but not A--B. The data are compatible with a microlensing demagnification of image D by at least a factor of 2. This demagnification translates into a mircrolensing corrected flux ratio $\Delta m_{\rm AD}$ = 0.25 mag. This estimate, however, may be affected by systematic errors caused by intrinsic variability. From the past light-curves of this system, we estimate that over the period corresponding to the time delay between images A and D, this systematic error could reach up to 0.2 mag. \edit{Therefore, we cannot \editth{definitively} attribute the observed ``flux-ratio anomaly'' to microlensing. In summary, whereas we cannot find strong evidence for the existence of the \editth{potential} substructure, we also cannot rule out its existence.} \editth{Since the presence or the impact of the dark substructure is not significant in our analysis, we omit it in our lens models for simplicity.}

\section{Nested sampling settings} \label{app:dypolychord_setup}

In this Appendix, we provide our adopted settings for the nested sampling software \textsc{dypolychord} and \editnn{validate that the numerical requirements for our analysis are met}. 

\begin{figure}
	\includegraphics[width=\columnwidth]{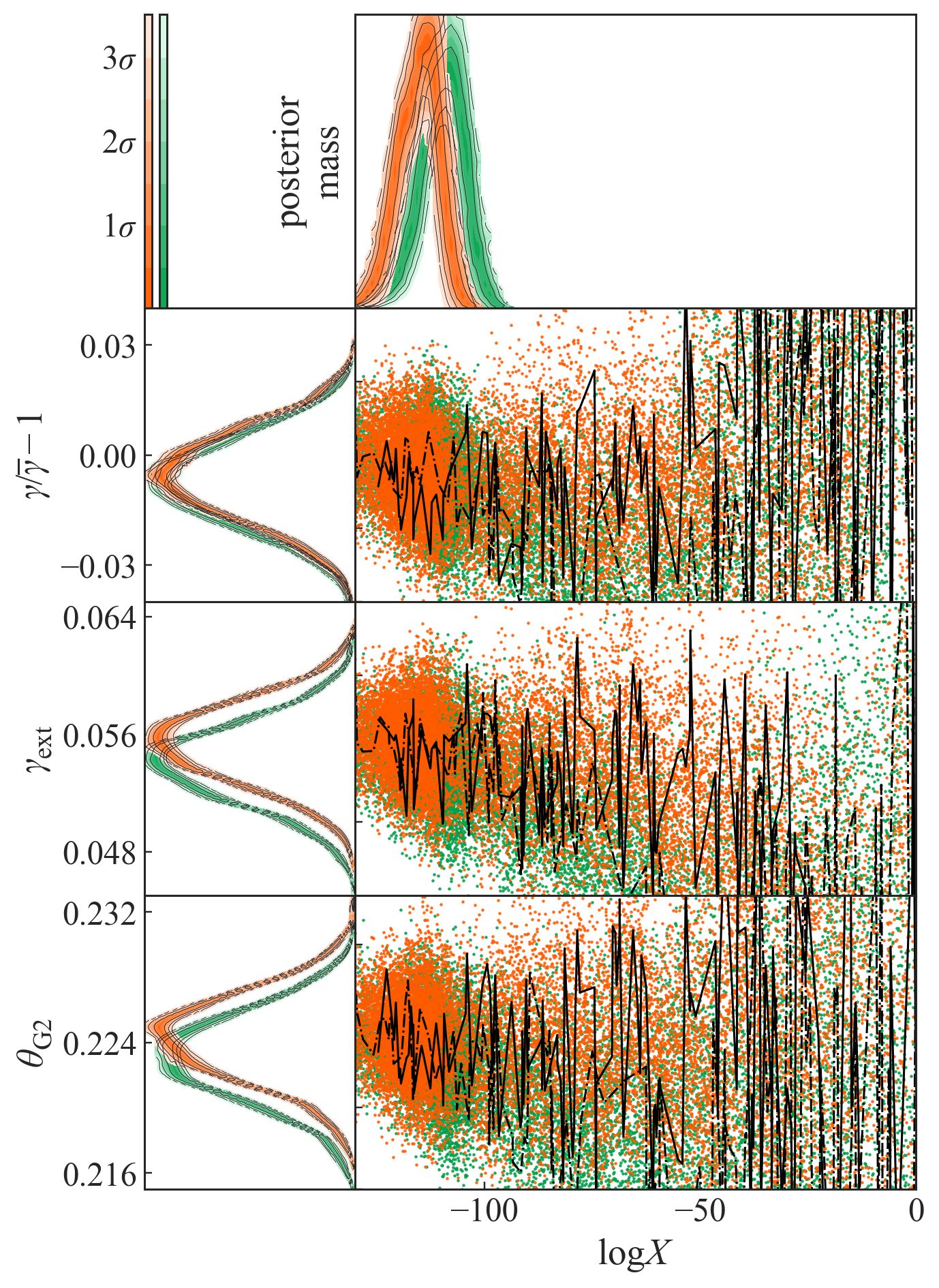}	
	\caption{\label{fig:compare_sampling_run}
	\edit{\textsc{nestcheck} diagnostic plot showing samples' distributions in two different runs with the same lens model for our chosen sampling settings. The top right-hand panel shows the relative posterior mass as a function of $\log X$, where $X$ is the prior volume fraction. The next three panels show the progressions of sampling (right to left) for three of the main lens mass parameters. The black solid and dashed lines show the evolution of one random thread of the dynamic nested sampler. The left column shows the posterior distributions of each parameter. The coloured contours represent the iso-probability credible regions of the probability density functions. The posteriors of the parameters are consistent within 1$\sigma$, therefore we accept that the chosen settings allow for sufficient exploration of the prior volume.}
	}
\end{figure}

We choose the \textsc{dypolychord} settings \texttt{ninit} = 100, \texttt{nrepeats} = 30, \texttt{nconst\_live} = 140, , \texttt{dynamic\_goal} = 0.9, \texttt{precision\_criterion} = 0.001 [see \citet{Higson19} for explanation of these settings]. \edit{To check the appropriateness of these settings, we run two sampling runs with the same lens model and sampler settings (Fig. \ref{fig:compare_sampling_run}). We find that the posteriors PDFs of the parameters are consistent within 1$\sigma$ between the two runs, therefore we accept the chosen settings to be appropriate for sufficient exploration of the prior space. However, we find the estimated evidence values to differ by more than the estimated statistical uncertainty. This difference indicates that there is a systematic scatter in the computed evidence value. To estimate this scatter, we run a second set of nested sampling runs for 17 different lens models with \texttt{precision\_criterion} = 0.01. We choose a lower \texttt{precision\_criterion} for this second set of runs to make the sampling run terminate faster. \editn{A lower \texttt{precision\_criterion} does not largely impact the evidence values, although it may affect the posterior estimation \citep{Higson18}. As we are only interested to obtain a conservative estimate of the scatter present in the computed $\log \mathcal{Z}$ values, this lower \texttt{precision\_criterion} is sufficient for this purpose.} By taking the mean of the evidence difference between runs from the two sets with the same lens model, we estimate the scatter in the evidence value as 24. Therefore, we take $\sigma^{\rm numeric}_{\log \mathcal{Z}}$ = 24 as the numerical error in the computed evidence values. Albeit, if we increase \texttt{nlive\_const} or \texttt{nrepeats}, we can decrease the error in the computed $\log \mathcal{Z}$ values in the exchange of a higher computational cost. However, as we down-weight the relative evidence ratios to account for sparse sampling of our models from the model space, this numerical error in $\log \mathcal{Z}$ is a subdominant factor (Section \ref{sec:modelling_workflow}).}

\section{Summary of uncertainty budget and systematic checks} \label{app:systematic_summary}
\editref{
In this appendix, we summarize the uncertainty budgets from different parts of the cosmographic analysis (Table \ref{tab:uncertainty_budget}) and the systematic checks performed in our analysis with their impacts on the inferred \Ho\ in Table \ref{tab:systematic_summary}.}

\begin{table}
	\caption{\label{tab:uncertainty_budget}
	\editref{Uncertainty contributions from different parts of the cosmographic analysis.}
	}
	\begin{tabular}{ll}
	\hline 
	Analysis component & Uncertainty budget (per cent) \\
	\hline 
	Time-delay measurement & 1.8 \\
	$\kappa_{\rm ext}$ estimation & 3.3 \\
	Lens modelling and other sources & 1.0 \\
	\textbf{Total uncertainty} & \textbf{3.9} \\
	\hline
	\end{tabular}
	
\end{table}

\begin{table}
	\caption{\label{tab:systematic_summary}
	\editref{Summary of systematic checks and their impacts on the inferred \Ho. The relevant section in the paper is referenced for each systematic check.}
	}
	\begin{tabular}{p{0.77\columnwidth}l}
	\hline 
	Systematic check &  $\Delta H_0$ \\
	& (per cent) \\
	\hline 
	Fixing $\Omega_{\rm m}=0.10$ in fiducial cosmology (App.~\ref{app:effect_cosmology}) & $-$0.06 \\
	Fixing $\Omega_{\rm m}=0.45$ in fiducial cosmology (App.~\ref{app:effect_cosmology}) & $+$0.94 \\
	Choosing a larger likelihood computation region (App.~\ref{app:effect_mask}) & $+$0.10 \\
	Explicitly accounting for the galaxy group (App.~\ref{app:impact_group}) & $-$0.40 \\
	Accounting for a possible dark subhalo near S3 (App.~\ref{app:effect_dark_perturber}) & $-$0.46 \\
	Not accounting for the deflection by S2's mass (Note, we included S2's mass in our models, thus already marginalizing in our quoted posterior; Section \ref{sec:souce_components}) & $+$0.95 \\
	Accounting for \citet{Tie18}'s microlensing time-delay (Section \ref{sec:microlensing}) & $-$0.10 \\
	\textbf{Total systematic shift} & \textbf{$-$0.92--0.08} \\
	\hline
	\end{tabular}
	
\end{table}

\section{Model summary and parameter priors} \label{app:param_prior}

\edit{In this appendix, we \editref{summarize the adopted models (Table \ref{tab:model_summary}) and} provide the priors for the parameters in our lens models (Table \ref{tab:param_prior}). }

To make the nested sampler efficiently explore and integrate over the high-dimensional ($\sim$60D) prior volume in our models, we narrow down the width of the uniform priors for some of the parameters more than that would be known purely \textit{a priori}. We choose these prior bounds by looking at the posterior PDFs for the lens models from the initial exploratory phase of this study and we further adjust these prior widths through trial and error. We check that the posterior PDFs of the parameters are fully contained within the chosen bounds for our final lens models, unless we specifically set the bound using an empirical or physical prior. Below we explain some of the parameters from Table \ref{tab:param_prior} that were not introduced within the main \editsx{body} of this paper.

\edit{The amplitude of the Chameleon convergence profile is parameterized with the deflection angle at 1 arcsec, $\alpha_1^{\rm Chm}$. The ellipticity parameters $e_1$ and $e_2$ in the relevant profiles are related to the axis ratio $q$ and the position angle $\phi$ as
\begin{linenomath}
	\begin{equation}
		\begin{split}
			&q = \frac{1 - \sqrt{e_1^2 + e_2^2}}{1 + \sqrt{e_1^2 + e_2^2}}, \\
			&\tan 2 \phi = \frac{e_2}{e_1}.
		\end{split}
	\end{equation}	
\end{linenomath}
Parameterizing the ellipticity with $e_1$ and $e_2$ avoids the periodicity in the polar coordinate $\phi$ and makes the sampling more efficient. The centroids $(\theta_1^{\rm c}, \theta_2^{\rm c})$ of the relevant profiles are relative to the coordinate RA 04:08:21.71 and Dec -53:53:59.34. We take a uniform prior $\mathcal{U}(-5\times10^4, 5\times10^4)$ for the amplitudes of the shapelet components -- which are linear parameters -- to compute the evidence using equation (\ref{eq:full_evidence_eqn}).
}

\begin{table}
	\caption{\label{tab:model_summary}
	\editref{Summary of model components. Alternative choices for some components are shown with bullet point lists.}
	}
	{
\begin{tabular}{p{0.2\columnwidth}p{0.7\columnwidth}}
	\hline 
	Component & Model description \\
	\hline
	G1 & 
	\begin{minipage}[t]{\linewidth}
	\begin{enumerate}[nosep,leftmargin=10pt,after=\strut]
	\item Elliptical power-law convergence, double elliptical S\'ersic light, external shear
	\item Elliptical NFW potential for dark matter, double elliptical Chameleon for luminous matter and light, external shear
	\end{enumerate} 
	\end{minipage}\\ \\
	G2 & SIS mass, S\'ersic light \\ \\
	G3 & SIS mass \\ \\
	G4, G5, G6 & {
		\begin{minipage}[t]{\linewidth}
		\begin{enumerate}[leftmargin=10pt]
 			\item SIS mass
 			\item NFW mass
 		\end{enumerate}
 		\end{minipage}
 		} \\ \\
	S1 & Elliptical S\'ersic light and shapelets on source plane  \\ \\
	S2 & Elliptical S\'ersic light and shapelets on source plane  \\ \\
	S3 & Elliptical S\'ersic light on source plane \\ \\
	Quasar images & Point sources on image plane \\ \\
	Cosmology &
	\begin{minipage}[t]{\linewidth}
	\lcdm \\ \Ho $\sim \mathcal{U}(0, 150)$ km s\textsuperscript{$-$1} Mpc\textsuperscript{$-$1} \\ $\Omega_{\rm m} \sim \mathcal{U}(0.05, 0.5)$
	\end{minipage} \\ \\
	\hline
\end{tabular}}
\end{table}

\onecolumn
\edit{
\setlength\LTleft{0pt}
\setlength\LTright{0pt}
\begin{longtable}{lll}
\caption{\label{tab:param_prior}
Prior for the model parameters.
} \\
\hline
Model component & Parameter & Prior \\
\hline
\endfirsthead
\multicolumn{3}{c}%
{\tablename\ \thetable\ -- \textit{Continued from previous page}} \\
\hline
Model component & Parameter & Prior \\
\hline
\endhead
\hline \multicolumn{3}{r}{\textit{Continued on next page}} \\
\endfoot
\hline
\endlastfoot
G1 mass, power law 		 & $\gamma$ 		 & $\mathcal{U}(1.80,\ 2.15)$ \\
G1 mass, composite 		 & $r_{\rm s}^{\rm NFW}$ (arcsec) 		 & $\mathcal{N}$(12.42, 2.94) with bound [5.6,\ 19.3) (Section \ref{sec:g1_mass_light}) \\
G1 mass, composite 		 & $\alpha_{1}^{\rm Chm}$ (arcsec) 		 & $\mathcal{U}(0.2,\ 1.5)$ \\
External shear 		 & $\gamma_{\rm ext}$ 		 & $\mathcal{U}(0.01,\ 0.09)$ \\
External shear 		 & $\psi_{\rm ext}$ (rad) 		 & $\mathcal{U}(0,\uppi)$ \\
G2 mass, SIS 		 & $\theta_{\rm E}$ (arcsec) 		 & $\mathcal{U}(0.19,\ 0.27)$ \\
G3 mass, SIS 		 & $\theta_{\rm E}$ (arcsec) 		 & $\mathcal{N}$(0.772, 0.024) with bound [0.3,\ 1.6) (Section \ref{sec:perturber_sis}, Fig. \ref{fig:perturber_prior}) \\
G4 mass, SIS 		 & $\theta_{\rm E}$ (arcsec) 		 & $\mathcal{N}$(0.353, 0.012) with bound [0.0,\ 1.0) (Section \ref{sec:perturber_sis}, Fig. \ref{fig:perturber_prior}) \\
G4 mass, NFW 		 & $\log_{10} (M_{200}/M_{\odot})$ 		 & Empirical prior with bound [11.3,\ 13.4) (Section \ref{sec:perturber_nfw}, Fig. \ref{fig:perturber_prior}) \\
G4 mass, NFW 		 & $c_{\rm 200}$ 		 & Empirical prior with bound [0.0,\ 16.0) (Section \ref{sec:perturber_nfw}, Fig. \ref{fig:perturber_prior}) \\
G5 mass, SIS 		 & $\theta_{\rm E}$ (arcsec) 		 & $\mathcal{N}$(0.046, 0.002) with bound [0.0,\ 0.2) (Section \ref{sec:perturber_sis}, Fig. \ref{fig:perturber_prior}) \\
G5 mass, NFW 		 & $\log_{10} (M_{200}/M_{\odot})$ 		 & Empirical prior with bound [10.8,\ 12.3) (Section \ref{sec:perturber_nfw}, Fig. \ref{fig:perturber_prior}) \\
G5 mass, NFW 		 & $c_{\rm 200}$ 		 & Empirical prior with bound [0.0,\ 16.5) (Section \ref{sec:perturber_nfw}, Fig. \ref{fig:perturber_prior}) \\
G6 mass, SIS 		 & $\theta_{\rm E}$ (arcsec) 		 & $\mathcal{N}$(0.070, 0.004) with bound [0.0,\ 0.3) (Section \ref{sec:perturber_sis}, Fig. \ref{fig:perturber_prior}) \\
G6 mass, NFW 		 & $\log_{10} (M_{200}/M_{\odot})$ 		 & Empirical prior with bound [11.4,\ 12.5) (Section \ref{sec:perturber_nfw}, Fig. \ref{fig:perturber_prior}) \\
G6 mass, NFW 		 & $c_{\rm 200}$ 		 & Empirical prior with bound [0.0,\ 20.0) (Section \ref{sec:perturber_nfw}, Fig. \ref{fig:perturber_prior}) \\
S2 mass, SIS 		 & $\theta_{\rm E}$ (arcsec) 		 & $\mathcal{N}$(0.0022, $9.98\times10^{-6}$) with bound [0.000,\ 0.022) (Section \ref{sec:perturber_sis}) \\
S1 light, S\'ersic 		 & $\theta_{\rm eff}$ (arcsec) 		 & $\mathcal{U}(0.04,\ 0.15)$ \\
S1 light, S\'ersic 		 & $n_{\rm s}$ 		 & $\mathcal{U}(0.6,\ 5.0)$ \\
S1 light, S\'ersic 		 & $e_1$ 		 & $\mathcal{U}(-0.05,\ 0.35)$ \\
S1 light, S\'ersic 		 & $e_2$ 		 & $\mathcal{U}(-0.16,\ 0.20)$ \\
S1 light, F814W+F475X, shapelets 		 & $\varsigma$ (arcsec) 		 & $\mathcal{U}(0.06,\ 0.11)$ \\
S1 light, F160W, shapelets 		 & $\varsigma$ (arcsec) 		 & $\mathcal{U}(0.08,\ 0.15)$ \\
S2 light, S\'ersic 		 & $\theta_{\rm eff}$ (arcsec) 		 & $\mathcal{U}(0.08,\ 0.40)$ \\
S2 light, S\'ersic 		 & $n_{\rm s}$ 		 & $\mathcal{U}(1.0,\ 5.0)$ \\
S2 light, S\'ersic 		 & $e_1$ 		 & $\mathcal{U}(0.04,\ 0.37)$ \\
S2 light, S\'ersic 		 & $e_2$ 		 & $\mathcal{U}(-0.20,\ 0.00)$ \\
S2 light 		 & $\theta_1^{\rm c}$ (arcsec) 		 & $\mathcal{U}(-0.47,\ 0.30)$ \\
S2 light 		 & $\theta_2^{\rm c}$ (arcsec) 		 & $\mathcal{U}(-2.48,\ -1.48)$ \\
S2 light, shapelets 		 & $\varsigma$ (arcsec) 		 & $\mathcal{U}(0.06,\ 0.12)$ \\
S3 light, S\'ersic 		 & $\theta_{\rm eff}$ (arcsec) 		 & $\mathcal{U}(0.18,\ 0.90)$ \\
S3 light, S\'ersic 		 & $n_{\rm s}$ 		 & $\mathcal{U}(0.6,\ 2.5)$ \\
S3 light, S\'ersic 		 & $e_1$ 		 & $\mathcal{U}(0.20,\ 0.42)$ \\
S3 light, S\'ersic 		 & $e_2$ 		 & $\mathcal{U}(0.00,\ 0.35)$ \\
S3 light 		 & $\theta_1^{\rm c}$ (arcsec) 		 & $\mathcal{U}(0.75,\ 1.80)$ \\
S3 light 		 & $\theta_2^{\rm c}$ (arcsec) 		 & $\mathcal{U}(1.70,\ 2.30)$ \\
G1 light, F814W 		 & $\theta_{\rm eff}$ (arcsec) 		 & $\mathcal{N}$(0.61, 0.27) with bound [0.1,\ 2.7) (Section \ref{sec:g1_mass_light}) \\
G1 light, F814W 		 & $n_{\rm s}$ 		 & $\mathcal{U}(2.0,\ 8.0)$ \\
G1 light, F814W+F475X 		 & $e_1$ 		 & $\mathcal{U}(-0.20,\ 0.00)$ \\
G1 light, F814W+F475X 		 & $e_2$ 		 & $\mathcal{U}(-0.05,\ 0.13)$ \\
G1 light 		 & $\theta_1^{\rm c}$ (arcsec) 		 & $\mathcal{U}(0.023,\ 0.035)$ \\
G1 light 		 & $\theta_1^{\rm c}$ (arcsec) 		 & $\mathcal{U}(-0.010,\ 0.003)$ \\
G1 light, F475X 		 & $\theta_{\rm eff}$ (arcsec) 		 & $\mathcal{N}$(0.61, 0.27) with bound [0.1,\ 2.7) (Section \ref{sec:g1_mass_light}) \\
G1 light, F475X 		 & $n_{\rm s}$ 		 & $\mathcal{U}(1.0,\ 5.0)$ \\
G1 light, F160W, S\'ersic 1 		 & $\theta_{\rm eff}$ (arcsec) 		 & $\mathcal{N}$(0.61, 0.27) with bound [0.1,\ 2.7) (Section \ref{sec:g1_mass_light}) \\
G1 light, F160W, S\'ersic 1 		 & $n_{\rm s}$ 		 & $\mathcal{U}(1.0,\ 5.0)$ \\
G1 light, F160W, S\'ersic 1 		 & $e_1$ 		 & $\mathcal{U}(-0.25,\ 0.00)$ \\
G1 light, F160W, S\'ersic 1 		 & $e_2$ 		 & $\mathcal{U}(-0.10,\ 0.12)$ \\
G1 light, F160W, S\'ersic 2 		 & $\theta_{\rm eff}$ (arcsec) 		 & $\mathcal{N}$(0.61, 0.27) with bound [0.1,\ 2.7) (Section \ref{sec:g1_mass_light}) \\
G1 light, F160W, S\'ersic 2 		 & $n_{\rm s}$ 		 & $\mathcal{U}(0.6,\ 6.0)$ \\
G1 light, F160W, S\'ersic 2 		 & $e_1$ 		 & $\mathcal{U}(-0.05,\ 0.10)$ \\
G1 light, F160W, S\'ersic 2 		 & $e_2$ 		 & $\mathcal{U}(0.07,\ 0.25)$ \\
G1 light, double Chameleon 		 & $I_{\rm 0, Chm1}/I_{\rm 0, Chm2}$ 		 & $\mathcal{U}(0.2,\ 9.5)$ \\
G1 light, Chameleon 1 		 & $w_{\rm c}$ (arcsec) 		 & $\mathcal{U}(0.00,\ 0.10)$ \\
G1 light, Chameleon 1 		 & $w_{\rm t}$ (arcsec) 		 & $\mathcal{U}(0.20,\ 1.00)$ \\
G1 light, Chameleon 1 		 & $e_1$ 		 & $\mathcal{U}(-0.25,\ 0.25)$ \\
G1 light, Chameleon 1 		 & $e_2$ 		 & $\mathcal{U}(-0.25,\ 0.25)$ \\
G1 light, Chameleon 2 		 & $w_{\rm c}$ (arcsec) 		 & $\mathcal{U}(0.01,\ 1.50)$ \\
G1 light, Chameleon 2 		 & $w_{\rm t}$ (arcsec) 		 & $\mathcal{U}(2.50,\ 9.00)$ \\
G1 light, Chameleon 2 		 & $e_1$ 		 & $\mathcal{U}(-0.20,\ 0.20)$ \\
G1 light, Chameleon 2 		 & $e_2$ 		 & $\mathcal{U}(-0.20,\ 0.20)$ \\
G2 light 		 & $\theta_{\rm eff}$ (arcsec) 		 & $\mathcal{U}(0.25,\ 1.10)$ \\
G2 light 		 & $n_{\rm s}$ 		 & $\mathcal{U}(2.0,\ 6.0)$ \\
G2 light 		 & $\theta_1^{\rm c}$ (arcsec) 		 & $\mathcal{U}(-1.60,\ -1.56)$ \\
G2 light 		 & $\theta_2^{\rm c}$ (arcsec) 		 & $\mathcal{U}(-0.97,\ -0.93)$ \\
Image A 		 & $\Delta \alpha$ (arcsec) 		 & $\mathcal{U}(1.940,\ 1.948)$ \\
Image A 		 & $\Delta \delta$ (arcsec) 		 & $\mathcal{U}(-1.576,\ -1.568)$ \\
Image B 		 & $\Delta \alpha$ (arcsec) 		 & $\mathcal{U}(-1.819,\ -1.809)$ \\
Image B 		 & $\Delta \delta$ (arcsec) 		 & $\mathcal{U}(0.263,\ 0.290)$ \\
Image C 		 & $\Delta \alpha$ (arcsec) 		 & $\mathcal{U}(-1.935,\ -1.926)$ \\
Image C 		 & $\Delta \delta$ (arcsec) 		 & $\mathcal{U}(-0.954,\ -0.940)$ \\
Image D 		 & $\Delta \alpha$ (arcsec) 		 & $\mathcal{U}(0.096,\ 0.110)$ \\
Image D 		 & $\Delta \delta$ (arcsec) 		 & $\mathcal{U}(1.385,\ 1.392)$ \\
Differential dust extinction 		 & $\tau_{0}^{\rm F814W}$ 		 & $\mathcal{U}(0.1,\ 2.0)$ \\
\hline
\end{longtable}
}
\twocolumn

\noindent
\textbf{Affiliations} \\
$^{1}$Department of Physics and Astronomy, University of California, Los Angeles, CA 90095-1547, USA (Email: \href{mailto:ajshajib@astro.ucla.edu}{ajshajib@astro.ucla.edu}) \\
$^{2}$Kavli Institute for Particle Astrophysics and Cosmology and Department of Physics, Stanford University, Stanford, CA 94305, USA \\
$^{3}$Packard Fellow \\ 
$^{4}$DARK, Niels Bohr Institute, University of Copenhagen, Lyngbyvej 2, DK-2100 Copenhagen, Denmark \\
$^{5}$Fermi National Accelerator Laboratory, P. O. Box 500, Batavia, IL 60510, USA \\
$^{6}$Institute of Physics, Laboratoire d'Astrophysique, Ecole Polytechnique F\'ed\'erale de Lausanne (EPFL), Observatoire de Sauverny, CH-1290 Versoix, Switzerland \\
$^{7}$Department of Astronomy \& Astrophysics, University of Chicago, Chicago, IL 60637 \\
$^{8}$Kavli Institute for Cosmological Physics, University of Chicago, Chicago, IL 60637\\
$^{9}$National Astronomical Observatory of Japan, 2-21-1 Osawa, Mitaka, Tokyo 181-8588, Japan \\
$^{10}$STAR Institute, Quartier Agora - All\'ee du six Ao\^ut, 19c B-4000 Li\`ege, Belgium\\
$^{11}$INAF - Osservatorio Astronomico di Capodimonte, Salita Moiariello, 16, I-80131 Napoli, Italy \\
$^{12}$European Southern Observatory, Karl-Schwarschild-Str. 2, 85748 Garching, Germany \\
$^{13}$Physics Department, UC Davis, 1 Shields Ave., Davis, CA 95616 \\
$^{14}$Institute of Cosmology and Gravitation, University of Portsmouth, Portsmouth, PO1 3FX, UK \\
$^{15}$Departamento de Ciencias Fisicas, Universidad Andres Bello, Fernandez Concha 700, Las Condes, Santiago, Chile \\
$^{16}$Kavli IPMU (WPI), UTIAS, The University of Tokyo, Kashiwa, Chiba 277-8583, Japan \\
$^{17}$Cerro Tololo Inter-American Observatory, National Optical Astronomy Observatory, Casilla 603, La Serena, Chile \\
$^{18}$Instituto de Fisica Teorica UAM/CSIC, Universidad Autonoma de Madrid, 28049 Madrid, Spain \\
$^{19}$LSST, 933 North Cherry Avenue, Tucson, AZ 85721, USA \\
$^{20}$Physics Department, 2320 Chamberlin Hall, University of Wisconsin-Madison, 1150 University Avenue Madison, WI  53706-1390 \\
$^{21}$Department of Physics \& Astronomy, University College London, Gower Street, London, WC1E 6BT, UK \\
$^{22}$Department of Physics and Astronomy, University of Pennsylvania, Philadelphia, PA 19104, USA \\
$^{23}$Kavli Institute for Particle Astrophysics \& Cosmology, P. O. Box 2450, Stanford University, Stanford, CA 94305, USA \\
$^{24}$SLAC National Accelerator Laboratory, Menlo Park, CA 94025, USA \\
$^{25}$Centro de Investigaciones Energ\'eticas, Medioambientales y Tecnol\'ogicas (CIEMAT), Madrid, Spain \\
$^{26}$Laborat\'orio Interinstitucional de e-Astronomia - LIneA, Rua Gal. Jos\'e Cristino 77, Rio de Janeiro, RJ - 20921-400, Brazil \\
$^{27}$Department of Astronomy, University of Illinois at Urbana-Champaign, 1002 W. Green Street, Urbana, IL 61801, USA \\
$^{28}$National Center for Supercomputing Applications, 1205 West Clark St., Urbana, IL 61801, USA \\
$^{29}$Institut de F\'{\i}sica d'Altes Energies (IFAE), The Barcelona Institute of Science and Technology, Campus UAB, 08193 Bellaterra (Barcelona) Spain \\
$^{30}$Institut d'Estudis Espacials de Catalunya (IEEC), 08034 Barcelona, Spain \\
$^{31}$Institute of Space Sciences (ICE, CSIC),  Campus UAB, Carrer de Can Magrans, s/n,  08193 Barcelona, Spain \\
$^{32}$INAF-Osservatorio Astronomico di Trieste, via G. B. Tiepolo 11, I-34143 Trieste, Italy \\
$^{33}$Institute for Fundamental Physics of the Universe, Via Beirut 2, 34014 Trieste, Italy \\
$^{34}$Observat\'orio Nacional, Rua Gal. Jos\'e Cristino 77, Rio de Janeiro, RJ - 20921-400, Brazil \\
$^{35}$Department of Physics, IIT Hyderabad, Kandi, Telangana 502285, India \\
$^{36}$Excellence Cluster Origins, Boltzmannstr.\ 2, 85748 Garching, Germany \\
$^{37}$Faculty of Physics, Ludwig-Maximilians-Universit\"at, Scheinerstr. 1, 81679 Munich, Germany \\
$^{38}$Department of Astronomy, University of Michigan, Ann Arbor, MI 48109, USA \\
$^{39}$Department of Physics, University of Michigan, Ann Arbor, MI 48109, USA \\
$^{40}$Instituto de Fisica Teorica UAM/CSIC, Universidad Autonoma de Madrid, 28049 Madrid, Spain \\
$^{41}$Department of Physics, Stanford University, 382 Via Pueblo Mall, Stanford, CA 94305, USA
$^{42}$Santa Cruz Institute for Particle Physics, Santa Cruz, CA 95064, USA \\
$^{43}$Center for Cosmology and Astro-Particle Physics, The Ohio State University, Columbus, OH 43210, USA \\
$^{44}$Department of Physics, The Ohio State University, Columbus, OH 43210, USA \\
$^{45}$Center for Astrophysics $\vert$ Harvard \& Smithsonian, 60 Garden Street, Cambridge, MA 02138, USA \\
$^{46}$Department of Astronomy/Steward Observatory, University of Arizona, 933 North Cherry Avenue, Tucson, AZ 85721-0065, USA \\
$^{47}$Department of Astrophysical Sciences, Princeton University, Peyton Hall, Princeton, NJ 08544, USA \\
$^{48}$Observatories of the Carnegie Institution for Science, 813 Santa Barbara St., Pasadena, CA 91101, USA \\
$^{49}$Departamento de F\'isica Matem\'atica, Instituto de F\'isica, Universidade de S\~ao Paulo, CP 66318, S\~ao Paulo, SP, 05314-970, Brazil \\
$^{50}$George P. and Cynthia Woods Mitchell Institute for Fundamental Physics and Astronomy, and Department of Physics and Astronomy, Texas A\&M University, College Station, TX 77843,  USA \\
$^{51}$Instituci\'o Catalana de Recerca i Estudis Avan\c{c}ats, E-08010 Barcelona, Spain \\
$^{52}$Department of Physics and Astronomy, Pevensey Building, University of Sussex, Brighton, BN1 9QH, UK
$^{53}$Instituto de F\'\i sica, UFRGS, Caixa Postal 15051, Porto Alegre, RS - 91501-970, Brazil \\
$^{54}$Department of Physics, Duke University Durham, NC 27708, USA \\
$^{55}$Institut d'Estudis Espacials de Catalunya (IEEC), 08034 Barcelona, Spain \\
$^{56}$School of Physics and Astronomy, University of Southampton,  Southampton, SO17 1BJ, UK \\
$^{57}$Brandeis University, Physics Department, 415 South Street, Waltham MA 02453 \\
$^{58}$Computer Science and Mathematics Division, Oak Ridge National Laboratory, Oak Ridge, TN 37831 \\
$^{59}$The Inter-University Center for Astronomy and Astrophysics, Post bag 4, Ganeshkhind, Pune, 411007, India \\
$^{60}$Institute of Astronomy, Madingley Rd, Cambridge, CB3 0HA, UK \\
$^{61}$Kavli Institute for Cosmology, University of Cambridge, Madingley Road, Cambridge, CB3 0HA, UK


\label{lastpage}
\end{document}